%% file: NSoliton.tex
\documentclass{amsart} 
\usepackage{fullpage,amssymb,amsmath,color,graphicx}
\newtheorem{proposition}{Proposition}
\newtheorem{theorem}{Theorem}

\title{The $N$-Soliton of the Focusing Nonlinear Schr\"odinger
Equation for $N$ Large} 

\author{Gregory Lyng}
\email{glyng@umich.edu}
\author{Peter D. Miller}
\email{millerpd@umich.edu}
\address{\begin{tabular}{l}Department of Mathematics\\University of Michigan\\Ann Arbor,
  MI 48109\end{tabular}} 
\date{July 22, 2005.}

\begin{document} 
\begin{abstract}
  We present a detailed analysis of the solution of the focusing
  nonlinear Schr\"odinger equation with initial condition
  $\psi(x,0)=N\,{\rm sech}(x)$ in the limit $N\rightarrow\infty$.  We
  begin by presenting new and more accurate numerical reconstructions
  of the $N$-soliton by inverse scattering (numerical linear algebra)
  for $N=5$, $10$, $20$, and $40$.  We then recast the
  inverse-scattering problem as a Riemann-Hilbert problem and provide
  a rigorous asymptotic analysis of this problem in the large-$N$
  limit.  For those $(x,t)$ where results have been obtained by other
  authors, we improve the error estimates from $O(N^{-1/3})$ to
  $O(N^{-1})$.  We also analyze the Fourier power spectrum in this
  regime and relate the results to the optical phenomenon of
  supercontinuum generation.  We then study the $N$-soliton for values
  of $(x,t)$ where analysis has not been carried out before, and we
  uncover new phenomena.  The main discovery of this paper is the
  mathematical mechanism for a secondary caustic (phase transition),
  which turns out to differ from the mechanism that generates the
  primary caustic.  The mechanism for the generation of the secondary
  caustic depends essentially on the discrete nature of the spectrum
  for the $N$-soliton, and more significantly, cannot be recovered
  from an analysis of an ostensibly similar Riemann-Hilbert problem in
  the conditions of which a certain formal continuum limit is taken on
  an {\em ad hoc} basis.
\end{abstract}

\maketitle 

\section{Introduction}
\input{section1}



\section{Linear Systems Describing the $N$-Soliton}\label{sec:linearsystem}
\subsection{Theory.}
The spectral data
\eqref{eq:evals}--\eqref{eq:proportionalityconstants} can be used to
set up a system of linear equations whose solution yields the
$N$-soliton for arbitrary $(x,t)$.  (See, for example, the discussion
in Chapter 2 of \cite{KMM}.)  To arrive at such a square inhomogeneous
system, we begin by defining expressions
\begin{align}
A(x,t,\lambda)& :=\sum_{p=0}^{N-1}A_p(x,t)\lambda^p, \\
C(x,t,\lambda)& :=\lambda^N+\sum_{p=0}^{N-1}C_p(x,t)\lambda^p, \\
\intertext{and}
F_{N,k}(x,t)&:=\exp\big(-2i(\lambda_{N,k}x+\lambda_{N,k}^2t)\hbar_N\big).
\end{align}
Here, the unknown coefficients $A_0(x,t),A_1(x,t),\ldots,A_{N-1}(x,t)$ 
and $C_0(x,t),C_1(x,t),\ldots,C_{N-1}(x,t)$
are determined by the spectral data according to the relations
\begin{align}
A(x,t,\lambda_{N,k})F_{N,k}(x,t)&=\gamma_{N,k}C(x,t,\lambda_{N,k}),\;k=0\ldots,N-1,\label{neweq1}\\
C(x,t,\lambda_{N,k}^*)F_{N,k}(x,t)^* & =
-\gamma_{N,k}^*A(x,t,\lambda_{N,k}^*),\;k=0,\ldots,N-1.\label{neweq2}
\end{align}
The $N$-soliton solution of the focusing nonlinear Schr\"odinger
equation is then given by 
\begin{equation}
\psi_N(x,t)=2i A_{N-1}(x,t).
\label{psidef}
\end{equation}
Clearly, the equations \eqref{neweq1} and \eqref{neweq2} amount to a
$2N\times 2N$ system of linear equations for the coefficients
$A_p(x,t)$ and $C_p(x,t)$, with coefficient matrices of
block-Vandermonde type.  One approach to computing the value of
$\psi_N(x,t)$ for fixed $(x,t)$ is then to solve this linear algebra
problem and then obtain $\psi_N(x,t)$ from \eqref{psidef}.  This
approach was used in \cite{MK} to find $\psi_N(x,t)$ on a grid of 
values of $(x,t)$ for $N=5$, $N=10$, and $N=20$.  There is an advantage
here over direct numerical simulation of the focusing nonlinear Schr\"odinger
equation in that numerical errors do not propagate from one value of $t$ to
another.  

While the above approach is attractive, it can be made more so by
reducing the problem explicitly to a linear algebra problem involving
matrices of size only $N\times N$.  We use Lagrange interpolation and
\eqref{neweq1} to express $C(x,t,\lambda)$ in terms of the values
$A(x,t,\lambda_{N,k}),\,k=0,\ldots,N-1$. Thus,
\begin{equation}
C(x,t,\lambda)  = \lambda^N+\sum_{n=0}^{N-1}
\left[\left[\frac{F_{N,n}(x,t)}{\gamma_{N,n}}A(x,t,\lambda_{N,n})-\lambda_{N,n}^N\right]
\prod_{\stackrel{\scriptstyle j=0}{j\neq n}}^{N-1}
\frac{\lambda-\lambda_{N,j}}{\lambda_{N,n}-\lambda_{N,j}}\right].
\label{Cinterp}
\end{equation}
Now we note that
\begin{equation}
\lambda^N-\sum_{n=0}^{N-1}\lambda_{N,n}^N
\prod_{\stackrel{\scriptstyle j=0}{j\neq
    n}}^{N-1}\frac{\lambda-\lambda_{N,j}}{\lambda_{N,n}-\lambda_{N,j}}
=\prod_{j=0}^{N-1}(\lambda-\lambda_{N,j}),
\label{identity}
\end{equation}
since the left-hand side is a monic polynomial of degree $N$ which
vanishes precisely at each the values $\lambda_{N,0},\ldots,\lambda_{N,N-1}$.
Therefore, from \eqref{Cinterp} and \eqref{identity} we write 
\begin{equation}
C(x,t,\lambda)=\sum_{n=0}^{N-1}\left[\left[\frac{F_{N,n}(x,t)}{\gamma_{N,n}}
A(x,t,\lambda_{N,n})\right]\prod_{\stackrel{\scriptstyle j=0}{j\neq n}}^{N-1}
\frac{\lambda-\lambda_{N,j}}{\lambda_{N,n}-\lambda_{N,j}}\right]
+\prod_{j=0}^{N-1}(\lambda-\lambda_{N,j}).
\label{cexpr}
\end{equation}
We evaluate \eqref{cexpr} at $\lambda_{N,k}^*$ and substitute into
\eqref{neweq2}:
\begin{multline}
\left\{\sum_{n=0}^{N-1}\left[\left[\frac{F_{N,n}(x,t)}{\gamma_{N,n}}
A(x,t,\lambda_{N,n})\right]\prod_{\stackrel{\scriptstyle j=0}{j\neq n}}^{N-1}
\frac{\lambda_{N,k}^*-\lambda_{N,j}}{\lambda_{N,n}-\lambda_{N,j}}\right]
+\prod_{j=0}^{N-1}(\lambda_{N,k}^*-\lambda_{N,j})\right\}F_{N,k}(x,t)^*=\\
-\gamma_{N,k}^*A(x,t,\lambda_{N,k}^*),\;k=0,\ldots,N-1.\label{Aeq}
\end{multline}
Using the Lagrange interpolation formula once again, we write $A$ in
terms of its values at $\lambda_{N,0}^*,\ldots,\lambda_{N,N-1}^*$:
\begin{equation}
A(x,t,\lambda)=\sum_{m=0}^{N-1}\left[A(x,t,\lambda_{N,m}^*)
\prod_{\stackrel{\scriptstyle j=0}{j\neq m}}^{N-1}
\frac{\lambda-\lambda_{N,j}^*}{\lambda_{N,m}^*-\lambda_{N,j}^*}\right].
\label{Ainterp}
\end{equation}
Evaluating \eqref{Ainterp} at $\lambda_{N,n}$ and substituting into
\eqref{Aeq}, we obtain 
\begin{equation}
A(x,t,\lambda_{N,k}^*)
+\sum_{m=0}^{N-1}w_{N,km}A(x,t,\lambda_{N,k}^*)=\phi_{N,k},\;
k=0,\ldots,N-1,
\label{linearsytem}
\end{equation}
where 
\begin{align*}
G_{N,k}(x,t) & :=\frac{F_{N,k}(x,t)}{\gamma_{N,k}},\\
\phi_{N,k} & : = -G_{N,k}(x,t)^*\prod_{j=0}^{N-1}(\lambda_{N,k}^*-\lambda_{N,j}),\\
p_{N,kj} & := \prod_{\stackrel{\scriptstyle m=0}{m\neq j}}^{N-1}
\frac{\lambda_{N,k}-\lambda_{N,m}^*}{\lambda_{N,j}^*-\lambda_{N,m}^*},\\
\intertext{and}
w_{N,km} & : =  G_{N,k}(x,t)^*\sum_{n=0}^{N-1}G_{N,n}(x,t)p_{N,nm}p_{N,kn}^{*}.
\end{align*}
We denote by $\mathbf{W}_N$ the $N\times N$ matrix with entries
$w_{N,km}$. The matrix $\mathbf{W}_N$ has a
representation as 
$$\mathbf{W}_N=\mathbf{B}_N^*\mathbf{B}_N,$$
where $B_{N,jk}:=G_{N,j}(x,t)p_{N,jk}$. The linear system to solve is thus
\begin{equation}
(\mathbb{I}+\mathbf{W}_N)\vec{A}=(\mathbb{I}+\mathbf{B}_N^*\mathbf{B}_N)\vec{A}=\vec{\Phi},
\label{system}
\end{equation}
where
$\vec{A}:=(A(x,t,\lambda_{N,0}^*),\ldots,A(x,t,\lambda_{N,N-1}^*))^{T}$
and $\vec{\Phi}:=(\phi_{N,0},\ldots,\phi_{N,N-1})^T$.
Given a solution $\vec{A}$ of \eqref{system}, we recover the
solution $\psi_N$ using \eqref{psidef} via 
\begin{equation}
\psi_N=2i\vec{A}\cdot\vec{V},  
\label{findpsi}
\end{equation}
where $\vec{V}:=(V_{N,0},\ldots,V_{N,N-1})^T$ and 
\begin{equation}
V_{N,k}:=\prod_{\stackrel{\scriptstyle n=0}{ n\neq k}}^{N-1}\frac{1}{\lambda_{N,k}^*-\lambda_{N,n}^*}.
\end{equation}
This follows from \eqref{Ainterp}, since
\begin{align*}
A(x,t,\lambda)&=\sum_{k=0}^{N-1}A(x,t,\lambda_{N,k}^*)\left[\prod_{\stackrel{\scriptstyle
    n=0}{
    n\neq k}}^{N-1}\frac{\lambda-\lambda_{N,n}^*}{\lambda_{N,k}^*-\lambda_{N,n}}\right] \\
           &=\sum_{k=0}^{N-1}A(x,t,\lambda_{N,k}^*)\left[\prod_{\stackrel{\scriptstyle n=0}{
    n\neq k}}^{N-1}\frac{1}{\lambda_{N,k}^*-\lambda_{N,n}}\lambda^{N-1}+O(\lambda^{N-2})\right],\;\text{as}\;\lambda\to\infty.
\end{align*}
\subsection{Numerical implementation.}
\input{section2-2}

\begin{figure}[htbp]
\begin{center}
\includegraphics[width=3 in]{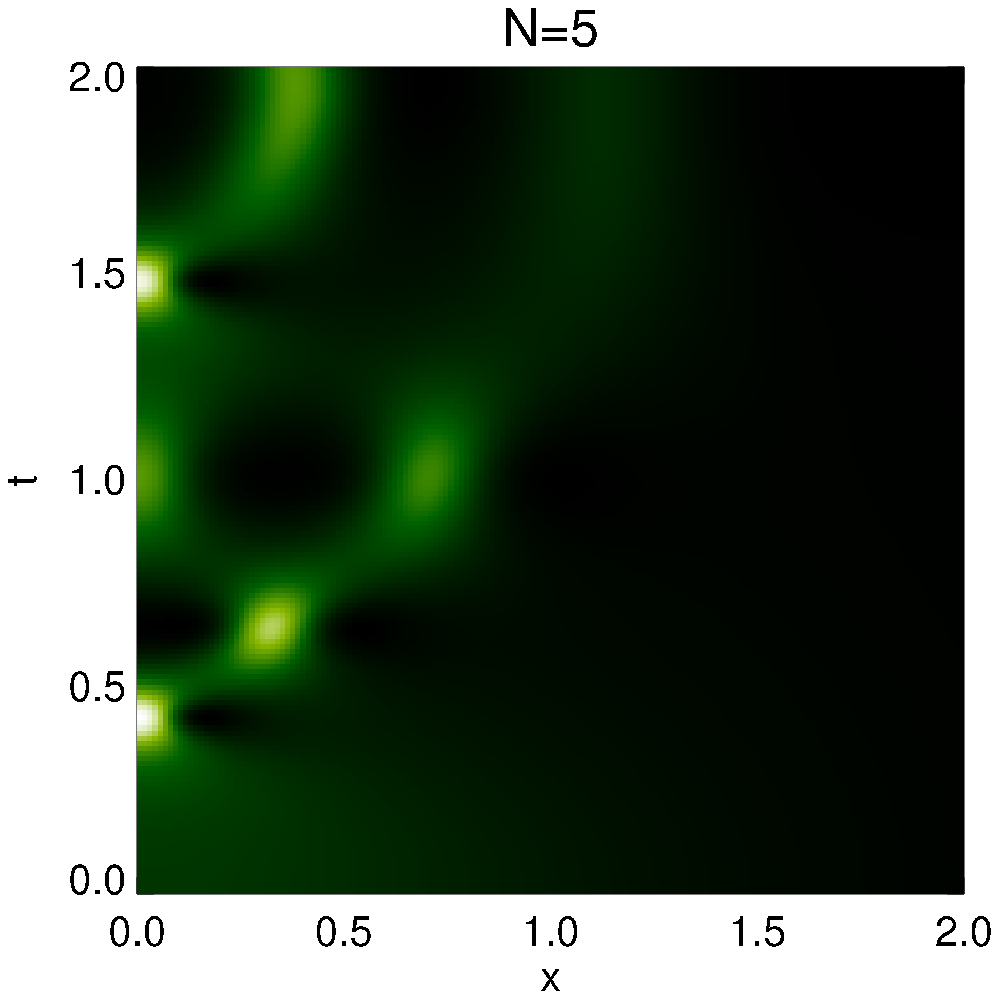}
\end{center}
\caption{\em The square modulus of $\psi_5(x,t)$ plotted over a part of the
positive quadrant of the $(x,t)$-plane.}
\label{fig:5soliton}
\end{figure}
\begin{figure}[htbp]
\begin{center}
\includegraphics[width=3 in]{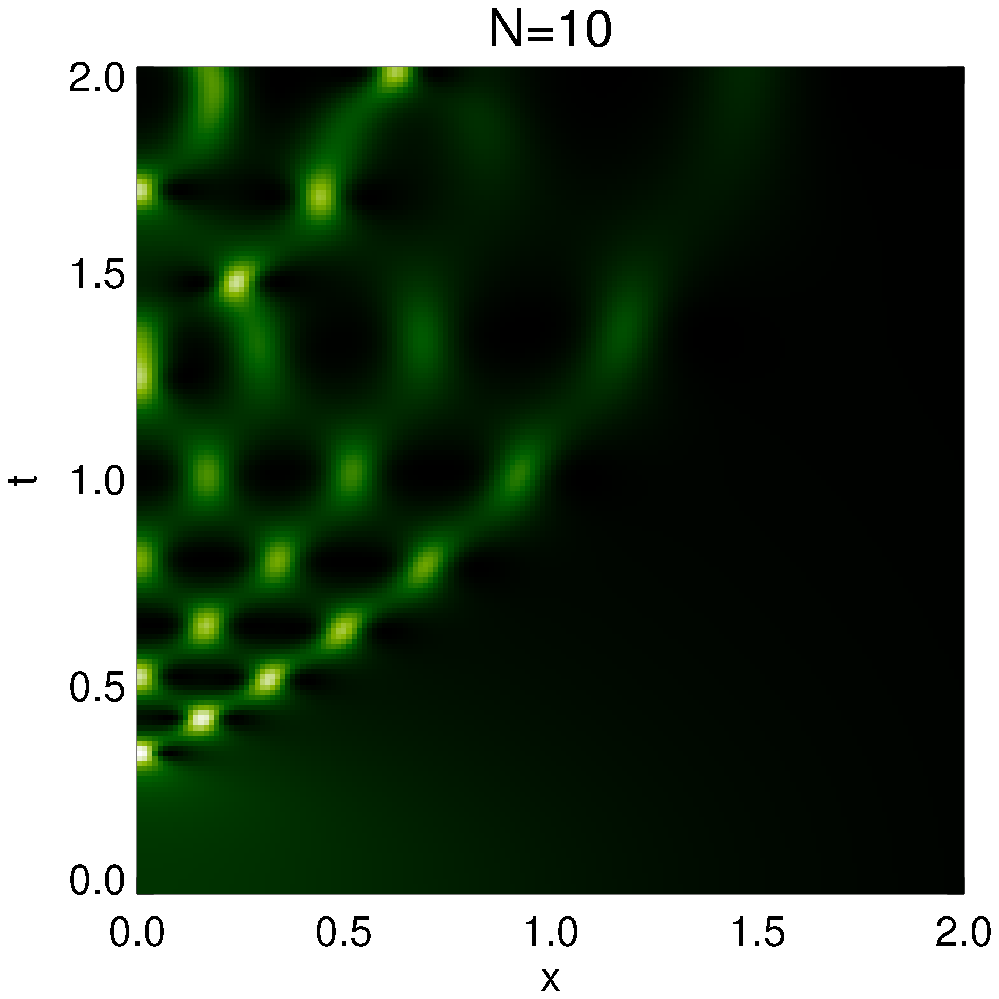}
\end{center}
\caption{\em The square modulus of $\psi_{10}(x,t)$ plotted over a part of the
positive quadrant of the $(x,t)$-plane.}
\label{fig:10soliton}
\end{figure}
\begin{figure}[htbp]
\begin{center}
\includegraphics[width=3 in]{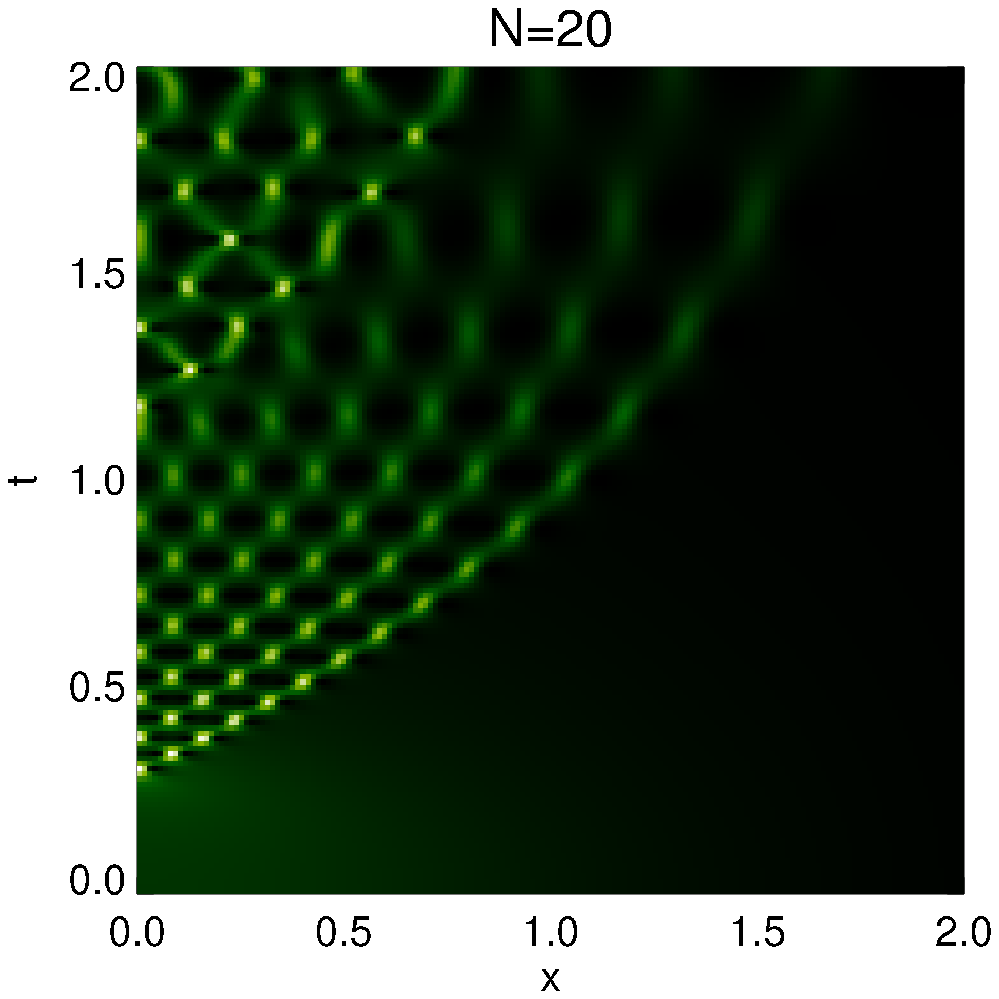}
\end{center}
\caption{\em The square modulus of $\psi_{20}(x,t)$ plotted over a part of the
positive quadrant of the $(x,t)$-plane.}
\label{fig:20soliton}
\end{figure}
\begin{figure}[htbp]
\begin{center}
\includegraphics[width=3 in]{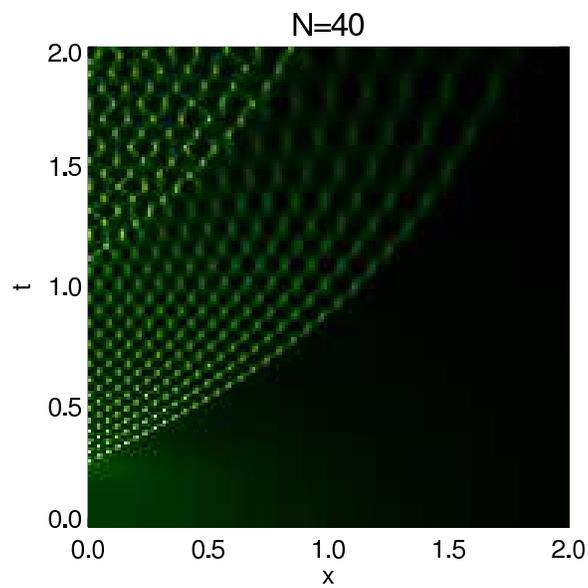}
\end{center}
\caption{\em The square modulus of $\psi_{40}(x,t)$ plotted over a part of the
positive quadrant of the $(x,t)$-plane.}
\label{fig:40soliton}
\end{figure}
\subsection{Phenomenology of the $N$-soliton for large $N$.}
\input{section2-3}

\clearpage

\section{Method of Asymptotic Analysis} 
\label{sec:asymptoticanalysis}
\subsection{Discrete Riemann-Hilbert problem.}\label{sec:drhp}
For each fixed positive integer $N$, and for fixed real values of $x$
and $t$, consider solving the following problem: find a $2\times 2$
matrix ${\bf m}(\lambda;N,x,t)$ with entries that are rational
functions of $\lambda$ such that
\begin{itemize}
\item The poles are all simple and are confined to the points $\{\lambda_{N,k}\}_{k=0}^{N-1}$ and $\{\lambda_{N,k}^*\}_{k=0}^{N-1}$, such that
\begin{equation}
\mathop{\text{Res}}_{\lambda_{N,k}}{\bf m}(\lambda;N,x,t)=
\lim_{\lambda\rightarrow
\lambda_{N,k}}{\bf m}(\lambda;N,x,t)
\left[\begin{array}{cc} 0 & 0\\c_{N,k}(x,t) & 0\end{array}\right]\,,
\end{equation}
and
\begin{equation}
\mathop{\text{Res}}_{\lambda_{N,k}^*}{\bf m}(\lambda;N,x,t)=
\lim_{\lambda\rightarrow
\lambda_{N,k}^*}{\bf m}(\lambda;N,x,t)
\left[\begin{array}{cc} 0 & -c_{N,k}(x,t)^*\\0 & 0\end{array}
\right]\,,
\end{equation}
hold for $k=0,\dots,N-1$, where $c_{N,k}(x,t):=
c_{N,k}^0e^{2iQ(\lambda_{N,k};x,t)/\hbar_N}$,
and,
\begin{equation}
c_{N,k}^0:=\frac{1}{\gamma_{N,k}}
\frac{\displaystyle\prod_{n=0}^{N-1}\lambda_{N,k}-\lambda_{N,n}^*}
{\displaystyle\mathop{\prod_{n=0}^{N-1}}_{n\neq k}\lambda_{N,k}-\lambda_{N,n}}
\,,\hspace{0.2 in}
Q(\lambda;x,t):=\lambda x+\lambda^2t\,.
\end{equation}
\item The matrix ${\bf m}(\lambda;N,x,t)$ is normalized so that 
\begin{equation}
\lim_{\lambda\rightarrow\infty}{\bf m}(\lambda;N,x,t)=\mathbb{I}\,.
\end{equation}
\end{itemize}
Then, the function defined in terms of ${\bf m}(\lambda;N,x,t)$ by the limit
\begin{equation}
\psi_N(x,t):=2i\lim_{\lambda\rightarrow\infty}\lambda 
m_{12}(\lambda;N,x,t)
\end{equation}
is the $N$-soliton solution of the semiclassically scaled focusing
nonlinear Schr\"odinger equation.

This Riemann-Hilbert problem essentially encodes the linear equations
introduced in \S~\ref{sec:linearsystem} describing the $N$-soliton, in
a way that is conducive to asymptotic analysis in the limit
$N\rightarrow\infty$. That the function $\psi_N(x,t)$ solves the
semiclassically scaled focusing nonlinear Schr\"odinger equation
\eqref{eq:semiclassical-fnls} is easy to show.  Indeed, let ${\bf
  n}(\lambda;N,x,t)={\bf
  m}(\lambda;N,x,t)e^{-iQ(\lambda;x,t)\sigma_3/\hbar_N}$.  Then, it is
easy to check that the residue conditions on ${\bf m}(\lambda;N,x,t)$
translate into analogous conditions on ${\bf n}(\lambda;N,x,t)$ that
are independent of $x$ and $t$:
\begin{equation}
\mathop{\text{Res}}_{\lambda_{N,k}}{\bf n}(\lambda;N,x,t)=
\lim_{\lambda\rightarrow\lambda_{N,k}} {\bf
  n}(\lambda;N,x,t)\left[\begin{array}{cc} 0 & 0\\ c_{N,k}^0 &
    0\end{array}\right]\,,
\end{equation}
and
\begin{equation}
\mathop{\text{Res}}_{\lambda_{N,k}^*}{\bf n}(\lambda;N,x,t)=
\lim_{\lambda\rightarrow\lambda_{N,k}^*} {\bf
  n}(\lambda;N,x,t)\left[\begin{array}{cc} 0 & -c_{N,k}^{0*}\\0  &
    0\end{array}\right]\,.
\end{equation}
It is easy to see that $\det({\bf n}(\lambda;N,x,t))\equiv 1$, because
clearly the determinant is a meromorphic function, with possible
simple poles only at the points
$\{\lambda_{N,k}\}\cup\{\lambda_{N,k}^*\}$, that tends to one as
$\lambda\rightarrow\infty$.  But from the residue conditions it is
easy to verify that $\det({\bf n}(\lambda;N,x,t))$ is regular at the
possible poles, and so is an entire function tending to one at
infinity that by Liouville's Theorem must be constant.  In particular,
${\bf n}(\lambda;N,x,t)$ is nonsingular for all $\lambda$, so it follows
that $\partial_x{\bf n}(\lambda;N,x,t){\bf n}(\lambda;N,x,t)^{-1}$ and
$\partial_t{\bf n}(\lambda;N,x,t){\bf n}(\lambda;N,x,t)^{-1}$ are both entire
functions of $\lambda$.  Moreover, they are polynomials in $\lambda$,
as can be deduced from their growth at infinity.  Indeed,
${\bf m}(\lambda;N,x,t)$ necessarily has an expansion
\begin{equation}
{\bf m}(\lambda;N,x,t)=\mathbb{I} + \sum_{p=1}^\infty \lambda^{-p}{\bf m}_p\,,
\hspace{0.2 in}{\bf m}_p={\bf m}_p(N,x,t)\,,
\end{equation}
which is uniformly convergent for $|\lambda|$ sufficiently large
(larger than the modulus of any prescribed singularity of ${\bf
  m}(\lambda;N,x,t)$ is enough), and differentiable term-by-term with
respect to $x$ and $t$.  Therefore,
\begin{equation}
\begin{array}{rcl}
\displaystyle
\partial_x{\bf n}(\lambda;N,x,t){\bf n}(\lambda;N,x,t)^{-1} &=&\displaystyle 
\partial_x\left[\left(\mathbb{I}+\lambda^{-1}{\bf m}_1 + \lambda^{-2}{\bf m}_2
+\cdots\right)e^{-iQ(\lambda;x,t)\sigma_3/\hbar_N}\right]\\\\
&&\displaystyle
\cdot e^{iQ(\lambda;x,t)\sigma_3/\hbar_N}\left(\mathbb{I}-\lambda^{-1}{\bf m}_1
+\lambda^{-2}({\bf m}_1^2-{\bf m}_2)+\cdots\right)\\\\
&=&\displaystyle
\left(-i\hbar_N^{-1}\lambda\sigma_3-i\hbar_N^{-1}{\bf m}_1\sigma_3
+\cdots\right)
\cdot
(\mathbb{I}-\lambda^{-1}{\bf m}_1 + \cdots)
\\\\
&=&\displaystyle -i\hbar_N^{-1}\left(\lambda\sigma_3 +
[{\bf m}_1,\sigma_3]\right)\,.
\end{array}
\end{equation}
Similarly,
\begin{equation}
\begin{array}{rcl}
\displaystyle
\partial_t{\bf n}(\lambda;N,x,t){\bf n}(\lambda;N,x,t)^{-1} &=&\displaystyle 
\partial_t\left[\left(\mathbb{I}+\lambda^{-1}{\bf m}_1 + \lambda^{-2}{\bf m}_2
+\cdots\right)e^{-iQ(\lambda;x,t)\sigma_3/\hbar_N}\right]\\\\
&&\displaystyle
\cdot e^{iQ(\lambda;x,t)\sigma_3/\hbar_N}\left(\mathbb{I}-\lambda^{-1}{\bf m}_1
+\lambda^{-2}({\bf m}_1^2-{\bf m}_2)+\cdots\right)\\\\
&=&\displaystyle
\left(-i\hbar_N^{-1}\lambda^2\sigma_3-i\hbar_N^{-1}\lambda{\bf m}_1\sigma_3
-i\hbar_N^{-1}{\bf m}_2\sigma_3+\cdots\right)\\\\
&&\displaystyle \cdot
(\mathbb{I}-\lambda^{-1}{\bf m}_1 + \lambda^{-2}({\bf m}_1^2-{\bf m}_2)+\cdots)
\\\\
&=&\displaystyle -i\hbar_N^{-1}\left(\lambda^2\sigma_3 +
\lambda[{\bf m}_1,\sigma_3] + 
[{\bf m}_2,\sigma_3]-[{\bf m}_1,\sigma_3]{\bf m}_1\right)\,.
\end{array}
\end{equation}
Consequently, ${\bf n}(\lambda;N,x,t)$ is a simultaneous fundamental
solution matrix for general $\lambda$ of the linear differential equations
\begin{equation}
\begin{array}{rcl}
\displaystyle i\hbar_N\partial_x{\bf n}(\lambda;N,x,t) &=&\displaystyle 
\left(\lambda\sigma_3+[{\bf m}_1,\sigma_3]\right){\bf n}(\lambda;N,x,t)\\\\
\displaystyle i\hbar_N\partial_t{\bf n}(\lambda;N,x,t)&=&\displaystyle
\left(\lambda^2\sigma_3+\lambda[{\bf m}_1,\sigma_3]+
[{\bf m}_2,\sigma_3]-[{\bf m}_1,\sigma_3]{\bf m}_1\right)
{\bf n}(\lambda;N,x,t)\,.
\end{array}
\end{equation}
The coefficient matrices therefore satisfy the zero-curvature compatibility
condition
\begin{equation}
\begin{array}{l}\displaystyle
i\hbar_N\partial_t\left(\lambda\sigma_3+[{\bf m}_1,\sigma_3]\right)
-i\hbar_N\partial_x\left(\lambda^2\sigma_3+\lambda[{\bf m}_1,\sigma_3]
+[{\bf m}_2,\sigma_3]-[{\bf m}_1,\sigma_3]{\bf m}_1\right) \\\\
\displaystyle\hspace{1 in}+\,\,\, 
[\lambda\sigma_3+[{\bf m}_1,\sigma_3],\lambda^2\sigma_3+
\lambda[{\bf m}_1,\sigma_3]+[{\bf m}_2,\sigma_3]-[{\bf m}_1,\sigma_3]{\bf m}_1]=0\,.
\end{array}
\end{equation}
Separating out the coefficients of the powers of $\lambda$ we obtain
two nontrivial equations:
\begin{equation}
-i\hbar_N\partial_x[{\bf m}_1,\sigma_3] - 
[[{\bf m}_2,\sigma_3],\sigma_3] +[[{\bf m}_1,\sigma_3]{\bf m}_1,\sigma_3]=0\,,
\end{equation}
and
\begin{equation}
i\hbar_N\partial_t[{\bf m}_1,\sigma_3] - i\hbar_N\partial_x
\left([{\bf m}_2,\sigma_3]-[{\bf m}_1,\sigma_3]{\bf m}_1\right) + 
[[{\bf m}_1,\sigma_3],[{\bf m}_2,\sigma_3]]-
[[{\bf m}_1,\sigma_3],[{\bf m}_1,\sigma_3]{\bf m}_1]=0\,.
\end{equation}
Introducing the notation 
\begin{equation}
{\bf A}={\bf A}^D+{\bf A}^{OD}\,,
\end{equation}
for separating a $2\times 2$ matrix into its diagonal and off-diagonal parts,
we have for any ${\bf A}$:
\begin{equation}
[{\bf A},\sigma_3]=2{\bf A}^{OD}\sigma_3\,,
\end{equation}
The first equation becomes
\begin{equation}
-i\hbar_N\partial_x{\bf m}_1^{OD}\sigma_3  -2{\bf m}_2^{OD} + 2{\bf m}_1^{OD}{\bf m}_1^D=0\,,
\end{equation}
which is purely off-diagonal, while the second equation has both
diagonal parts:
\begin{equation}
i\hbar_N\partial_x({\bf m}_1^{OD}\sigma_3{\bf m}_1^{OD}) +
2[{\bf m}_1^{OD}\sigma_3,{\bf m}_2^{OD}\sigma_3]
-2[{\bf m}_1^{OD}\sigma_3,{\bf m}_1^{OD}\sigma_3{\bf m}_1^{D}]=0\,,
\end{equation}
and off-diagonal parts:
\begin{equation}
i\hbar_N\partial_t({\bf m}_1^{OD}\sigma_3)-\frac{iA}{M}\partial_x(
{\bf m}_2^{OD}\sigma_3)+\frac{iA}{M}\partial_x({\bf m}_1^{OD}{\bf m}_1^{D}
\sigma_3)-2[{\bf m}_1^{OD}\sigma_3,{\bf m}_1^{OD}\sigma_3{\bf m}_1^{OD}]=0\,.
\end{equation}
Eliminating ${\bf m}_2^{OD}$ using the first equation, the off-diagonal part
of the second equation becomes:
\begin{equation}
i\hbar_N\sigma_3\partial_t{\bf m}_1^{OD} + \frac{\hbar_N^2}{2}
\partial_x^2{\bf m}_1^{OD} - 4{\bf m}_1^{OD3}=0\,.
\end{equation}
In other words, if we introduce notation for the off-diagonal elements
of ${\bf m}_1$ as follows,
\begin{equation}
{\bf m}_1^{OD}=\left[\begin{array}{cc} 0 & q\\r & 0\end{array}\right]\,,
\end{equation}
then $q$ and $r$ satisfy a coupled system of partial differential equations:
\begin{equation}
i\hbar_N\partial_tq + \frac{\hbar_N^2}{2}\partial_x^2q-4rq^2=0\,,\hspace{0.2 in}
-i\hbar_N\partial_tr + \frac{\hbar_N^2}{2}\partial_x^2r-4qr^2=0\,.
\end{equation}
If for all $x$ and $t$ we have $r=-q^*$, then these become
the focusing nonlinear Schr\"odinger equation 
\begin{equation}
i\hbar_N\partial_t\psi + \frac{\hbar_N^2}{2}\partial_x^2\psi +
|\psi|^2\psi=0\,, \hspace{0.2 in}\psi = 2iq = -2ir^*\,.
\end{equation}
Therefore, to complete the proof, it remains only to show that indeed
$r=-q^*$.  To do this we consider along with the solution ${\bf
  m}(\lambda;N,x,t)$ of the discrete Riemann-Hilbert problem the
corresponding matrix $\tilde{\bf m}(\lambda;N,x,t):=\sigma_2{\bf
  m}(\lambda^*;N,x,t)^*\sigma_2$, where the star denotes componentwise
complex conjugation.  Clearly $\tilde{\bf m}(\lambda;N,x,t)$ is analytic in
$\lambda$ with simple poles at $\{\lambda_{N,k}\}\cup\{\lambda_{N,k}^*\}$
(because the pole set is complex-conjugate invariant) and tends to the
identity as $\lambda\rightarrow\infty$ (because $\sigma_2^2=\mathbb{I}$).
Furthermore, 
\begin{equation}
\begin{array}{rcl}
\displaystyle
\mathop{\text{Res}}_{\lambda_{N,k}}\tilde{\bf m}(\lambda;N,x,t)&=&\displaystyle
\sigma_2\left(\mathop{\text{Res}}_{\lambda_{N,k}^*}
{\bf m}(\lambda;N,x,t)\right)^*\sigma_2 \\\\
&=&\displaystyle 
\sigma_2\left(\lim_{\lambda\rightarrow\lambda_{N,k}^*}{\bf m}(\lambda;N,x,t)
\left[
\begin{array}{cc}0 & -c_{N,k}(x,t)^*\\0 & 0\end{array}\right]\right)^*\sigma_2
\\\\ &=&\displaystyle
\lim_{\lambda\rightarrow\lambda_{N,k}}\tilde{\bf m}(\lambda;N,x,t)
\sigma_2\left[\begin{array}{cc}0 & -c_{N,k}(x,t)\\0 & 0\end{array}\right]\sigma_2
\\\\ &=&\displaystyle
\lim_{\lambda\rightarrow\lambda_{N,k}}\tilde{\bf m}(\lambda;N,x,t)
\left[\begin{array}{cc}
0 & 0\\c_{N,k}(x,t) & 0\end{array}\right]\,.
\end{array}
\end{equation}
By a similar calculation,
\begin{equation}
\mathop{\text{Res}}_{\lambda_{N,k}^*}\tilde{\bf m}(\lambda;N,x,t)=
\lim_{\lambda\rightarrow\lambda_{N,k}^*}\tilde{\bf m}(\lambda;N,x,t)
\left[\begin{array}{cc} 0 & -c_{N,k}(x,t)^* \\ 0 & 0\end{array}\right]\,.
\end{equation}
It follows that ${\bf m}(\lambda;N,x,t)\tilde{\bf
  m}(\lambda;N,x,t)^{-1}$ is an entire function of $\lambda$ that
tends to the identity matrix as $\lambda\rightarrow\infty$.  By
Liouville's Theorem, we thus have $\tilde{\bf m}(\lambda;N,x,t)\equiv
{\bf m}(\lambda;N,x,t)$.  Expanding both sides of this identity near
$\lambda=\infty$, we get
\begin{equation}
\mathbb{I}+\lambda^{-1}{\bf m}_1  = \mathbb{I}+\lambda^{-1}\sigma_2{\bf m}_1^*\sigma_2 + O(\lambda^{-2})\,,
\end{equation}
so in particular,
\begin{equation}
{\bf m}_1^{OD} = \sigma_2{\bf m}_1^{OD*}\sigma_2\,,
\end{equation}
which gives $r=-q^*$.

That the function $\psi_N(x,t)$ satisfies
$\psi_N(x,0)=A\,\text{sech}(x)$ for all $N$ is more difficult to show by
studying properties of the matrix ${\bf m}(\lambda;N,x,t)$.  In
general, this follows from solving the corresponding direct-scattering
problem which was done with the help of hypergeometric functions by
Satsuma and Yajima \cite{SY}.  Here we illustrate the corresponding
inverse-scattering calculation in the most tractable case of $N=1$.
When $N=1$, ${\bf m}(\lambda;1,x,0)$ may be sought in the form
\begin{equation}
{\bf m}(\lambda;1,x,0)=\left[\begin{array}{cc}
\displaystyle\frac{\lambda+\alpha_{11}(x)}{\lambda-iA/2} &
\displaystyle\frac{\alpha_{12}(x)}{\lambda+iA/2}\\\\
\displaystyle\frac{\alpha_{21}(x)}{\lambda-iA/2} &
\displaystyle\frac{\lambda+\alpha_{22}(x)}{\lambda+iA/2}
\end{array}\right]\,.
\end{equation}
The residue relations then say that
\begin{equation}
\begin{array}{rcl}
\displaystyle iA/2+\alpha_{11}(x) &=&\displaystyle
 -iAe^{-x}\cdot\frac{\alpha_{12}(x)}{iA}\,,\\\\
\displaystyle
\alpha_{21}(x) &=&\displaystyle -iAe^{-x}\cdot \frac{iA/2 + \alpha_{22}(x)}{iA}\,,\\\\
\displaystyle \alpha_{12}(x) &=&\displaystyle -iAe^{-x}\cdot\frac{iA/2-\alpha_{11}(x)}{iA}\,,\\\\
\displaystyle -iA/2 +\alpha_{22}(x) &=&\displaystyle -iAe^{-x}\cdot\frac{-\alpha_{21}(x)}{iA}\,,
\end{array}
\end{equation}
It follows that $\alpha_{12}(x)=-iA\,\text{sech}(x)/2$, from which we
indeed find that $\psi_1(x,0)=A\,\text{sech}(x)$.

\subsection{First modification:  removal
  of poles.}  
\label{sec:nopoles}
\subsubsection{Interpolation of residues.}
\begin{figure}[htbp]
\begin{center}
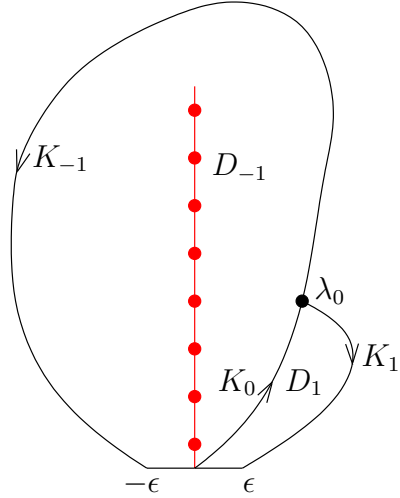
\end{center}
\caption{\em The regions $D_1$ and $D_{-1}$ in the upper half-plane,
  and the oriented boundary arcs $K_{-1}$, $K_0$, and $K_1$.  The $N$
  soliton eigenvalues $\{\lambda_{N,k}\}_{k=0}^{N-1}$ and the
  imaginary interval $[0,iA]$ in which they accumulate as
  $N\rightarrow\infty$ are shown in red.  The number $\epsilon>0$ will
  be taken later to be sufficiently small, but independent of $N$.}
\label{fig:InitialConfiguration}
\end{figure}

In the following, to keep the notation as simple as
possible, we suppress the parametric dependence on $N$, $x$, and $t$.  We
first modify the matrix unknown ${\bf m}(\lambda):={\bf
  m}(\lambda;N,x,t)$ by multiplication on the right by an explicit
matrix factor which differs from the identity matrix in the regions
$D_1$, $D_{-1}$, and their complex conjugates, as shown in
Figure~\ref{fig:InitialConfiguration}.
Let $K_0$ denote the common boundary arc of $D_{1}$ and $D_{-1}$
oriented in the direction away from the origin.  Let $K_1$ denote the
remaining part of the boundary of $D_1$ lying in the open upper
half-plane, oriented in the direction toward $\lambda=\epsilon$.  Let
$K_{-1}$ denote the remaining part of the boundary of $D_{-1}$
lying in the open upper half-plane, oriented in the direction toward
$\lambda=-\epsilon$.  We will call the point where the three contour
arcs $K_j$ meet $\lambda_0$.

Note that the region $D_{-1}$ contains the soliton eigenvalues
$\{\lambda_{N,k}\}_{k=0}^{N-1}$ for all $N$.  The region $D_1$ is
needed for a technical reason; with its help we will be able to
ultimately remove some jump discontinuities from the neighborhood of
$\lambda=0$.  We set
\begin{equation}
{\bf M}(\lambda):={\bf m}(\lambda)
\left[\begin{array}{cc}
1 & 0 \\
-iP(\lambda)e^{[2iQ(\lambda)+i\theta^0(\lambda)]/\hbar} & 1
\end{array}\right]\,,\hspace{0.2 in}\text{for $\lambda\in D_1$,}
\end{equation}
\begin{equation}
{\bf M}(\lambda):={\bf m}(\lambda)
\left[\begin{array}{cc}
1 & 0 \\
iP(\lambda)e^{[2iQ(\lambda)-i\theta^0(\lambda)]/\hbar} & 1
\end{array}\right]\,,\hspace{0.2 in}\text{for $\lambda\in D_{-1}$,}
\end{equation}
For all $\lambda$ in the
upper half-plane outside the closure of $D_1\cup D_{-1}$, we set ${\bf
  M}(\lambda)={\bf m}(\lambda)$.  For $\lambda$ in the
lower half-plane, we define ${\bf M}(\lambda):=\sigma_2{\bf
  M}(\lambda^*)^*\sigma_2$, where the star denotes componentwise
complex-conjugation.  Here, we are using the notation
\begin{equation}
P(\lambda):=\prod_{k=0}^{N-1}\frac{\lambda-\lambda_{N,k}^*}{\lambda-\lambda_{N,k}}\,,
\end{equation}
and
\begin{equation}
\theta^0(\lambda):=i\pi\lambda+\pi A\,,
\end{equation}
and, $\hbar=\hbar_N$.  It is easy to see that the matrix ${\bf
  M}(\lambda)$ is holomorphic at the soliton eigenvalues
$\lambda=\lambda_{N,k}$ where ${\bf m}(\lambda)$ has its poles in the
upper half-plane.
\subsubsection{Aside: even symmetry of $\psi_N(x,t)$ in $x$ and the formal
  continuum limit.}  
\label{sec:aside}
Since ${\bf M}(\lambda)={\bf m}(\lambda)$ in a
neighborhood of $\lambda=\infty$, the $N$-soliton is equivalently defined
by the formula
\begin{equation}
\psi_N(x,t)=2i\lim_{\lambda\rightarrow\infty}\lambda M_{12}(\lambda)\,.
\label{eq:psiNM}
\end{equation}
From the conditions determining ${\bf M}(\lambda)$ is is easy to show
that $\psi_N(x,t)$ is, for each $N$ and each $t$, an even function of
$x$.  For this purpose, we may suppose without any modification of
$\psi_N(x,t)$ that $\epsilon=0$, that $D_1=\emptyset$, and that
$D_{-1}$ is symmetric about the imaginary axis.  We also temporarily
re-introduce the explicit parametric dependence on $x$, and suppose
that $D_{-1}$ is fixed as $x\in\mathbb{R}$ varies.  Then the matrix
${\bf M}(-\lambda;x)$ has the same domain of analyticity as ${\bf
  M}(\lambda;x)$, and therefore we may compare ${\bf M}(\lambda;x)$
with the matrix ${\bf M}^\sharp(\lambda;x)$ defined by setting
\begin{equation}
{\bf M}^\sharp(\lambda;x):=\sigma_3{\bf M}(-\lambda;x)\sigma_3
\left[\begin{array}{cc} 0 & 
ie^{[2i(\lambda x-\lambda^2 t)+i\theta^0(\lambda)]/\hbar}\\
ie^{-[2i(\lambda x-\lambda^2 t)+i\theta^0(\lambda)]/\hbar} &
P(\lambda)^{-1}\end{array}\right]\,,\hspace{0.2 in}\lambda\in D_{-1}\,,
\label{eq:hatMinside}
\end{equation}
while elsewhere in the upper half-plane we set
\begin{equation}
{\bf M}^\sharp(\lambda;x):=\sigma_3{\bf M}(-\lambda;x)\sigma_3P(\lambda)^{\sigma_3}\,,
\label{eq:hatMoutside}
\end{equation}
and then we define ${\bf M}^\sharp(\lambda;x)$ for $\lambda$ in the
lower half-plane by setting ${\bf M}^\sharp(\lambda;x):=\sigma_2{\bf
  M}^\sharp(\lambda^*;x)^*\sigma_2$ for $\Im(\lambda)<0$.  Note that
because the zeros of $P(\lambda)$ are confined to $D_{-1}^*$ while the
poles are confined to $D_{-1}$, this defines ${\bf
  M}^\sharp(\lambda;x)$ as a sectionally holomorphic function of
$\lambda$ with the same ($x$-independent) domain of analyticity as
${\bf M}(\lambda;x)$.

Now, if we use a subscript ``$+$'' (respectively ``$-$'') to denote a
boundary value taken on the boundary of $D_{-1}$ from inside
(respectively outside), then it is an easy exercise to check that
\begin{equation}
{\bf M}^\sharp_-(\lambda;x)^{-1}{\bf M}^\sharp_+(\lambda;x) = 
{\bf M}_-(\lambda;-x)^{-1}{\bf M}_+(\lambda;-x)\,,\hspace{0.2 in}
\text{for $x\in\mathbb{R}$ and $\lambda\in \partial D_{-1}\cup\partial D_{-1}^*$\,.}
\end{equation}
Also, since $P(\lambda)\rightarrow 1$ as $\lambda\rightarrow\infty$,
it follows that ${\bf M}^\sharp(\lambda;x)\rightarrow\mathbb{I}$ as
$\lambda\rightarrow\infty$ for each $x\in\mathbb{R}$.  By Liouville's
Theorem, which assures the uniqueness of ${\bf M}(\lambda;x)$ given
the jump condition across $\partial D_{-1}\cup\partial D_{-1}^*$ and
the normalization condition at $\lambda=\infty$, it follows that
\begin{equation}
{\bf M}^\sharp(\lambda;x) = {\bf M}(\lambda;-x)\,.
\label{eq:hatMM}
\end{equation}
Using this relation, we clearly have that
\begin{equation}
2i\lim_{\lambda\rightarrow\infty}\lambda{M}^\sharp_{12}(\lambda;x) = 
2i\lim_{\lambda\rightarrow\infty}\lambda M_{12}(\lambda;-x) = 
\psi_N(-x,t)\,,
\end{equation}
where the last equality follows from \eqref{eq:psiNM}.  On the other
hand, directly from the definition \eqref{eq:hatMoutside} valid for
sufficiently large $|\lambda|$, we may use the fact that
$P(\lambda)\rightarrow 1$ as $\lambda\rightarrow\infty$ to conclude
that
\begin{equation}
2i\lim_{\lambda\rightarrow\infty}\lambda{M}^\sharp_{12}(\lambda;x) = 
-2i\lim_{\lambda\rightarrow\infty}\lambda M_{12}(-\lambda;x)
= 2i\lim_{\mu\rightarrow\infty} \mu M_{12}(\mu;x) = \psi_N(x,t)\,.
\end{equation}
Therefore we learn that $\psi_N(-x,t)=\psi_N(x,t)$ holds for all real
$x$, so the $N$-soliton is an even function of $x$ for each $t$.
Therefore, in all remaining calculations in this paper we will suppose
without loss of generality that $x\ge 0$.  (Since from
\eqref{eq:semiclassical-fnls} we see that time reversal is equivalent
to complex conjugation of $\psi_N(x,t)$ because $\psi_N(x,0)$ is real,
we will also assume that $t\ge 0$.)  These choices lead to certain
asymmetries in the complex plane as already apparent in
Figure~\ref{fig:InitialConfiguration}; for a discussion of how choice
of signs of $x$ and $t$ relates to the parity of certain structures in
the complex spectral plane that are introduced to aid in asymptotic
analysis, see \cite{KMM}.

The even symmetry of $\psi_N(x,t)$ in $x$ is not an earth-shattering
result, of course, but what is interesting is that the same argument
fails completely if a natural continuum limit related to the limit
$N\rightarrow\infty$ is introduced on an {\em ad hoc} basis.  Indeed,
it is natural to consider replacing the product $P(\lambda)$ by a
formal continuum limit by ``condensing the poles'' as follows: define
\begin{equation}
\tilde{P}(\lambda):=\exp\left(\frac{1}{\hbar}\left[
i\int_0^{iA}\log(-i(\lambda-\eta))\,d\eta - i\int_{-iA}^0
\log(-i(\lambda-\eta))\,d\eta\right]\right)\,,
\end{equation}
which may be viewed as coming from interpreting the sums in the
exact formula
\begin{equation}
P(\lambda) = \exp\left(\frac{1}{\hbar}\left[\sum_{k=0}^{N-1}
\log(\lambda-\lambda_{N,k}^*)\hbar-\sum_{k=0}^{N-1}
\log(\lambda-\lambda_{N,k})\hbar
\right]\right)
\end{equation}
as Riemann sums and passing to the natural integral limit for
$\lambda$ fixed.  The function $\tilde{P}(\lambda)$ is analytic for
$\lambda\in\mathbb{C}\setminus [-iA,iA]$, and where $P(\lambda)$ has
accumulating poles and zeros, $\tilde{P}(\lambda)$ has a logarithmic
branch cut.  Then, while ${\bf M}(\lambda;x)$ is defined as being a
matrix with the symmetry ${\bf M}(\lambda;x) = \sigma_2{\bf
  M}(\lambda^*;x)^*\sigma_2$ that satisfies the normalization
condition ${\bf M}(\lambda;x)\rightarrow\mathbb{I}$ as
$\lambda\rightarrow\infty$ and is analytic except on $\partial
D_{-1}\cup\partial D_{-1}^*$ along which it takes continuous boundary
values related by
\begin{equation}
{\bf M}_+(\lambda;x) = {\bf M}_-(\lambda;x)
\left[\begin{array}{cc} 1 & 0 \\ iP(\lambda)e^{[2i(\lambda x+\lambda^2 t)-i\theta^0(\lambda)]/\hbar} & 1\end{array}\right]\,,\hspace{0.2 in}
\lambda\in \partial D_{-1}\,,
\end{equation}
we may define a matrix $\tilde{\bf M}(\lambda)$ that satisfies exactly
the same conditions as ${\bf M}(\lambda)$ but with $P(\lambda)$
replaced by $\tilde{P}(\lambda)$ in the jump condition.  It is a direct
matter to show that like $\psi_N(x,t)$, the function defined by
\begin{equation}
\tilde{\psi}_N(x,t):=2i\lim_{\lambda\rightarrow\infty}\lambda
\tilde{M}_{12}(\lambda;x)
\end{equation}
is also a solution of the semiclassically scaled focusing nonlinear
Schr\"odinger equation \eqref{eq:semiclassical-fnls} (virtually the
same arguments apply as was used to prove this about $\psi_N(x,t)$).
However, whether $\tilde{\psi}_N(x,t)$ is an even function of $x$ is
in our opinion an open question.  Indeed, if we try to mimic the above
proof of evenness of $\psi_N(x,t)$, we would be inclined to try to
define a matrix $\tilde{\bf M}^\sharp(\lambda;x)$ starting from
$\tilde{\bf M}(\lambda;x)$ by formulae analogous to
\eqref{eq:hatMinside} and \eqref{eq:hatMoutside}, but with
$P(\lambda)$ replaced everywhere by $\tilde{P}(\lambda)$.  Comparing
$\tilde{\bf M}^\sharp(\lambda;x)$ with $\tilde{\bf M}(\lambda;-x)$
then becomes a problem, because while $\tilde{\bf M}(\lambda;-x)$ is
analytic in $D_{-1}\cup D_{-1}^*$, the matrix $\tilde{\bf
  M}^\sharp(\lambda;x)$ has a jump discontinuity across the segment
$[-iA,iA]$.  Thus, in stark contrast with \eqref{eq:hatMM} we have
\begin{equation}
\tilde{\bf M}^\sharp(\lambda;x)\neq\tilde{\bf M}(\lambda;-x)\,,
\end{equation}
and we cannot conclude (by a completely analogous proof, anyway) any
evenness of $\tilde{\psi}(x,t)$.

The possibility that $\tilde{\psi}_N(x,t)$ may not be an even function
of $x$ while $\psi_N(x,t)$ most certainly is even casts some doubt on
the prospect that $\tilde{\psi}_N(x,t)$ might be a good approximation
to $\psi_N(x,t)$.  As it is $\tilde{\psi}_N(x,t)$ that is related to
the solution of the ``continuum-limit'' Riemann-Hilbert problem for a
matrix $\tilde{\bf P}(\lambda)$ to be introduced in
\S~\ref{sec:formallimit}, we have some reason to suspect at this point
that without careful accounting of the errors introduced by making the
{\em ad hoc} substitution $P(\lambda)\rightarrow\tilde{P}(\lambda)$, a
study of the ``condensed-pole'' problem may not be relevant at all to
the asymptotic analysis of the $N$-soliton $\psi_N(x,t)$, at least for
certain $x$ and $t$.  We will give evidence in this paper that such
suspicion is entirely justifiable.

\subsection{Second modification:  introduction of $g$-function.}  
\label{sec:gfunction}
Suppose that $g(\lambda)$ is a function analytic
for $\lambda\in \mathbb{C}\setminus (K_{-1}\cup K_0\cup K_{-1}^*\cup
K_0^*)$ that satisfies the symmetry condition
\begin{equation}
g(\lambda)+g(\lambda^*)^*=0\,,
\label{eq:gsymmetry}
\end{equation}
and
$g(\lambda)\rightarrow 0$ as $\lambda\rightarrow\infty$.  Note that in
particular $g(\lambda)$ is analytic in the real intervals
$(-\epsilon,0)$ and $(0,\epsilon)$.  We change variables again to a
matrix function ${\bf N}(\lambda)$ by setting
\begin{equation}
{\bf N}(\lambda):={\bf M}(\lambda)
e^{-g(\lambda)\sigma_3/\hbar}\,.
\end{equation}
Letting the subscript ``$+$'' (respectively ``$-$'') denote a boundary
value taken on one of the contours $K_j$ from the left (respectively right)
according to its orientation, we deduce from our definitions and the
continuity of ${\bf m}(\lambda)$ across each of these contours
that the following ``jump relations'' hold:
\begin{equation}
{\bf N}_+(\lambda)={\bf N}_-(\lambda)
\left[\begin{array}{cc} e^{-[g_+(\lambda)-g_-(\lambda)]/\hbar} & 0 \\
iP(\lambda)e^{[2iQ(\lambda)-i\theta^0(\lambda)-g_+(\lambda)-g_-(\lambda)]/
\hbar} & e^{[g_+(\lambda)-g_-(\lambda)]/\hbar}
\end{array}\right]\,,\hspace{0.2 in}\lambda\in K_{-1}\,,
\label{eq:jumpNKm1}
\end{equation}
\begin{equation}
{\bf N}_+(\lambda)={\bf N}_-(\lambda)
\left[\begin{array}{cc} 1 & 0 \\
iP(\lambda)e^{[2iQ(\lambda)+i\theta^0(\lambda)-2g(\lambda)]/\hbar} & 
1
\end{array}\right]\,,\hspace{0.2 in}\lambda\in K_{1}\,,
\end{equation}
\begin{equation}
{\bf N}_+(\lambda)={\bf N}_-(\lambda)
\left[\begin{array}{cc} e^{-[g_+(\lambda)-g_-(\lambda)]/\hbar} & 0 \\
2iP(\lambda)\cos(\theta^0(\lambda)/\hbar)
e^{[2iQ(\lambda)-g_+(\lambda)-g_-(\lambda)]/\hbar} & 
e^{[g_+(\lambda)-g_-(\lambda)]/\hbar}
\end{array}\right]\,,\hspace{0.2 in}\lambda\in K_{0}\,.
\end{equation}
The jump relations holding on the conjugate contours in the lower
half-plane follow from the symmetry ${\bf N}(\lambda^*)=\sigma_2{\bf
  N}(\lambda)^*\sigma_2$.  Finally, there are also jump
discontinuities across the real intervals $(-\epsilon,0)$ and
$(0,\epsilon)$, both of which we assign an orientation from left to
right.  Using the above symmetry relation along with the facts
(holding for $x$ and $t$ real and $\hbar=A/N$ with $N\in
\mathbb{Z}$):
\begin{equation}
P(\lambda^*)^*=P(\lambda)^{-1}\,,\hspace{0.2 in}
Q(\lambda^*)^* = Q(\lambda)\,,\hspace{0.2 in}
\theta^0(\lambda^*)^* = -\theta^0(\lambda)+2\pi \hbar N\,,
\label{eq:PQthetasymmetry}
\end{equation}
we find that
\begin{equation}
{\bf N}_+(\lambda)={\bf N}_-(\lambda)
\left[\begin{array}{cc} 1+e^{-2i\theta_0(\lambda)/\hbar} &
-iP(\lambda)^{-1}e^{[-2iQ(\lambda)-i\theta^0(\lambda)+2g(\lambda)]/\hbar}\\
iP(\lambda)e^{[2iQ(\lambda)-i\theta^0(\lambda)-2g(\lambda)]/\hbar} & 1
\end{array}\right]\,,\hspace{0.2 in}\lambda\in (-\epsilon,0)\,,
\label{eq:Njumprealneg}
\end{equation}
\begin{equation}
{\bf N}_+(\lambda)={\bf N}_-(\lambda)
\left[\begin{array}{cc} 1+e^{2i\theta_0(\lambda)/\hbar} &
iP(\lambda)^{-1}e^{[-2iQ(\lambda)+i\theta^0(\lambda)+2g(\lambda)]/\hbar}\\
-iP(\lambda)e^{[2iQ(\lambda)+i\theta^0(\lambda)-2g(\lambda)]/\hbar} & 1
\end{array}\right]\,,\hspace{0.2 in}\lambda\in (0,\epsilon)\,.
\label{eq:Njumprealpos}
\end{equation}

Let $C$ denote a simple contour lying in $D_{-1}$ starting from the
origin and terminating at $\lambda=iA$.  There is a unique function
$L(\lambda)$ with the properties that (i) it is analytic for
$\lambda\in \mathbb{C}\setminus(C\cup C^*)$, (ii) it takes continuous
boundary values on each side of the contour $C\cup C^*$ satisfying
\begin{equation}
L_+(\lambda)-L_-(\lambda)=\left\{\begin{array}{ll} -2i\theta^0(\lambda)\,,&
\hspace{0.2 in}\text{for $\lambda\in C$,}\\
-2i\theta^0(\lambda^*)^*\,,&\hspace{0.2 in}\text{for $\lambda\in C^*$,}
\end{array}\right.
\label{eq:Ljump}
\end{equation}
where $L_+(\lambda)$ (respectively $L_-(\lambda)$) refers to the
boundary value taken on $C\cup C^*$ from the left (respectively right)
as it is traversed from $-iA$ to $iA$, and (iii) it is normalized so
that $L(\lambda)$ tends to zero as $\lambda\rightarrow\infty$.
Indeed, the function $L(\lambda)$ is easily seen to be unique from
these conditions (by Liouville's theorem and continuity of the
boundary values as is compatible with the conditions \eqref{eq:Ljump})
and we will give an explicit construction later (see
\eqref{eq:fU}--\eqref{eq:fL}).  The function $L(\lambda)$ enjoys the
following symmetry property:
\begin{equation}
L(\lambda)+L(\lambda^*)^* =0\,.
\label{eq:Lsymmetry}
\end{equation}
Related to $L(\lambda)$ is another function
$\overline{L}(\lambda)$ defined as follows.  Let $C_\infty$ denote an
infinite simple contour in the upper half-plane emanating from
$\lambda=iA$ and tending to infinity in the upper half-plane, avoiding
the domain $D_1$.  Note that the union of contours $C\cup C_\infty\cup
C^*\cup C_\infty^*$ divides the complex plane in half.  We say that
the left (right) half-plane according to $C\cup C_\infty\cup C^*\cup
C_\infty^*$ is the half containing the negative (positive) real axis.
For $\Im(\lambda)>0$ we then define
\begin{equation}
\overline{L}(\lambda):=\left\{\begin{array}{ll}L(\lambda)+i\theta^0(\lambda)\,,
&\hspace{0.2 in}\text{for $\lambda$ in the left half-plane according to
$C\cup C_\infty\cup C^*\cup C_\infty^*$,}\\
L(\lambda)-i\theta^0(\lambda)\,,&\hspace{0.2 in}\text{for $\lambda$ in the
right half-plane according to $C\cup C_\infty\cup C^*\cup C_\infty^*$.}
\end{array}\right.
\label{eq:Lbardef}
\end{equation}
We note from
\eqref{eq:Ljump} that $\overline{L}(\lambda)$ extends continuously to
$C$ and thus may be viewed as a function analytic for
$\lambda\in \mathbb{C}_+\setminus C_\infty$, where $\mathbb{C}_+$ denotes
the upper half-plane.  We introduce the notation
\begin{equation}
T(\lambda):=2e^{-\overline{L}(\lambda)/\hbar}
P(\lambda)\cos(\theta^0(\lambda)/\hbar)\,.
\end{equation}
so the jump relation holding on $K_0$ may be equivalently written in the form
\begin{equation}
{\bf N}_+(\lambda)={\bf N}_-(\lambda)
\left[\begin{array}{cc} e^{-[g_+(\lambda)-g_-(\lambda)]/\hbar} & 0 \\
iT(\lambda)e^{[2iQ(\lambda)+\overline{L}(\lambda)-g_+(\lambda)-g_-(\lambda)]/
\hbar} & e^{[g_+(\lambda)-g_-(\lambda)]/\hbar}
\end{array}\right]\,,\hspace{0.2 in}\lambda\in K_{0}\,.
\label{eq:K0jump}
\end{equation}

The jump matrix in \eqref{eq:K0jump} can be factored as follows:
\begin{equation}
\begin{array}{l}
\displaystyle
\left[\begin{array}{cc} e^{-[g_+(\lambda)-g_-(\lambda)]/\hbar} & 0 \\
iT(\lambda)e^{[2iQ(\lambda)+\overline{L}(\lambda)-g_+(\lambda)-g_-(\lambda)]/
\hbar} & e^{[g_+(\lambda)-g_-(\lambda)]/\hbar}
\end{array}\right]=\\\\
\displaystyle\hspace{0.5 in}
\left[\begin{array}{cc} T(\lambda)^{-1/2} & -iT(\lambda)^{-1/2}
e^{[-2iQ(\lambda)-\overline{L}(\lambda)+2g_-(\lambda)]/\hbar}\\
0 & T(\lambda)^{1/2}\end{array}\right]\times\\\\
\displaystyle\hspace{0.5 in}
\left[\begin{array}{cc} 0 & i
e^{[-2iQ(\lambda)-\overline{L}(\lambda)+g_+(\lambda)+g_-(\lambda)]/\hbar}\\
ie^{[2iQ(\lambda)+\overline{L}(\lambda)-g_+(\lambda)-g_-(\lambda)]
/\hbar} & 0\end{array}\right]
\times \\\\\displaystyle\hspace{0.5 in}
\left[\begin{array}{cc} T(\lambda)^{1/2} & -iT(\lambda)^{-1/2}
e^{[-2iQ(\lambda)-\overline{L}(\lambda)+2g_+(\lambda)]/\hbar} \\
0 & T(\lambda)^{-1/2}\end{array}\right]\,.
\end{array}
\label{eq:K0factorization}
\end{equation}
Such a factorization makes sense because we will see later (in
\S~\ref{sec:error}) that $T(\lambda)\rightarrow 1$ as
$N\rightarrow\infty$, so the fractional powers are well-defined for
large enough $N$ as having similar asymptotics, converging to $1$ as
$N\rightarrow\infty$.  The left-most (respectively right-most) matrix
factor is the boundary value on $K_0$ taken by a function analytic on
the ``minus'' (respectively ``plus'') side of $K_0$.  Let $K_L$ denote
a contour arc in $D_{-1}$ connecting the point $\lambda_0$ with the
point $\lambda=-\epsilon$, and oriented in the direction away from
$\lambda_0$.  Let $D_L$ denote the region enclosed by $K_0$, $K_L$,
and the interval $[-\epsilon,0]$.  We introduce a new unknown ${\bf
  O}(\lambda)$ based on the above factorization as follows:
\begin{equation}
{\bf O}(\lambda):={\bf N}(\lambda)
\left[\begin{array}{cc} T(\lambda)^{-1/2} & -iT(\lambda)^{-1/2}
e^{[-2iQ(\lambda)-\overline{L}(\lambda)+2g(\lambda)]/\hbar}\\
0 & T(\lambda)^{1/2}\end{array}\right]\,,\hspace{0.2 in}
\lambda\in D_1\,,
\label{eq:OD1}
\end{equation}
\begin{equation}
{\bf O}(\lambda):={\bf N}(\lambda)
\left[\begin{array}{cc} T(\lambda)^{-1/2} & iT(\lambda)^{-1/2}
e^{[-2iQ(\lambda)-\overline{L}(\lambda)+2g(\lambda)]/\hbar}\\
0 & T(\lambda)^{1/2}\end{array}\right]\,,\hspace{0.2 in}
\lambda\in D_L\,,
\label{eq:ODL}
\end{equation}
and elsewhere in the upper half-plane we set ${\bf O}(\lambda):=
{\bf N}(\lambda)$.  For $\lambda$ in the lower half-plane, we
define ${\bf O}(\lambda)$ so as to preserve the symmetry ${\bf
  O}(\lambda^*)=\sigma_2{\bf O}(\lambda)^*\sigma_2$.

\begin{proposition}
  The matrix ${\bf O}(\lambda)$ has no jump discontinuity across
  the real intervals $(-\epsilon,0)$ or $(0,\epsilon)$, and thus may
  be viewed as an analytic function for $\lambda\in
  \mathbb{C}\setminus (K_{-1}\cup K_0\cup K_L\cup K_1\cup K_{-1}^*\cup
  K_0^*\cup K_L^*\cup K_1^*)$ that takes continuous boundary values
  from each region where it is analytic.
\end{proposition}

\begin{proof}
The boundary value taken by ${\bf O}(\lambda)$ on $(0,\epsilon)$ from 
the upper half-plane is
\begin{equation}
{\bf O}_+(\lambda)={\bf N}_+(\lambda)
\left[\begin{array}{cc} T(\lambda)^{-1/2} & -iT(\lambda)^{-1/2}
e^{[-2iQ(\lambda)-L(\lambda)+i\theta^0(\lambda)+2g(\lambda)]/\hbar} \\
0 & T(\lambda)^{-1/2}\cdot P(\lambda)e^{[-L(\lambda)+i\theta^0(\lambda)]/\hbar}(e^{i\theta^0(\lambda)/\hbar}+e^{-i\theta^0(\lambda)/\hbar})
\end{array}\right]\,,
\label{eq:OplusNplus}
\end{equation}
which follows from \eqref{eq:OD1}, where we used the fact that
$(0,\epsilon)$ is in the right half-plane according to $C\cup
C_\infty\cup C^*\cup C_\infty^*$ to write $\overline{L}(\lambda)$ in
terms of $L(\lambda)$ with the help of \eqref{eq:Lbardef}.  Here all
of the quantities in the exponent are analytic functions on
$(0,\epsilon)$, and $T(\lambda)^{-1/2}$ is interpreted in the sense of
its boundary value taken on $(0,\epsilon)$ from the upper half-plane.
From the conjugation symmetry relations satisfied by ${\bf
  O}(\lambda)$ and ${\bf N}(\lambda)$, it then follows that
the boundary value taken by ${\bf O}(\lambda)$ on $(0,\epsilon)$
from the lower half-plane is
\begin{equation}
{\bf O}_-(\lambda)={\bf N}_-(\lambda)
\left[\begin{array}{cc} (T(\lambda^*)^*)^{-1/2}\cdot 
P(\lambda)^{-1}e^{[L(\lambda)+i\theta^0(\lambda)]/\hbar}
(e^{i\theta^0(\lambda)/\hbar}+e^{-i\theta^0(\lambda)/\hbar}) & 0\\
-i(T(\lambda^*)^*)^{-1/2}
e^{[2iQ(\lambda)+L(\lambda)+i\theta^0(\lambda)-2g(\lambda)]/\hbar}  & (T(\lambda^*)^*)^{-1/2} 
\end{array}\right]\,.
\label{eq:NminusOminus}
\end{equation}
Here we have used the relations \eqref{eq:gsymmetry},
\eqref{eq:Lsymmetry}, and \eqref{eq:PQthetasymmetry} to simplify the
exponents.  Since the matrix factor appearing in
\eqref{eq:NminusOminus} has determinant one, to compute the jump
relation for ${\bf O}(\lambda)$ across the interval
$(0,\epsilon)$ we will need to know the boundary value of the
product $T(\lambda)^{-1/2}(T(\lambda^*)^*)^{-1/2}$ as $\lambda$
approaches $(0,\epsilon)$ from the upper half-plane.  First, note that
for $\Im(\lambda)>0$ in the right half-plane according to $C\cup
C_\infty\cup C^*\cup C_\infty^*$ we may use \eqref{eq:Lsymmetry} and
\eqref{eq:PQthetasymmetry} to find that
\begin{equation}
T(\lambda)T(\lambda^*)^* = (1+e^{2i\theta^0(\lambda)/\hbar})^2\,.
\end{equation}
(Similarly, if $\Im(\lambda)>0$ and $\lambda$ is in the left half-plane
according to $C\cup C_\infty\cup C^*\cup C_\infty^*$, then the identity
\begin{equation}
T(\lambda)T(\lambda^*)^*=(1+e^{-2i\theta^0(\lambda)/\hbar})^2
\end{equation}
holds.)  The pointwise asymptotic (see \S~\ref{sec:error}) that
$T(\lambda)^{-1/2}\rightarrow 1$ as $N\rightarrow\infty$ then gives
that
\begin{equation}
T(\lambda)^{-1/2}(T(\lambda^*)^*)^{-1/2} = \left\{\begin{array}{ll}
(1+e^{2i\theta^0(\lambda)/\hbar})^{-1}\,,&
\hspace{0.2 in}\lambda\in (0,\epsilon)\,,\\
(1+e^{-2i\theta^0(\lambda)/\hbar})^{-1}\,,&
\hspace{0.2 in}\lambda\in (-\epsilon,0)\,.
\end{array}\right.
\label{eq:Tonehalfproduct}
\end{equation}
Using this fact, one substitutes \eqref{eq:NminusOminus} into
\eqref{eq:Njumprealpos}, and then \eqref{eq:Njumprealpos} into
\eqref{eq:OplusNplus}.  It is then an elementary calculation to deduce
that ${\bf O}_+(\lambda)={\bf O}_-(\lambda)$ for
$\lambda\in (0,\epsilon)$.

Starting with \eqref{eq:ODL}, and proceeding in a similar way as we
did to arrive at \eqref{eq:OplusNplus}, we find that the boundary value
taken by ${\bf O}(\lambda)$ on the real interval $(-\epsilon,0)$
(which lies in the left half-plane according to $C\cup C_\infty\cup
C^*\cup C_\infty^*$) from the upper half-plane is
\begin{equation}
{\bf O}_+(\lambda)={\bf N}_+(\lambda)
\left[\begin{array}{cc}
T(\lambda)^{-1/2} & iT(\lambda)^{-1/2}
e^{[-2iQ(\lambda)-L(\lambda)-i\theta^0(\lambda)+2g(\lambda)]/\hbar}\\
0 & T(\lambda)^{-1/2}\cdot P(\lambda)
e^{[-L(\lambda)-i\theta^0(\lambda)]/\hbar}
(e^{i\theta^0(\lambda)/\hbar}+e^{-i\theta^0(\lambda)/\hbar})
\end{array}\right]\,.
\label{eq:OplusNplusLeft}
\end{equation}
By conjugation symmetry, we then find
\begin{equation}
{\bf O}_-(\lambda)={\bf N}_-(\lambda)
\left[\begin{array}{cc}
(T(\lambda^*)^*)^{-1/2}P(\lambda)^{-1}
e^{[L(\lambda)-i\theta^0(\lambda)]/\hbar}(e^{i\theta^0(\lambda)/\hbar}+
e^{-i\theta^0(\lambda)/\hbar}) &
0\\
i(T(\lambda^*)^*)^{-1/2}
e^{[2iQ(\lambda)+L(\lambda)-i\theta^0(\lambda)-2g(\lambda)]/\hbar} & 
(T(\lambda^*)^*)^{-1/2}\end{array}\right]\,.
\label{eq:NminusOminusLeft}
\end{equation}
Combining \eqref{eq:NminusOminusLeft}, \eqref{eq:Njumprealneg},
and \eqref{eq:OplusNplusLeft} with the help of \eqref{eq:Tonehalfproduct}
then shows that ${\bf O}_+(\lambda)={\bf O}_-(\lambda)$
for $\lambda\in (-\epsilon,0)$ as well.

That the boundary values taken by ${\bf O}(\lambda)$ from each
component of the complex plane where it is analytic are in fact
continuous functions along the boundary even at self-intersection
points follows from the fact that this was initially true for ${\bf
  m}(\lambda)$, and is preserved by each of our substitutions to
arrive at the matrix ${\bf O}(\lambda)$.  (But it is also
straightforward to verify this directly, even at the points
$\lambda=-\epsilon$, $\lambda=0$, and $\lambda=\epsilon$.)
\end{proof}

\begin{figure}[htbp]
\begin{center}
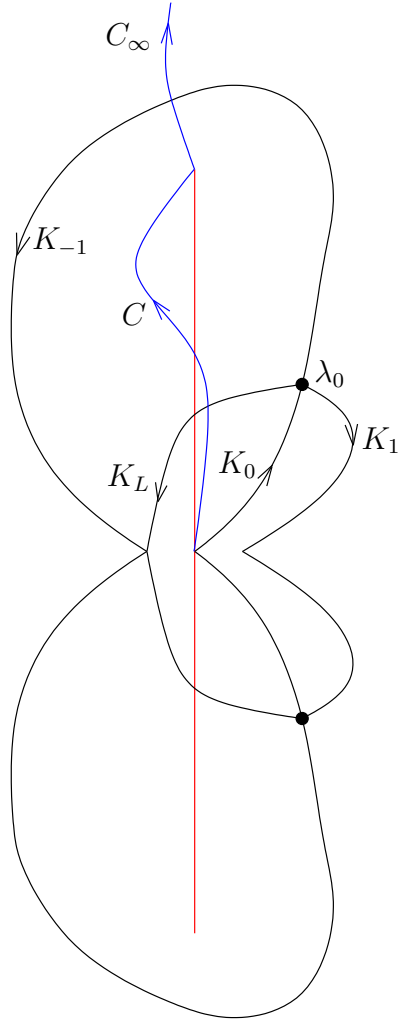
\end{center}
\caption{\em The contours $K_1$, $K_0$, $K_{-1}$, and $K_L$, and their
  conjugates are the contours of discontinuity of ${\bf
    O}(\lambda)$ and are shown in black.  Superimposed in blue
  are the curves $C$ and $C_\infty$.}
\label{fig:Configuration2}
\end{figure}

The contours where ${\bf O}(\lambda)$ has discontinuities in the complex
plane are illustrated with black curves in Figure~\ref{fig:Configuration2}.

\subsection{The choice of $g(\lambda)$.  Bands and gaps.}  
\label{sec:choiceofg}
Up until this point, the contours have been more or less arbitrary,
with the only conditions being that the region $D_{-1}$ contain the
imaginary interval $[0,iA]$ and the branch cut $C$.  Likewise, the
function $g(\lambda)$ remains undetermined aside from its analyticity
properties relative to the contours and the symmetry property
\eqref{eq:gsymmetry}.  We now will describe how to choose both the
system of contours and the function $g(\lambda)$ to render the
construction of ${\bf O}(\lambda)$ asymptotically tractable in the
limit $N\rightarrow\infty$.

Recall that on the contour $K_0$ we have the jump relation
\eqref{eq:K0jump} for ${\bf N}(\lambda)$.  Since $K_0$ lies in the
right-half plane according to $C\cup C_\infty\cup C^*\cup C_\infty^*$,
we may write the corresponding jump relation for ${\bf O}(\lambda)$ in
the form
\begin{equation}
{\bf O}_+(\lambda)={\bf O}_-(\lambda)\left[\begin{array}{cc}
0 & ie^{-[2iQ(\lambda)+L(\lambda)-i\theta^0(\lambda)-g_+(\lambda)-g_-(\lambda)]/\hbar}\\
ie^{[2iQ(\lambda)+L(\lambda)-i\theta^0(\lambda)
-g_+(\lambda)-g_-(\lambda)]/\hbar} &
0\end{array}\right]\,,
\hspace{0.2 in}\lambda\in K_0\,.
\label{eq:OjumpK0}
\end{equation}
Similarly, the jump relation satisfied by ${\bf O}(\lambda)$ for
$\lambda$ on the contour $K_{-1}$ follows from \eqref{eq:jumpNKm1} and
can be written in the form
\begin{equation}
{\bf O}_+(\lambda)={\bf O}_-(\lambda)\left[\begin{array}{cc}
e^{-[g_+(\lambda)-g_-(\lambda)]/\hbar} & 0 \\
iS(\lambda)e^{[2iQ(\lambda)+L(\lambda)
-i\theta^0(\lambda)-g_+(\lambda)-g_-(\lambda)]/
\hbar} & e^{[g_+(\lambda)-g_-(\lambda)]/\hbar}\end{array}\right]\,,\hspace{0.2 in}\lambda\in K_{-1}\,,
\label{eq:OjumpKm1}
\end{equation}
where we define
\begin{equation}
S(\lambda):=P(\lambda)e^{-L(\lambda)/\hbar}\,.
\end{equation}
The key observation at this point is that whether $\lambda\in K_0$
or $\lambda\in K_{-1}$, the exponents of the entries in the jump matrix
have the same form.  

We suppose that the contour loop $K_0\cup K_{-1}$ across which
$g(\lambda)$ is allowed discontinuities in the upper half-plane is
divided into two complementary systems of sub-arcs called {\em bands}
and {\em gaps}.  The characteristics of bands and gaps are most easily
phrased in terms of two auxiliary functions defined along $K_0\cup
K_{-1}$ that are related to the boundary values of $g(\lambda)$:
\begin{equation}
\theta(\lambda):= i(g_+(\lambda)-g_-(\lambda))\,,
\label{eq:thetadef}
\end{equation}
\begin{equation}
\phi(\lambda):= 2iQ(\lambda)+L(\lambda)-i\theta^0(\lambda)
-g_+(\lambda)-g_-(\lambda)\,.
\label{eq:phidef}
\end{equation}
In terms of these functions, we have the following definitions.
\begin{itemize}
\item A band is an arc along which the following two conditions hold:
\begin{equation}
\phi(\lambda)\equiv \text{imaginary constant,}\hspace{0.2 in}
\theta(\lambda) = \text{real decreasing function.}
\label{eq:bandconditions}
\end{equation}
We will require that the arc $K_0$ is a band, along (possibly) with
some sub-arcs of $K_{-1}$.
\item
A gap is an arc along which the following two conditions hold:
\begin{equation}
\Re(\phi(\lambda))<0\,,\hspace{0.2 in}
\theta(\lambda)\equiv\text{real constant.}
\label{eq:gapconditions}
\end{equation}
We will require that the terminal arc of $K_{-1}$ (that meets the real axis
at the point $\lambda=-\epsilon$) is a gap, along (possibly) with
some other sub-arcs of $K_{-1}$.
\end{itemize}
By ``decreasing'' in \eqref{eq:bandconditions} we mean in the
direction of orientation, namely counterclockwise.  Note however, that
while the orientation of a contour containing a band is essentially an
arbitrary choice, the condition that $\theta(\lambda)$ be decreasing
has intrinsic meaning because $\theta(\lambda)$ is by definition
proportional to $g_+(\lambda)-g_-(\lambda)$ which changes sign upon
reversal of orientation.  

Given an arbitrary set of contours, it may be the case that for no
function $g(\lambda)$ defined relative to these contours can the
conditions \eqref{eq:bandconditions} and \eqref{eq:gapconditions} both
be satisfied by choice of systems of bands and gaps.  Specification of
the contours is thus part of the problem of finding $g(\lambda)$.

To find $g(\lambda)$ and the contours along which its discontinuities
occur, we suppose that the number of bands along $K_0\cup K_{-1}$ is
known in advance, and show that certain conditions are then necessary
for the existence of $g(\lambda)$ for which the conditions
\eqref{eq:bandconditions} and \eqref{eq:gapconditions} both hold.
Taking a derivative, we see that the function $g'(\lambda)$ must satisfy
the following four conditions
\begin{equation}
g'_+(\lambda)+g'_-(\lambda)=2iQ'(\lambda)+L'(\lambda)
-i\theta^{0\prime}(\lambda)\,,\hspace{0.2 in}
\text{for $\lambda$ in the bands,}
\label{eq:gprimebands}
\end{equation}
\begin{equation}
g'_+(\lambda)-g'_-(\lambda)=0\,,\hspace{0.2 in}
\text{for $\lambda$ in the gaps,}
\label{eq:gprimegaps}
\end{equation}
\begin{equation}
g'(\lambda)+g'(\lambda^*)^* = 0\,,
\label{eq:gprimesymmetry}
\end{equation}
and
\begin{equation}
g'(\lambda)=O(\lambda^{-2})\,,\hspace{0.2 in}\text{as $\lambda
\rightarrow\infty$.}
\label{eq:gprimedecay}
\end{equation}
(Here we are assuming that differentiation with respect to $\lambda$
commutes with the limit of taking boundary values.  That this is
justified will be seen shortly.)  These four conditions amount to a
scalar Riemann-Hilbert problem for the function $g'(\lambda)$.

To solve for $g'(\lambda)$, it is most convenient to first modify the
unknown to eliminate the $L'(\lambda)$ term from
\eqref{eq:gprimebands}.  Let $C_f$ denote the contour $K_0\cup
K_{-1}\cup (-\epsilon,0)\cup C$ oriented in the direction from
$\lambda=0$ to $\lambda=iA$.  Set 
\begin{equation}
f_U(\lambda):=\frac{i}{2}\int_{C_f}\frac{d\eta}{\eta-\lambda}\,,\hspace{0.2 in}
\lambda\in\mathbb{C}\setminus C_f\,,
\label{eq:fU}
\end{equation}
and then 
\begin{equation}
f(\lambda):=f_U(\lambda)-f_U(\lambda^*)^*\,,\hspace{0.2 in}\lambda\in
\mathbb{C}\setminus (C_f\cup C_f^*)\,.
\label{eq:ffromfU}
\end{equation}
The change of variables we make is
\begin{equation}
h(\lambda):=g'(\lambda)+f(\lambda)\,.
\end{equation}
Note that if $\lambda\not\in D_{-1}$, then since $K_0\cup K_{-1}\cup
(-\epsilon,0) = \partial D_{-1}$, we may write $f_U(\lambda)$ in the
equivalent form
\begin{equation}
f_U(\lambda) = \frac{i}{2}\int_C\frac{d\eta}{\eta-\lambda} = 
\frac{i}{2}\int_0^{iA}\frac{d\eta}{\eta-\lambda}\,,\hspace{0.2 in}
\lambda\not\in D_{-1}\,,
\label{eq:fUequivoutside}
\end{equation}
from which it is clear that $f_U(\lambda) = A/2\lambda +
O(\lambda^{-2})$ as $\lambda\rightarrow\infty$.  Therefore
$f(\lambda)=O(\lambda^{-2})$ in this limit.  Also, from
\eqref{eq:ffromfU}, we have 
\begin{equation}
f(\lambda)+f(\lambda^*)^*=0\,.
\label{eq:fsymmetry}
\end{equation}  
The formula \eqref{eq:fUequivoutside} also shows that
\begin{equation}
f(\lambda) = -\frac{1}{2}L'(\lambda)\,,\hspace{0.2 in}\lambda\in\mathbb{C}\setminus (D_{-1}\cup D_{-1}^*)\,,
\label{eq:fL}
\end{equation}
and from the Plemelj formula, at each point of $K_0\cup K_{-1}$ and $C$ we have
\begin{equation}
f_+(\lambda)-f_-(\lambda) = -\pi=i\theta^{0\prime}(\lambda)\,,
\hspace{0.2 in}\lambda\in K_0\cup K_{-1}\cup C\,,
\end{equation}
while from the condition \eqref{eq:fsymmetry} and \eqref{eq:Lsymmetry}
we get that $f(\lambda)$ 
has a jump discontinuity across the real interval $(-\epsilon,0)$ given by
\begin{equation}
f_+(\lambda)-f_-(\lambda) = -2\pi\,,\hspace{0.2 in}\lambda\in (-\epsilon,0)\,,
\end{equation}
where orientation from left to right is understood (so $f_+$ is a
boundary value from the upper half-plane).  It follows that
$h(\lambda)$ is a function analytic for $\lambda\in\mathbb{C}\setminus
K_0\cup K_{-1}\cup C\cup K_0^*\cup K_{-1}^*\cup C^*\cup (-\epsilon,0)$
whose boundary values are related to those of $g'(\lambda)$ as
follows:
\begin{equation}
h_+(\lambda)+h_-(\lambda) = g'_+(\lambda)+g'_-(\lambda) - L'(\lambda)+i\theta^{0\prime}(\lambda)\,,\hspace{0.2 in}\lambda\in K_0\cup K_{-1}\,,
\label{eq:hpmgppm}
\end{equation}
\begin{equation}
h_+(\lambda)-h_-(\lambda) = g'_+(\lambda)-g'_-(\lambda)
-\pi \,,\hspace{0.2 in}\lambda\in K_0\cup K_{-1}\cup C\,,
\label{eq:hdiffgpdiff}
\end{equation}
and
\begin{equation}
h_+(\lambda)-h_-(\lambda) = -2\pi\,,\hspace{0.2 in}\lambda\in (-\epsilon,0)\,.
\label{eq:hjumpreal}
\end{equation}

Consequently, $h(\lambda)$ is characterized by \eqref{eq:hjumpreal}
and the following four conditions:
\begin{equation}
h_+(\lambda)+h_-(\lambda) = 2iQ'(\lambda)\,,\hspace{0.2 in}\text{for $\lambda$
in the bands of $K_0\cup K_{-1}$,}
\label{eq:hbands}
\end{equation}
\begin{equation}
h_+(\lambda)-h_-(\lambda) = -\pi\,,\hspace{0.2 in}
\text{for $\lambda$ in the gaps of $K_0\cup K_{-1}$ and on $C$,}
\label{eq:hgaps}
\end{equation}
\begin{equation}
h(\lambda)+h(\lambda^*)^* = 0\,,
\label{eq:hsymmmetry}
\end{equation}
and
\begin{equation}
h(\lambda) = O(\lambda^{-2})\,,\hspace{0.2 in}
\text{as $\lambda\rightarrow\infty$.}
\label{eq:hdecay}
\end{equation}
These conditions make up a scalar Riemann-Hilbert problem for
$h(\lambda)$.  Note that the boundary values taken by $h(\lambda)$ are
continuous except at $\lambda=0$, where a logarithmic singularity is
allowed (and necessary, to cancel the corresponding singularity of
$f(\lambda)$).

To solve for $h(\lambda)$ and thus obtain $g'(\lambda)$, we suppose
that the endpoints of the bands along $K_0\cup K_{-1}$ are given:
\begin{equation}
\lambda_0,\lambda_1,\dots,\lambda_G\,,\hspace{0.2 in}
\text{in order along $K_0\cup K_{-1}$, for $G$ even.}
\end{equation}
(Despite similar notation, these are not directly related to the soliton
eigenvalues $\lambda_{N,k}$.)  We then introduce
the square-root function $R(\lambda)$ defined by the equation
\begin{equation}
R(\lambda)^2 = \prod_{n=0}^G (\lambda-\lambda_n)(\lambda-\lambda_n^*)\,,
\end{equation}
and the condition that the branch cuts are the bands of $K_0\cup
K_{-1}$ and their complex conjugates, and that $R(\lambda) =
\lambda^{G+1} + O(\lambda^G)$ as $\lambda\rightarrow\infty$.  We attempt
to solve for $h(\lambda)$ by writing it in the form
\begin{equation}
h(\lambda)=R(\lambda)k(\lambda)\,,
\label{eq:hRk}
\end{equation}
for some other unknown function $k(\lambda)$.  Since
$R(\lambda)=R(\lambda^*)^*$ and since the jump discontinuities of
$R(\lambda)$ are restricted to the bands and their conjugates, 
where they satisfy $R_+(\lambda)=-R_-(\lambda)$ for $\lambda$ in the
bands of $K_0\cup K_{-1}$, the conditions imposed on $h(\lambda)$ take
the form of conditions on $k(\lambda)$ as follows:
\begin{equation}
k_+(\lambda)-k_-(\lambda)=\frac{2iQ'(\lambda)}{R_+(\lambda)}\,,\hspace{0.2 in}
\text{for $\lambda$ in the bands of $K_0\cup K_{-1}$,}
\label{eq:kcond1}
\end{equation}
\begin{equation}
k_+(\lambda)-k_-(\lambda)= -\frac{\pi}{R(\lambda)}\,,\hspace{0.2 in}
\text{for $\lambda$ in the gaps of $K_0\cup K_{-1}$ and on $C$,}
\label{eq:kcond2}
\end{equation}
\begin{equation}
k_+(\lambda)-k_-(\lambda) = -\frac{2\pi}{R(\lambda)}\,,\hspace{0.2 in}
\text{for $\lambda\in (-\epsilon,0)$,}
\label{eq:kcond3}
\end{equation}
\begin{equation}
k(\lambda)+k(\lambda^*)^* = 0\,,
\label{eq:kcond4}
\end{equation}
and
\begin{equation}
k(\lambda) = O(\lambda^{-(G+3)})\,,\hspace{0.2 in}\text{as $\lambda
\rightarrow\infty$.}
\label{eq:kdecay}
\end{equation}
We allow $k(\lambda)$ to have singularities at the band endpoints, as
long as the product $R(\lambda)k(\lambda)$ is regular there, and to
have a logarithmic singularity at $\lambda=0$.

Since the differences of boundary values of $k(\lambda)$ are known, it
is easy to see that it is necessary given this information that $k(\lambda)$
has the following form:
\begin{equation}
k_U(\lambda):=k_U^{(1)}(\lambda)+k_U^{(2)}(\lambda)\,,\hspace{0.2 in}
k_U^{(1)}(\lambda):= \frac{1}{\pi}\int_{\text{bands}\subset C_f}\frac{Q'(\eta)\,d\eta}{R_+(\eta)(\eta-\lambda)}\,,\hspace{0.2 in}
k_U^{(2)}(\lambda):= -\frac{1}{2i}\int_{C_f \setminus \text{bands}}
\frac{d\eta}{R(\eta)(\eta-\lambda)}\,,
\end{equation}
and
\begin{equation}
k(\lambda):=k_U(\lambda)-k_U(\lambda^*)^*\,.
\end{equation}
Because a residue calculation gives
\begin{equation}
k_U^{(1)}(\lambda)-k_U^{(1)}(\lambda^*)^* = \frac{iQ'(\lambda)}{R(\lambda)} - 2it\delta_{G,0}\,,
\end{equation}
the formula for $k(\lambda)$ becomes
\begin{equation}
k(\lambda)=
\frac{iQ'(\lambda)}{R(\lambda)} - 2it\delta_{G,0} + 
k_U^{(2)}(\lambda)-k_U^{(2)}(\lambda^*)^*\,.
\label{eq:kformula}
\end{equation}
The required properties \eqref{eq:kcond1}--\eqref{eq:kcond4} are
satisfied by this expression, and its singularities are easily seen to
be of the required types.  Here we can see also that as a consequence
of the analyticity of the densities of the Cauchy integrals used to
define $k(\lambda)$ and hence $g(\lambda)$, differentiation of
$g(\lambda)$ clearly commutes with taking boundary values, at least
away from the endpoints of the bands.

On the other hand, the decay condition \eqref{eq:kdecay} is not
necessarily satisfied by our formula for $k(\lambda)$.  By explicit
expansion of \eqref{eq:kformula} for large $\lambda$, we see that for
\eqref{eq:kdecay} to hold the following additional conditions are
required.  For any integer $p\ge 0$, define
\begin{equation}
M_p(\lambda_0,\dots,\lambda_G):=
\Re\left(\int_{C_f\setminus\text{bands}}\frac{\eta^p\,d\eta}{R(\eta)}
\right)\,.
\label{eq:Mpdef}
\end{equation}
If $G=0$, then the condition \eqref{eq:kdecay} requires that
\begin{equation}
\begin{array}{rcl}
\displaystyle
M_0(\lambda_0)
&= & x+2ta_0\,,\\\\
\displaystyle M_1(\lambda_0)
&=&\displaystyle xa_0 + 2t\left(a_0^2-\frac{1}{2}b_0^2\right)\,,
\end{array}
\label{eq:moments0}
\end{equation}
while for even integers $G\ge 2$, \eqref{eq:kdecay} requires that
\begin{equation}
\begin{array}{rcl}
\displaystyle
M_p(\lambda_0,\dots,\lambda_G)
&=&
0\,,\hspace{0.2 in}\text{for $0\le p\le G-2$,}\\\\
\displaystyle M_{G-1}(\lambda_0,\dots,\lambda_G)
&=& 2t\,,\\\\
\displaystyle M_{G}(\lambda_0,\dots,\lambda_G)
&=&
\displaystyle x+2t\sum_{n=0}^Ga_n\,,\\\\
\displaystyle M_{G+1}(\lambda_0,\dots,\lambda_G)
&=&\displaystyle
x\sum_{n=0}^G a_n + 2t\sum_{n=0}^G\left(a_n^2-\frac{1}{2}b_n^2\right) +
2t\sum_{n=0}^G\sum_{m=n+1}^G a_ma_n\,.
\end{array}
\label{eq:momentsG}
\end{equation}
In these formulae, $a_n:=\Re(\lambda_n)$ and $b_n:=\Im(\lambda_n)$.
These are $G+2$ real constraints that must be satisfied by choice of
the $G+1$ complex numbers $\lambda_0,\dots,\lambda_G$ in the upper
half-plane.

For each configuration of contours and band endpoints consistent with
the conditions \eqref{eq:moments0} or \eqref{eq:momentsG}, we
therefore obtain a candidate for the function $g(\lambda)$ by the
formula
\begin{equation}
g(\lambda):=-\int_\lambda^\infty g'(\eta)\,d\eta = -\int_\lambda^\infty
\left[R(\eta)k(\eta)-f(\eta)\right]\,d\eta\,,
\label{eq:gfromgprime}
\end{equation}
where the integration path is an arbitrary path from $\lambda$ to
infinity in the region $\mathbb{C}\setminus (D_{-1}\cup D_{-1}^*)$ if
$\lambda$ lies in this region as well, whereas if $\lambda\in
D_{-1}\cup D_{-1}^*$, then a first component of the path lies in this
region and connects $\lambda$ to $-\epsilon$, followed by a second
component that coincides with the real half-line
$(-\infty,-\epsilon)$.  We may then attempt to enforce on $g(\lambda)$
the conditions that $\Re(\phi(\lambda))=0$ within each band and
$\Im(\theta(\lambda))=0$ within each gap.  For $n=1,\dots,G/2$, let
$A_n$ denote a simple closed contour with positive orientation that
surrounds the band with endpoints $\lambda_{2n-1}$ and $\lambda_{2n}$
and no other discontinuities of $g'(\lambda)$.  The
definition \eqref{eq:thetadef} and the integral formula
\eqref{eq:gfromgprime} for $g(\lambda)$ shows that
$\theta(\lambda)\equiv 0$ in the terminal gap of $K_{-1}$ (from the
point $\lambda=\lambda_{G}$ along $K_{-1}$ to the point
$\lambda=-\epsilon$). Therefore, $\theta(\lambda)$, already made constant in
the remaining gaps via the jump conditions imposed on $g'(\lambda)$,
will have a purely real value in each gap if
\begin{equation}
\Re\left(\oint_{A_n}g'(\eta)\,d\eta\right)=0\,,\hspace{0.2 in}
n=1,\dots,G/2\,.
\label{eq:bandints}
\end{equation}
Similarly, for $n=1,\dots,G/2$, let $\Gamma_n$ be a contour arc
representing the gap in $K_{-1}$ between $\lambda_{2n-2}$ and
$\lambda_{2n-1}$.  The definition \eqref{eq:phidef} shows that (due to
the symmetries \eqref{eq:Lsymmetry} and \eqref{eq:gsymmetry}) $\phi(\lambda)$
is already purely imaginary in the band $K_0$.  Consequently, its constant
value in the remaining bands will be purely imaginary also if
\begin{equation}
\Re
\left(\int_{\Gamma_n}\left[2g'(\eta)-2iQ'(\eta)-L'(\eta)+
i\theta^{0\prime}(\eta)\right]\,d\eta\right)=0\,,
\hspace{0.2 in}n=1,\dots,G/2\,.
\label{eq:gapints}
\end{equation}
If $G$ is an even positive number, then the conditions
\eqref{eq:momentsG}, \eqref{eq:bandints}, and \eqref{eq:gapints} taken
together are real equations sufficient in number to determine the real
and imaginary parts of the complex endpoints
$\lambda_0,\dots,\lambda_G$ along $K_0\cup K_{-1}$.  If $G=0$, then
there are no conditions of the form \eqref{eq:bandints} or
\eqref{eq:gapints} and the equations \eqref{eq:moments0} are expected
to determine the single endpoint $\lambda_0$.

The conditions \eqref{eq:bandints} and \eqref{eq:gapints} can be written
in a common form that is useful for computation.  First, note that
since $g'_+(\eta)=g'_-(\eta)=g'(\eta)$ for $\eta$ in any gap 
$\Gamma_n$, from \eqref{eq:hpmgppm} we get
\begin{equation}
2g'(\eta)-2iQ'(\eta)-L'(\eta)+i\theta^{0\prime}(\eta) = 
h_+(\eta)+h_-(\eta)-2iQ'(\eta)\,,\hspace{0.2 in}\eta\in\Gamma_n\,.
\end{equation}
Then, using \eqref{eq:hRk}, 
\begin{equation}
2g'(\eta)-2iQ'(\eta)-L'(\eta)+i\theta^{0\prime}(\eta) = 
R(\eta)\left[k_+(\eta)+k_-(\eta) -\frac{2iQ'(\eta)}{R(\eta)}\right]\,,
\hspace{0.2 in}\eta\in \Gamma_n\,.
\end{equation}
Using \eqref{eq:kformula} in the case $G>0$, we then get
\begin{equation}
2g'(\eta)-2iQ'(\eta)-L'(\eta)+i\theta^{0\prime}(\eta) = 
R(\eta)\left[k^{(2)}_+(\eta)+k^{(2)}_-(\eta)\right]\,,
\hspace{0.2 in}\eta\in\Gamma_n\,,
\end{equation}
where $k^{(2)}(\eta):=k^{(2)}_U(\lambda)-k^{(2)}_U(\lambda^*)^*$.  For
$\eta$ in a gap $\Gamma_n$ of $K_{-1}$, an elementary contour
deformation argument shows that $k^{(2)}_+(\eta)+k^{(2)}_-(\eta) =
Y(\eta)$, where
\begin{equation}
Y(\eta):=Y_U(\eta)-Y_U(\eta^*)^*\,,\hspace{0.2 in}
\text{and}\hspace{0.2 in}
Y_U(\eta):=\frac{i}{2}\int_{-\infty}^\infty \frac{d\nu}
{R(\nu)(\nu-\eta)} +i\int_C \frac{d\nu}{R(\nu)(\nu-\eta)}\,.
\label{eq:Y}
\end{equation}
(Note that the integrand in $Y_U(\eta)$ has a jump discontinuity on
the real axis at $\nu=0$ due to the factor $R(\eta)$ in the
denominator.)
To simplify the conditions \eqref{eq:bandints}, we let $I_n$ denote the
band arc connecting $\lambda_{2n-1}$ to $\lambda_{2n}$, and note that
\eqref{eq:bandints} can be written in the equivalent form
\begin{equation}
\Re\left(\int_{I_n}\left[g'_+(\eta)-g'_-(\eta)\right]\,d\eta\right) = 0\,,
\hspace{0.2 in}n=1,\dots,G/2\,.
\end{equation}
Indeed, this form is more natural given the connection of these
conditions to the net change in the function $\theta(\lambda)$ as
$\lambda$ moves through the band $I_n$.  Using \eqref{eq:hdiffgpdiff},
we have
\begin{equation}
g'_+(\eta)-g'_-(\eta) = h_+(\eta)-h_-(\eta)+\pi\,,\hspace{0.2 in}\eta\in I_n\,,
\end{equation}
and then from \eqref{eq:hRk},
\begin{equation}
g'_+(\eta)-g'_-(\eta) = R_+(\eta)\left[k_+(\eta)+k_-(\eta)+
\frac{\pi}{R_+(\eta)}\right]\,,\hspace{0.2 in}\eta\in I_n\,,
\end{equation}
because $R(\eta)$ changes sign across the branch cut $I_n$.  Another
contour deformation argument then shows that the combination
$k_+(\eta)+k_-(\eta)+\pi/R_+(\eta)$ may, for $\eta\in I_n$, be
identified with the same function $Y(\eta)$ as defined by
\eqref{eq:Y}.  Therefore, the conditions \eqref{eq:bandints} and
\eqref{eq:gapints} may be written similarly as
\begin{equation}
\left.\begin{array}{rcccl}
\displaystyle
R_n(\lambda_0,\dots,\lambda_G)&:=&\displaystyle
\Re\left(\int_{\lambda_{2n-1}}^{\lambda_{2n}} 
R(\eta)Y(\eta)\,d\eta\right)&=&0
\\\\ \displaystyle
V_n(\lambda_0,\dots,\lambda_G)&:=&\displaystyle
\Re\left(\int_{\lambda_{2n-2}}^{\lambda_{2n-1}} 
R(\eta)Y(\eta)\,d\eta\right)&=&0
\end{array}
\right\}\hspace{0.2 in}n=1,\dots,G/2\,,
\label{eq:bandgapconds}
\end{equation}
where the paths of integration lie in the region of analyticity of the
integrand.  Now, strictly speaking, this does not amount to a
definition of functions $R_n(\lambda_0,\ldots,\lambda_G)$ and
$V_n(\lambda_0,\ldots,\lambda_G)$ because the integrals are not
individually independent of path due to monodromy about the branch
cuts of $R(\eta)$.  However, the totality of the conditions
\eqref{eq:bandgapconds} is clearly independent of any particular
choice of paths (for example, adding to the path from $\lambda_0$ to
$\lambda_1$ a circuit about the branch cut of $R(\eta)$ connecting
$\lambda_1$ and $\lambda_2$ amounts to adding to $V_1$ a multiple of
$R_1$, which is zero on a configuration satisfying
\eqref{eq:bandgapconds}).

\subsubsection{Whitham equations.}
\label{sec:Whitham}
The endpoints $\lambda_0,\dots,\lambda_G$ are determined implicitly as
functions of $x$ and $t$ through the equations $E_n(\vec{v})=0$ for
$n=1,\dots,2G+2$, where the unknowns are
$\vec{v}=(\lambda_0,\dots,\lambda_G,\lambda_0^*,\dots,\lambda_G^*)^T$
and the equations are
\begin{equation}
E_n(\vec{v}):=V_{(n+1)/2}(\vec{v})\,,\hspace{0.2 in}
\text{for $n$ odd,}
\end{equation}
and
\begin{equation}
E_n(\vec{v}):=R_{n/2}(\vec{v})\,,\hspace{0.2 in}
\text{for $n$ even,}
\end{equation}
for $n$ in the range $n=1,\dots,G$, and then
\begin{equation}
E_n(\vec{v}):=M_{n-G-1}(\vec{v})\,,\hspace{0.2 in}
\text{for $n=G+1,\dots,2G-1$,}
\end{equation}
\begin{equation}
E_{2G}(\vec{v}):=M_{G-1}(\vec{v})-2t\,,
\end{equation}
\begin{equation}
E_{2G+1}(\vec{v}):=M_G(\vec{v})-\left[x+2t\sum_{n=0}^Ga_n
\right]\,,
\end{equation}
and
\begin{equation}
E_{2G+2}(\vec{v}):=M_{G+1}(\vec{v})-
\left[x\sum_{n=0}^Ga_n+2t\sum_{n=0}^G\left(a_n^2-\frac{1}{2}b_n^2\right)
+2t\sum_{n=0}^G\sum_{m=n+1}^G a_ma_n\right]\,.
\end{equation}
In these equations, $a_k=(\lambda_k+\lambda_k^*)/2$,
$b_k=(\lambda_k-\lambda_k^*)/(2i)$, and $M_n(\vec{v})$,
$R_n(\vec{v})$, and $V_n(\vec{v})$ stand for the complexification of
the corresponding real quantities.  That is,
\begin{equation}
M_p(\vec{v}):=\frac{1}{2}\int_C\frac{\eta^p\,d\eta}{R(\eta)} - 
\frac{1}{2}\int_{C^*}\frac{\eta^p\,d\eta}{R(\eta)} + 
\frac{1}{2}\int_{-W}^W\frac{\eta^p\,d\eta}{R(\eta)} +\frac{iW^{p+1}}{4}
\int_{0}^\pi \frac{e^{i(p+1)\theta}\,d\theta}{R(We^{i\theta})} -
\frac{iW^{p+1}}{4}\int_{-\pi}^0\frac{e^{i(p+1)\theta}\,d\theta}{R(We^{i\theta})}\,,
\end{equation}
where the contour $C^*$ is taken to be oriented from $\eta=-iA$ to
$\eta=0$ and $W>0$ is a sufficiently large number (if $p<G$ then we
may pass to the limit $W\rightarrow\infty$ and drop the last two
integrals).  The complexified $M_p$ agrees with the expression defined
by \eqref{eq:Mpdef} when $\lambda_k=a_k+ib_k$ and
$\lambda_k^*=a_k-ib_k$ with $a_k$ and $b_k$ being restricted to real
values.  The complexified $M_p$ is a function of the independent
complex variables $\lambda_k$ and $\lambda_k^*$ through the
branch points of the function $R$ in the integrand.  Similarly,
\begin{equation}
R_n(\vec{v}):=\frac{1}{2}\int_{\lambda_{2n-1}}^{\lambda_{2n}}R(\eta)Y(\eta)\,d\eta -\frac{1}{2}\int_{\lambda_{2n}^*}^{\lambda_{2n-1}^*}R(\eta)Y(\eta)\,d\eta\,,
\end{equation}
and
\begin{equation}
V_n(\vec{v}):=\frac{1}{2}\int_{\lambda_{2n-2}}^{\lambda_{2n-1}}R(\eta)Y(\eta)\,d\eta -\frac{1}{2}\int_{\lambda_{2n-1}^*}^{\lambda_{2n-2}^*}R(\eta)Y(\eta)\,d\eta\,,
\end{equation}
where in both cases the integrals in the two terms are taken over
complex-conjugated paths.  Simple contour deformations near the
endpoints of integration show that in each case derivatives of
$R_n(\vec{v})$ and $V_n(\vec{v})$ with respect to any of the $v_k$ at
all can be calculated by differentiation under the integral sign (even
if $v_k$ is one of the endpoints of integration).  Once again, these
complexified quantities, while functions of the independent complex
variables $v_1,\dots,v_{2G+2}$, agree with the previous definitions
when $v_{k+G+1}=v_k^*$.  Of course we are only interested in those
solutions of the equations $E_n(\vec{v})=0$ that have this conjugation
symmetry.

If $\vec{v}(x,t)$ is a differentiable solution of the equations
$E_n(\vec{v})=0$ for $n=1,\dots,2G+2$, then we may calculate
$\partial\vec{v}/\partial x$ and $\partial\vec{v}/\partial t$ by implicit
differentiation.  Thus, 
\begin{equation}
\frac{\partial\vec{E}}{\partial\vec{v}}\cdot\frac{\partial\vec{v}}{\partial x}=
-\frac{\partial\vec{E}}{\partial x}\,,\hspace{0.2 in}
\text{and}\hspace{0.2 in}
\frac{\partial\vec{E}}{\partial\vec{v}}\cdot\frac{\partial\vec{v}}{\partial t}=
-\frac{\partial\vec{E}}{\partial t}\,,
\label{eq:implicitdifferentiation}
\end{equation}
where $\partial\vec{E}/\partial\vec{v}$ denotes the Jacobian matrix of
the $E_n$ with respect to the $v_k$ holding $x$ and $t$ fixed, while
$\partial\vec{E}/\partial x$ and $\partial\vec{E}/\partial t$ are the
corresponding vectors of partial derivatives with respect to $x$ and
$t$ (holding the $v_k$ fixed).  Clearly, these latter partial
derivatives contain no explicit $x$ and $t$ dependence because the
$E_n$ are all linear functions of $x$ and $t$.  In fact, it turns out
that when $\vec{v}=\vec{v}(x,t)$ satisfies the equations
$\vec{E}(\vec{v})=0$, the Jacobian matrix can also be expressed up to
a diagonal factor in terms of $\vec{v}$ alone ($x$ and $t$ may be
eliminated).  Indeed, direct calculations show that
\begin{equation}
\frac{\partial}{\partial v_k}\left(R(\eta)Y(\eta)\right)=-\frac{1}{2}Y(v_k)
\frac{R(\eta)}{\eta-v_k}\,,
\label{eq:RYderiv}
\end{equation}
and 
\begin{equation}
Y(v_k)=4i\frac{\partial M_0}{\partial v_k}\,,
\label{eq:YM0relation}
\end{equation}
so
\begin{equation}
\frac{\partial E_n}{\partial v_k} = -i\frac{\partial M_0}{\partial v_k}
\left[\int_{\lambda_{n-1}}^{\lambda_{n}}\frac{R(\eta)\,d\eta}{\eta-v_k}
-\int_{\lambda_{n}^*}^{\lambda_{n-1}^*}\frac{R(\eta)\,d\eta}{\eta-v_k}
\right]=:J_{nk}(\vec{v})\frac{\partial M_0}{\partial v_k}\,,\hspace{0.2 in}\text{for $1\le n\le G$}\,.
\label{eq:integralderivs}
\end{equation}
Also by direct calculation, 
\begin{equation}
\frac{\partial E_n}{\partial v_k} = \frac{1}{2}E_{n-1}+v_k
\frac{\partial E_{n-1}}{\partial v_k}\,,\hspace{0.2 in}
\text{for $G+2\le n\le 2G+2$}\,,
\label{eq:Momentderivsrelation}
\end{equation}
so on a solution $\vec{v}(x,t)$, 
\begin{equation}
\frac{\partial E_n}{\partial v_k} = v_k^{n-1}\frac{\partial E_{G+1}}
{\partial v_k}
= v_k^{n-1}\frac{\partial M_0}{\partial v_k}=:J_{nk}(\vec{v})\frac{\partial M_0}{\partial v_k}\,,\hspace{0.2 in}\text{for $G+1\le n\le 2G+2$\,,}
\label{eq:Momentderivs}
\end{equation}
assuming that $G>0$.  This proves that on a solution
$\vec{v}=\vec{v}(x,t)$, the Jacobian matrix can be expressed in the form
\begin{equation}
\frac{\partial \vec{E}}{\partial \vec{v}} = {\bf J}(\vec{v})\cdot
\text{diag}\left(\frac{\partial M_0}{\partial v_1},\dots,
\frac{\partial M_0}{\partial v_{2G+2}}\right)\,,
\end{equation}
where the matrix ${\bf J}(\vec{v})$ is an explicit function of
$\vec{v}$ whose elements are defined by \eqref{eq:integralderivs} and
\eqref{eq:Momentderivs}.  Its determinant is nonzero as long as the
$v_k$ are distinct.  Applying Cramer's rule to
\eqref{eq:implicitdifferentiation}, we then find that
\begin{equation}
\frac{\partial v_k}{\partial t} + c_k(\vec{v})\frac{\partial v_k}{\partial x}=0\,,\hspace{0.2 in}\text{for $k=1,\dots,2G+2$}\,,
\label{eq:WhithamI}
\end{equation}
where
\begin{equation}
c_k(\vec{v}) = -\frac{\det {\bf J}^{k,t}}{\det {\bf J}^{k,x}}\,,
\label{eq:WhithamIspeeds}
\end{equation}
where ${\bf J}^{k,t}$ (respectively ${\bf J}^{k,x}$) denotes the
matrix ${\bf J}$ with the $k$th column replaced by
$\partial\vec{E}/\partial t$ (respectively $\partial\vec{E}/\partial
x$).  The system of quasilinear partial differential equations
\eqref{eq:WhithamI} satisfied by the endpoints $\vec{v}(x,t)$ is
automatically in Riemann invariant (diagonal) form, regardless of how
large $G$ is.  These equations are frequently called the {\em Whitham
  equations}.  They clearly play a secondary role in our analysis, as
they were derived from the algebraic equations $E_n(\vec{v})=0$ which
are more fundamental (and in particular encode initial data).

\subsubsection{Inequalities and topological conditions.}
\label{sec:inequalities}
Subject to being able to solve for the endpoints
$\lambda_0,\dots,\lambda_G$, there is now a candidate for the function
$g(\lambda)$ associated with each even nonnegative integer $G$.  We
refer to such a guess for $g(\lambda)$ below as a {\em genus-$G$
  ansatz}.  The selection principle for the number $G$ is that it must
be chosen so that the inequalities
\begin{equation}
\begin{array}{l}
\text{$\Re(\phi(\lambda))<0$ for $\lambda$ 
in each gap of $K_0\cup K_{-1}$,}\\\\
\text{$\theta(\lambda)$ 
is strictly decreasing along bands of $K_0\cup K_{-1}$}
\end{array}
\end{equation}
both hold true.  Given that the actual band and gap arcs have not yet
been chosen, these conditions are really topological conditions on the
level curves of the real part of the integral
\begin{equation}
I(\lambda)=\int^\lambda R(\eta)Y(\eta)\,d\eta\,.
\end{equation}
(These level curves are also known in the literature as the orthogonal
trajectories of the quadratic differential
$R(\lambda)^2Y(\lambda)^2d\lambda^2$.)  A band arc of $K_0\cup K_{-1}$
must coincide with a level curve of $\Re(I(\lambda))$ connecting the origin
with $\lambda_0$, or $\lambda_{2n-1}$ with $\lambda_{2n}$ for $n=1,\dots,G/2$.
Furthermore, it must be possible to choose the remaining arcs (gaps) so that
they lie in the region where $\Re(I(\lambda))$ is less than at either
endpoint.  

\subsection{Third modification:  opening lenses around bands 
(steepest descent).}
\label{sec:steepestdescent}
In terms of the functions $\theta(\lambda)$ and $\phi(\lambda)$, the
jump discontinuity of ${\bf O}(\lambda)$ across $K_{-1}$ takes the
form
\begin{equation}
{\bf O}_+(\lambda)={\bf O}_-(\lambda)\left[\begin{array}{cc}
e^{i\theta(\lambda)/\hbar} & 0 \\
iS(\lambda)e^{\phi(\lambda)/\hbar} & e^{-i\theta(\lambda)/\hbar}
\end{array}\right]\,.
\label{eq:OKm1jump}
\end{equation}
Assuming that $K_{-1}$ remains bounded away from the imaginary
interval $[0,iA]$ of accumulation of poles for ${\bf m}(\lambda)$ by a
fixed distance, we have by a midpoint rule analysis for Riemann sums
that $S(\lambda) = 1+O(\hbar)$.  If $\lambda$ is a point in a gap
$\Gamma_j\subset K_{-1}$, then $\theta(\lambda)\equiv
\theta_j\in\mathbb{R}$, and $\Re(\phi(\lambda))<0$, so the jump matrix
in \eqref{eq:OKm1jump} is an exponentially small perturbation of the
constant (with respect to $\lambda\in\Gamma_j$) jump matrix
$e^{i\theta_j\sigma_3/\hbar}$.

\begin{figure}[htbp]
\begin{center}
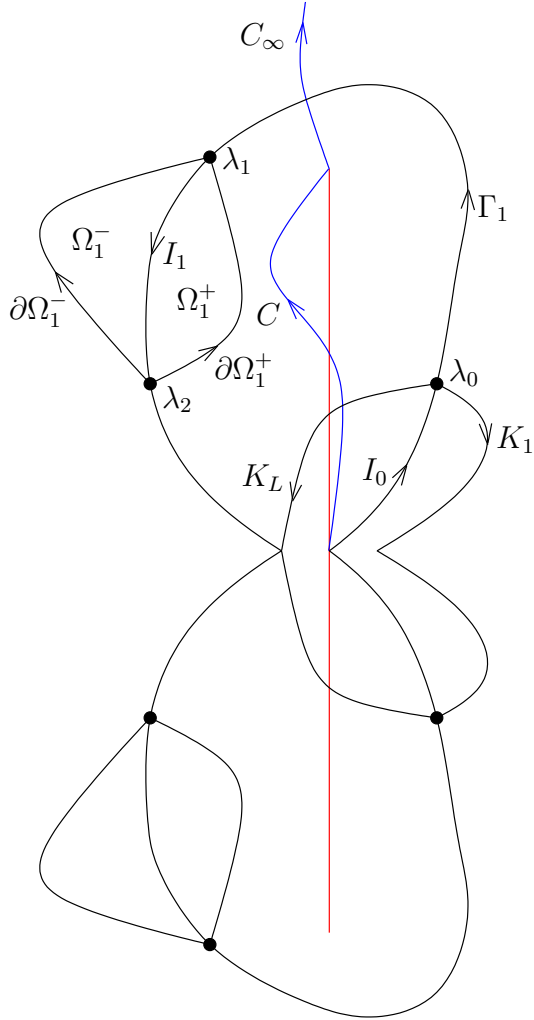
\end{center}
\caption{\em The discontinuity contours for ${\bf P}(\lambda)$.}
\label{fig:Configuration3}
\end{figure}

On the other hand, in direct analogy with the factorization
\eqref{eq:K0factorization}, the jump discontinuity of ${\bf
  O}(\lambda)$ across $K_{-1}$ can be written in factorized form:
\begin{equation}
\begin{array}{l}
\displaystyle
\left[\begin{array}{cc} e^{i\theta(\lambda)/\hbar} & 0 \\
iS(\lambda)e^{\phi(\lambda)/
\hbar} & e^{-i\theta(\lambda)/\hbar}
\end{array}\right]=\\\\
\displaystyle\hspace{0.5 in}
\left[\begin{array}{cc} S(\lambda)^{-1/2} & -iS(\lambda)^{-1/2}
e^{-\phi(\lambda)/\hbar}e^{i\theta(\lambda)/\hbar}\\
0 & S(\lambda)^{1/2}\end{array}\right]\times\\\\
\displaystyle\hspace{0.5 in}
\left[\begin{array}{cc} 0 & i
e^{-\phi(\lambda)/\hbar}
\\
ie^{\phi(\lambda)/\hbar}
& 0\end{array}\right]
\times \\\\\displaystyle\hspace{0.5 in}
\left[\begin{array}{cc} S(\lambda)^{1/2} & -iS(\lambda)^{-1/2}
e^{-\phi(\lambda)/\hbar}e^{-i\theta(\lambda)/\hbar}
\\
0 & S(\lambda)^{-1/2}\end{array}\right]\,.
\end{array}
\label{eq:Km1factorization}
\end{equation}
This factorization is useful for $\lambda$ in a band $I_j\subset
K_{-1}$.  (Recall that we are also assuming that the contour $K_0$ is
itself a band in its entirety; thus $K_0=I_0$).  Indeed, let
$\Omega_j^+$ (respectively $\Omega_j^-$) be a lens-shaped domain lying
to the left (respectively right) of the band $I_j\subset K_{-1}$.  Let
$i\kappa_j$ be the purely imaginary constant value of $\phi(\lambda)$
in the band $I_j$.  We introduce a new unknown ${\bf P}(\lambda)$
based on this factorization as follows:
\begin{equation}
{\bf P}(\lambda):={\bf O}(\lambda)\left[\begin{array}{cc}
S(\lambda)^{-1/2} & iS(\lambda)^{-1/2} e^{-i\kappa_j/\hbar}
e^{-i\theta(\lambda)/\hbar}\\0 & S(\lambda)^{1/2}\end{array}\right]\,,
\hspace{0.2 in}\lambda\in \Omega_j^+\,,
\end{equation}
\begin{equation}
{\bf P}(\lambda):={\bf O}(\lambda)\left[\begin{array}{cc}
S(\lambda)^{-1/2} & -iS(\lambda)^{-1/2}e^{-i\kappa_j/\hbar}
e^{i\theta(\lambda)/\hbar} \\ 0 & S(\lambda)^{1/2}\end{array}
\right]\,,\hspace{0.2 in}
\lambda\in\Omega_j^-\,,
\end{equation}
for all other $\lambda$ in the upper half-plane where ${\bf
  O}(\lambda)$ takes a definite value we set ${\bf P}(\lambda)={\bf
  O}(\lambda)$, and finally for all $\lambda$ in the lower half-plane
we set ${\bf P}(\lambda) = \sigma_2{\bf P}(\lambda^*)^*\sigma_2$.  In
writing down this change of variables we are making use of the fact,
apparent from the explicit formula for $g'(\lambda)$, that the
function $\theta(\lambda)$ has an analytic continuation from each band
$I_j\subset K_{-1}$ to the regions $\Omega_j^\pm$.  The contours in the
complex plane across which ${\bf P}(\lambda)$ has jump discontinuities
are shown with black curves in Figure~\ref{fig:Configuration3}.
A simple Cauchy-Riemann argument taking into account the monotonicity
of the real-valued analytic function $\theta(\lambda)$ in the bands
then shows that on the boundary contours $\partial\Omega_j^\pm$ (not
including the band $I_j$, as shown in Figure~\ref{fig:Configuration3})
the jump matrix converges to the identity matrix as $\hbar\rightarrow
0$.  The convergence is uniform away from the endpoints of the band,
with a rate of convergence $O(\hbar)$.

These heuristic arguments will be used to suggest in
\S~\ref{sec:parametrix} a model for ${\bf P}(\lambda)$ we call a {\em
  parametrix}, and then they will be recycled in \S~\ref{sec:error} to
prove that the parametrix is indeed an accurate model for ${\bf
  P}(\lambda)$.  For now, it suffices to note that the matrix ${\bf
  P}(\lambda)$ is the unique solution of a matrix Riemann-Hilbert problem
that by our explicit steps is equivalent to the discrete Riemann-Hilbert
problem satisfied by ${\bf m}(\lambda)$.  This problem is the following.
Seek a $2\times 2$ matrix ${\bf P}(\lambda)$ with entries that are piecewise
analytic functions of $\lambda$ in the complement of the contours $K_{-1}$,
$K_0$, $K_1$, $K_L$, the lens boundaries $\partial\Omega_j^\pm$, and their
complex conjugates such that
\begin{itemize}
\item The boundary values taken on each arc of the discontinuity contour
are continuous along the arc and have continuous extensions to the arc 
endpoints.  The boundary values are related by the following jump conditions.
For $\lambda\in I_j\subset K_{-1}$ or for $\lambda\in I_0=K_0$,
\begin{equation}
{\bf P}_+(\lambda)={\bf P}_-(\lambda)\left[\begin{array}{cc}
0 & ie^{-i\kappa_j/\hbar}\\
ie^{i\kappa_j/\hbar} & 0\end{array}\right]\,,
\hspace{0.2 in}\lambda\in I_j\,.
\end{equation}
On the arc $K_L$ we have
\begin{equation}
{\bf P}_+(\lambda)={\bf P}_-(\lambda)\left[\begin{array}{cc}
T(\lambda)^{-1/2} & iT(\lambda)^{-1/2}e^{-i\kappa_0/\hbar}e^{-i\theta(\lambda)/\hbar}\\0 & T(\lambda)^{1/2}\end{array}\right]\,,
\hspace{0.2 in}\lambda\in K_L\,,
\end{equation}
where $\theta(\lambda)$ refers to the value of the function
$\theta(\lambda)$ analytically continued from $I_0$ to $K_L$.  On the
arc $K_1$,
\begin{equation}
\begin{array}{rcl}
\displaystyle {\bf P}_+(\lambda)&=&\displaystyle
{\bf P}_-(\lambda)\left[\begin{array}{cc}
T(\lambda)^{1/2} & iT(\lambda)^{-1/2} e^{-i\kappa_0/\hbar}e^{i\theta(\lambda)/\hbar}\\0 & T(\lambda)^{-1/2}\end{array}\right]\\\\
&&\displaystyle\,\,\,\times\,\,\,
\left[\begin{array}{cc}
1 & 0 \\ iS(\lambda)e^{i\kappa_0/\hbar}e^{i(2\theta^0(\lambda)-\theta(\lambda))/\hbar} & 1
\end{array}\right]\,,
\hspace{0.2 in}\lambda\in K_1\,,
\end{array}
\end{equation}
where again $\theta(\lambda)$ refers to the value analytically continued from
$I_0$.
On a gap $\Gamma_j\subset K_{-1}$,
\begin{equation}
{\bf P}_+(\lambda)={\bf P}_-(\lambda)\left[\begin{array}{cc}
e^{i\theta_j/\hbar} & 0 \\
iS(\lambda)e^{\phi(\lambda)/\hbar} & e^{-i\theta_j/\hbar}\end{array}\right]\,,
\hspace{0.2 in}\lambda\in \Gamma_j\,.
\end{equation}
(This holds with $\theta_j=0$ on the final gap of $K_{-1}$ from
$\lambda_G$ to $-\epsilon$ because $g(\lambda)$ is analytic in this
gap so $\theta(\lambda)\equiv 0$.)  On the lens boundaries
$\partial\Omega_j^\pm$,
\begin{equation}
{\bf P}_+(\lambda)={\bf P}_-(\lambda)\left[\begin{array}{cc}
S(\lambda)^{\mp 1/2} & iS(\lambda)^{-1/2}e^{-i\kappa_j/\hbar}e^{\mp i\theta(\lambda)}\\0 & S(\lambda)^{\pm 1/2}\end{array}\right]\,,\hspace{0.2 in}
\lambda\in \partial\Omega_j^\pm\,,
\end{equation}
where here $\theta(\lambda)$ refers to the value analytically
continued from $I_j$.  Finally, the jump relations satisfied in the
lower half-plane are by definition consistent with the symmetry ${\bf
  P}(\lambda)=\sigma_2{\bf P}(\lambda^*)^*\sigma_2$.
\item The matrix ${\bf P}(\lambda)$ is normalized so that 
\begin{equation}
\lim_{\lambda\rightarrow\infty}{\bf P}(\lambda)=\mathbb{I}\,.
\end{equation}
\end{itemize}
The function defined in terms of ${\bf P}(\lambda)$ by the limit
\begin{equation}
\psi_N(x,t):=2i\lim_{\lambda\rightarrow\infty}\lambda 
P_{12}(\lambda)
\end{equation}
is the $N$-soliton solution of the semiclassically scaled focusing
nonlinear Schr\"odinger equation.

\subsection{Parametrix construction.}
\label{sec:parametrix}
Building a model for ${\bf P}(\lambda)$ consists of two steps: (i)
dealing first with the asymptotic behavior of the jump matrix away
from the endpoints of the bands, and (ii) local analysis near the
endpoints.  Both of these constructions have been described in detail
elsewhere, and we will give just an outline of the calculations.
\subsubsection{Pointwise asymptotics:  the outer model problem.}
As $N\rightarrow\infty$ (so in particular $\hbar=\hbar_N\rightarrow
0$) the jump matrix defining the ratio of boundary values in the
Riemann-Hilbert problem for ${\bf P}(\lambda)$ may be approximated by
a piecewise constant (with respect to $\lambda$) jump matrix that
differs from the identity matrix only in the bands of $K_0\cup K_{-1}$
and their complex conjugates and also in the nonterminal gaps of
$K_{-1}$.  Thus we may pose another Riemann-Hilbert problem whose
solution we hope we can prove is a good approximation in a certain
sense of ${\bf P}(\lambda)$.  We seek a $2\times 2$ matrix function
$\dot{\bf P}(\lambda)$ that is piecewise analytic in the complement
of the bands and nonterminal gaps and their complex conjugates such
that
\begin{itemize}
\item The boundary values taken on each band or gap of the
  discontinuity contour are continuous along the arc and have
  singularities of at worst inverse fourth-root type at the band/gap
  endpoints.  The boundary values are related by the following jump
  conditions.  For $\lambda\in I_j\subset K_{-1}$ or for $\lambda\in
  I_0=K_0$,
\begin{equation}
\dot{\bf P}_+(\lambda)=\dot{\bf P}_-(\lambda)\left[\begin{array}{cc}
0 & ie^{-i\kappa_j/\hbar}\\ie^{i\kappa_j/\hbar} & 0\end{array}\right]\,,
\hspace{0.2 in}\lambda\in I_j\,.
\end{equation}
For $\lambda$ in a nonterminal gap $\Gamma_j\subset K_{-1}$,
\begin{equation}
\dot{\bf P}_+(\lambda)=\dot{\bf P}_-(\lambda)\left[\begin{array}{cc}
e^{i\theta_j/\hbar} & 0 \\ 0 & e^{-i\theta_j/\hbar}\end{array}\right]\,,
\hspace{0.2 in}\lambda\in\Gamma_j\,.
\label{eq:Pdotgapjump}
\end{equation}
Finally, the jump relations satisfied by $\dot{\bf P}(\lambda)$ in the
lower half-plane are by definition consistent with the symmetry
$\dot{\bf P}(\lambda)=\sigma_2\dot{\bf P}(\lambda^*)^*\sigma_2$.
\item
The matrix $\dot{\bf P}(\lambda)$ is normalized so that
\begin{equation}
\lim_{\lambda\rightarrow\infty}\dot{\bf P}(\lambda)=\mathbb{I}\,.
\end{equation}
\end{itemize}
This Riemann-Hilbert problem can be solved by first introducing an
auxiliary scalar Riemann-Hilbert problem with the aim of removing the
jump discontinuities of $\dot{\bf P}(\lambda)$ across the nonterminal
gaps as expressed by the jump conditions \eqref{eq:Pdotgapjump} while
converting the jump matrix for $\dot{\bf P}(\lambda)$ in all of the
bands to the matrix $\sigma_1$.  The two columns of the resulting
matrix unknown thus may be considered to be restrictions of a single
vector-valued analytic function on a hyperelliptic Riemann surface $X$
constructed by identifying two copies of the complex $\lambda$-plane
across a system of cuts made in the bands and their complex
conjugates.  This Riemann surface has genus $G$, which explains our
terminology for the ``genus'' of a configuration of endpoints for the
$g$-function.  The vector-valued function defined on the surface $X$
that leads to the solution of the Riemann-Hilbert problem for
$\dot{\bf P}(\lambda)$ is known as a Baker-Akhiezer function.  It can
be expressed explicitly in terms of the Riemann theta function of $X$,
as can $\dot{\bf P}(\lambda)$.  The details of this construction can 
be found in \cite{KMM}.  

Two features of this solution are important for the subsequent steps
in the analysis.  First, the dependence of the solution on $\hbar$
enters through the real quantities $\theta_j/\hbar$ and
$\kappa_j/\hbar$, which determine a point in the real part of the
Jacobian variety of $X$, topologically a torus of dimension $G$.
Essentially, this point is a phase shift in the argument of the
Riemann theta functions used to construct the solution, and in
particular this implies that the matrix $\dot{\bf P}(\lambda)$ remains
uniformly bounded for $\lambda$ away from the band/gap endpoints in
the limit $\hbar\rightarrow 0$, even though the phase point oscillates
wildly in the Jacobian in this limit.  Next, for $\lambda$ near the
band/gap endpoints, the matrix $\dot{\bf P}(\lambda)$ exhibits a
singularity of a universal type that, while a poor model of ${\bf
  P}(\lambda)$ near the endpoints, nonetheless turns out to match well
onto another matrix function that is a better model.

\subsubsection{Endpoint asymptotics:  the Airy function local parametrix.}
To determine what this better model should be, it suffices to fix a
sufficiently small neighborhood $U_k$ of each band/gap endpoint
$\lambda_k$, and to find a matrix that {\em exactly} satisfies the
jump conditions of ${\bf P}(\lambda)$ in this neighborhood.  Such a
matrix can be found because the jump matrices restricted to $U_k$ can
be written in a canonical form with the use of an appropriate
conformal mapping (Langer transformation) taking $U_k$ to a
neighborhood of the origin.  Once the jump matrices in $U_k$ are
exhibited in canonical form, a piecewise analytic matrix function
satisfying the corresponding jump conditions can be written down
explicitly in terms of Airy functions.  Next one observes that in fact
there are many piecewise analytic matrices defined in $U_k$ satisfying
the exact jump conditions of ${\bf P}(\lambda)$, all differing only by
multiplication on the left by a matrix factor analytic in $U_k$.  The
choice of this factor can be used to single out a particular local
solution that is a good match onto the explicit matrix $\dot{\bf
  P}(\lambda)$ on the boundary $\partial U_k$ of $U_k$.  Specifically,
one chooses the factor so that the resulting local solution, which we
will call $\hat{\bf P}_k(\lambda)$, satisfies
\begin{equation}
\hat{\bf P}_k(\lambda)\dot{\bf P}(\lambda)^{-1} = \mathbb{I} + O(\hbar)\,,
\end{equation}
as $\hbar\rightarrow 0$, uniformly for $\lambda\in \partial U_k$.

The Airy function local parametrix is described in detail, for
example, in \cite{DOP}.  Here, we need a slight modification of the
construction of \cite{DOP}, because the jump matrices for ${\bf
  P}(\lambda)$ restricted to $U_k$ involve the function $S(\lambda)$,
and also the function $T(\lambda)$ in the case of $U_0$.  We can
easily remove these functions from the jump matrices by making a local
change of variables in $U_k$ as the first step in the construction of
$\hat{\bf P}_k(\lambda)$.  Suppose first that $k>0$ and in the part
of $U_k$ lying outside of the lenses we set
\begin{equation}
{\bf Q}(\lambda)={\bf P}(\lambda)\left[\begin{array}{cc} S(\lambda)^{-1/2} &
0 \\ 0 & S(\lambda)^{1/2}\end{array}\right]\,,
\end{equation}
while in the rest of $U_k$ we set ${\bf Q}(\lambda)={\bf P}(\lambda)$.
It is then easy to check that the matrix ${\bf Q}(\lambda)$ satisfies
the same jump conditions as does ${\bf P}(\lambda)$ but with
$S(\lambda)$ simply replaced by $1$.  This turns out to be a
near-identity transformation since $S(\lambda)=1+O(\hbar)$ uniformly
for $\lambda\in U_k$.  Next, consider the jump conditions satisfied by
${\bf P}(\lambda)$ in $U_0$.  In the part of $U_0$ common to $D_{-1}$
but outside of the lenses $D_1$ and $D_L$ we set
\begin{equation}
{\bf Q}(\lambda)={\bf P}(\lambda)\left[\begin{array}{cc} T(\lambda)^{-1/2} & 0
\\ 0 & T(\lambda)^{1/2}\end{array}\right]\,,
\end{equation}
while in the part of $U_0$ outside both the lenses and $D_{-1}$ we set
\begin{equation}
{\bf Q}(\lambda)={\bf P}(\lambda)\left[\begin{array}{cc}
1 & 0 \\ -iS(\lambda) e^{i\kappa_0/\hbar}e^{i(2\theta^0(\lambda)-\theta(\lambda))/\hbar} & 1\end{array}\right]
\left[\begin{array}{cc} T(\lambda)^{-1/2} & 0 \\ 0 & T(\lambda)^{1/2}
\end{array}\right]\,,
\end{equation}
and in the remaining parts of $U_0$ we set ${\bf Q}(\lambda)={\bf P}(\lambda)$.
Using the relationship between $L(\lambda)$ and $\overline{L}(\lambda)$ valid
in the right half-plane according to $C\cup C_\infty\cup C^*\cup C_\infty^*$
it then follows that on $\Gamma_1\cap U_0$, $K_L\cap U_0$, and $I_0\cap U_0$,
the jump conditions satisfied by ${\bf Q}(\lambda)$ are of the same form
as those satisfied by ${\bf P}(\lambda)$ but with $S(\lambda)$ and $T(\lambda)$
both replaced by $1$, while for $\lambda\in K_1\cap U_0$,
\begin{equation}
{\bf Q}_+(\lambda)={\bf Q}_-(\lambda)\left[\begin{array}{cc}
1 & ie^{-i\kappa_0/\hbar}e^{i\theta(\lambda)/\hbar}\\0 & 1\end{array}\right]\,.
\end{equation}
From this point, the construction follows that in \cite{DOP}
precisely, with ${\bf Q}(\lambda)$ being studied within each $U_k$ by
means of an appropriate Langer transformation.

\subsubsection{Global parametrix.}
We now propose the following global parametrix, $\hat{\bf
  P}(\lambda)$, as a model for ${\bf P}(\lambda)$ uniformly valid in
the whole complex plane.  The matrix is well-defined globally with the
exception of certain contours on which continuous boundary values are
taken from each side:
\begin{equation}
\hat{\bf P}(\lambda):=\hat{\bf P}_k(\lambda)\,,\hspace{0.2 in}
\text{for $\lambda\in U_k$, $k=0,\dots,G$,}
\end{equation}
\begin{equation}
\hat{\bf P}(\lambda):=\sigma_2\hat{\bf P}(\lambda^*)^*\sigma_2\,,
\hspace{0.2 in}\text{for $\lambda\in U_k^*$, $k=0,\dots,G$,}
\end{equation}
and
\begin{equation}
\hat{\bf P}(\lambda):=\dot{\bf P}(\lambda)\,,\hspace{0.2 in}
\text{for $\lambda$ outside all neighborhoods $U_k$ and their conjugates.}
\end{equation}

\subsection{Error analysis.}  \label{sec:error}
We now argue that ${\bf E}(\lambda):= {\bf P}(\lambda)\hat{\bf
  P}(\lambda)^{-1}$ satisfies
\begin{equation}
\lim_{\lambda\rightarrow\infty}
\lambda({\bf E}(\lambda)-\mathbb{I}) = O(\hbar)\,.
\label{eq:Easymp}
\end{equation}
The basic properties of the matrix ${\bf E}(\lambda)$ follow on the
one hand from the conditions of the Riemann-Hilbert problem satisfied
by the factor ${\bf P}(\lambda)$ and on the other from our explicit
knowledge of the global parametrix $\hat{\bf P}(\lambda)$.  Clearly,
${\bf E}(\lambda)$ is a piecewise analytic matrix in the complex
$\lambda$-plane, satisfying
\begin{equation}
\lim_{\lambda\rightarrow\infty}{\bf E}(\lambda)=\mathbb{I}\,,
\end{equation}
with jump discontinuities across the following contours:
\begin{itemize}
\item Across the boundaries $\partial U_k$ of neighborhoods of
  endpoints in the upper half-plane, taken with counterclockwise
  orientation, we have
\begin{equation}
{\bf E}_+(\lambda)={\bf E}_-(\lambda)\dot{\bf P}(\lambda)\hat{\bf P}_k(\lambda)^{-1}\,,\hspace{0.2 in}\lambda\in\partial U_k\,.
\end{equation}
\item Across the bands and nonterminal gaps of $K_0\cup K_{-1}$
  outside the neighborhoods $U_k$ in the upper half-plane, where both
  ${\bf P}$ and $\hat{\bf P}=\dot{\bf P}$ have jump discontinuities,
\begin{equation}
{\bf E}_+(\lambda)={\bf E}_-(\lambda)\dot{\bf P}_-(\lambda){\bf v}(\lambda)
\dot{\bf v}(\lambda)^{-1}\dot{\bf P}_-(\lambda)^{-1}\,,
\end{equation}
where across the same contour ${\bf P}_+(\lambda)={\bf
  P}_-(\lambda){\bf v}(\lambda)$ and $\dot{\bf P}_+(\lambda)=\dot{\bf
  P}_-(\lambda)\dot{\bf v}(\lambda)$.
\item Across the portions of $K_{1}$, $K_L$, the terminal gap of
  $K_{-1}$, and the lens boundaries $\partial\Omega_k^\pm$ that lie
  outside of the neighborhoods $U_k$ in the upper half-plane, where
  only the factor ${\bf P}(\lambda)$ is discontinuous, we have
\begin{equation}
{\bf E}_+(\lambda)={\bf E}_-(\lambda)\dot{\bf P}(\lambda){\bf v}(\lambda)
\dot{\bf P}(\lambda)^{-1}\,,
\end{equation}
where the jump matrix ${\bf v}(\lambda)$ is defined by ${\bf
  P}_+(\lambda)={\bf P}_-(\lambda){\bf v}(\lambda)$.
\end{itemize}
The jump discontinuities of ${\bf E}(\lambda)$ in the lower half-plane
are consistent with the symmetry ${\bf E}(\lambda)=\sigma_2{\bf
  E}(\lambda^*)^*\sigma_2$.  In particular, ${\bf E}(\lambda)$ is an
analytic function inside all neighborhoods $U_k$ of endpoints and
their complex conjugates because $\hat{\bf P}_k(\lambda)$ is chosen to
satisfy the jump conditions of ${\bf P}(\lambda)$ exactly within
$U_k$.  

This information means that ${\bf E}(\lambda)$ itself is the solution
of a matrix Riemann-Hilbert problem with given data.  By exploiting a
well-known connection with systems of singular integral equations with
Cauchy-type kernels it suffices to estimate the uniform difference
between the ratio of boundary values ${\bf E}_-(\lambda)^{-1}{\bf
  E}_+(\lambda)$ and the identity matrix $\mathbb{I}$.  In fact, we
will show that ${\bf E}_-(\lambda)^{-1}{\bf E}_+(\lambda)-\mathbb{I} =
O(\hbar)$ holds uniformly on the $\hbar$-independent contour of
discontinuity for ${\bf E}(\lambda)$.  From this estimate, the estimate
\eqref{eq:Easymp} follows from the connection to integral equations.

To show that ${\bf E}_-(\lambda)^{-1}{\bf E}_+(\lambda)=\mathbb{I}+O(\hbar)$
requires only a little more than the properties of $g(\lambda)$ and the
global parametrix $\hat{\bf P}(\lambda)$ already established.  We also
need the asymptotic behavior of the functions $S(\lambda)$ and $T(\lambda)$.
Analogous functions are analyzed carefully in \cite{DOP}, so we just
quote the results:
\begin{itemize}
\item The function $S(\lambda)$ is analytic for $\lambda$ in the upper
  half-plane outside the region bounded by the curve $C$ and the
  imaginary interval with the same endpoints $[0,iA]$.  Uniformly on
  compact subsets of the open domain of analyticity we have
  $S(\lambda)=1+O(\hbar)$.
\item The function $T(\lambda)$ is analytic for $\lambda$ in the upper
  half-plane outside the region bounded by the curve $C_\infty$ and
  the imaginary axis above $iA$.  Uniformly on compact subsets of the
  open domain of analyticity we have $T(\lambda)=1+O(\hbar)$.
\end{itemize}
These facts are enough to prove that ${\bf E}_-(\lambda)^{-1}{\bf
  E}_+(\lambda)=\mathbb{I}+O(\hbar)$ on all contours with the
exception of $K_1$ restricted to a neighborhood of $\lambda=\epsilon$
and $K_{-1}$ restricted to a neighborhood of $\lambda=-\epsilon$.  The
jump matrix for ${\bf P}(\lambda)$ on $K_1$ involves both
$e^{i\theta(\lambda)/\hbar}$ and also
$e^{(2i\theta^0(\lambda)-i\theta(\lambda))/\hbar}$.  The former is
exponentially small as $\hbar\downarrow 0$ by a Cauchy-Riemann
argument for $K_1$ sufficiently close to $K_0$; the choice of a
sufficiently small but positive $\epsilon$ is crucial to provide the
decay where $K_1$ meets the real axis. The latter is also
exponentially small on the parts of $K_1$ that are bounded away from
the imaginary axis, because
$\Re(2i\theta^0(\lambda))=-2\pi\Re(\lambda)$, which dominates
$\Re(i\theta(\lambda))$ for $K_1$ close to $K_0$.  But it is not
immediately clear that an $\epsilon>0$ can be found so that the
inequality $\Re(2i\theta^0(\lambda)-i\theta(\lambda))<0$ persists
along $K_1$ to the real axis.  However, taking a limit of
$g'_+(\lambda)-g'_-(\lambda)$ as $\lambda\rightarrow 0$ along $K_0$ shows that
\begin{equation}
\Re(-i\theta'(0)) = \pi\,,
\end{equation}
so $\Re(2i\theta^0(\lambda)-i\theta(\lambda)) = -\pi\Re(\lambda) +
O(|\lambda|^2)$, which means that the inequality persists along $K_1$
to $\lambda=\epsilon>0$, for $\epsilon$ sufficiently small.  A similar
explicit calculation involving $g'(\lambda)$ near $\lambda=0$ shows
also that $\Re(\phi(\lambda))$ is decreasing linearly away from the
origin along the negative real axis, which proves that while the limit
of $\phi(\lambda)$ as $\lambda$ approaches the origin along $K_{-1}$
is purely imaginary, the inequality $\Re(\phi(\lambda))<0$ is
satisfied strictly throughout the terminal gap of $K_{-1}$ as long as
$\epsilon>0$ is sufficiently small.

This concludes our discussion of the error matrix ${\bf E}(\lambda)$.
We only note two things at this point.  Firstly, the bound \eqref{eq:Easymp}
proves that 
\begin{equation}
\psi_N(x,t) = 2i\lim_{\lambda\rightarrow\infty}\lambda\dot{P}_{12}(\lambda) + O(N^{-1})\,,
\label{eq:psiasymp}
\end{equation}
as $N\rightarrow\infty$ because ${\bf P}(\lambda)={\bf
  E}(\lambda)\dot{\bf P}(\lambda)$ for $|\lambda|$ sufficiently large,
and $\hbar = \hbar_N=A/N$.  Therefore, the strong asymptotics of the
$N$-soliton are provided by the modulated multiphase wavetrain that
arises from the solution of the outer model problem for $\dot{\bf
  P}(\lambda)$ in terms of Riemann theta functions of genus $G$.  In
particular, the curves in the $(x,t)$-plane along which the genus
changes abruptly are the {\em caustic curves} seen in
Figures~\ref{fig:5soliton}--\ref{fig:40soliton}.  Secondly, we want to
point out that the error estimate of $O(\hbar)=O(N^{-1})$ in
\eqref{eq:psiasymp} is an improvement over the error bound obtained
for the same problem in \cite{KMM}.  The improvement comes from (i)
the $\epsilon$-modifications of the contours near $\lambda=0$ which
obviates the need for a local parametrix near the origin (this was
also used to handle ``transition points'' in \cite{DOP}) and (ii) the
careful tracing of the influence of the functions $S(\lambda)$ and
$T(\lambda)$ through the asymptotics, especially their explicit
removal near the band/gap endpoints via the near-identity
transformations ${\bf P}\rightarrow{\bf Q}$.

\subsection{The formal continuum-limit problem.}
\label{sec:formallimit}
Much of the above analysis is based on the facts that $S(\lambda) =
1+O(\hbar)$ and $T(\lambda)=1+O(\hbar)$ under the assumptions in force
on the relation between the contours on which these functions appear
in the jump matrix and the contours $C$ and $C_\infty$.  These two
functions measure the difference between the discreteness of the
eigenvalue distribution and a natural continuum limit thereof (a weak
limit of a sequence of sums of point masses).  It is tempting to
notice the role played by the approximations $S(\lambda)\approx 1$ and
$T(\lambda)\approx 1$ in the rigorous analysis and propose in place of
the problem for ${\bf P}(\lambda)$ an {\em ad hoc} ``continuum-limit''
Riemann-Hilbert problem for a matrix $\tilde{\bf P}(\lambda)$; the
conditions of this Riemann-Hilbert problem are precisely the same as
those of the Riemann-Hilbert problem governing ${\bf P}(\lambda)$
except that in all cases one makes the substitutions
$S(\lambda)\rightarrow 1$ and $T(\lambda)\rightarrow 1$.  

The rigorous analysis described earlier proves that, {\em as long as
  the contour geometry admits the approximations $S(\lambda)\approx 1$
  and $T(\lambda)\approx 1$} one may also compute the asymptotic
behavior of the $N$-soliton by studying the formal continuum-limit
Riemann-Hilbert problem for $\tilde{\bf P}(\lambda)$.  However, it
turns out that there are some circumstances in which the conditions
that constrain the contours of the Riemann-Hilbert problem for ${\bf
  P}(\lambda)$ are inconsistent with the approximations
$S(\lambda)\approx 1$ and $T(\lambda)\approx 1$.  We will show that
this is not merely a technical inconvenience standing in the way of
analyzing the large $N$ limit with the help of the formal continuum-limit
Riemann-Hilbert problem for $\tilde{\bf P}(\lambda)$, but that the
modifications necessary to complete the analysis of ${\bf P}(\lambda)$
rigorously under these circumstances introduce new mathematical
features that ultimately provide the correct description of the
secondary caustic.

Note also that making the substitutions $S(\lambda)\rightarrow 1$ and
$T(\lambda)\rightarrow 1$ is completely analogous to the {\em ad hoc}
substitution $P(\lambda)\rightarrow \tilde{P}(\lambda)$, some of the
consequences of which were described in \S~\ref{sec:aside}.  These
arguments suggest that extreme care must be taken in relating
asymptotic properties of the matrix $\tilde{\bf P}(\lambda)$ with
those of the matrix ${\bf P}(\lambda)$.  Probably it is better to
avoid analyzing $\tilde{\bf P}(\lambda)$ and instead keep track of the
errors by working (as we do in this paper) with ${\bf P}(\lambda)$
directly.

\subsection{Dual interpolant modification necessary for contours passing 
  through the branch cut.}  
\label{sec:dual}
The nonlinear equations that determine the endpoints only involve
quantities related to the continuum limit of the distribution of
eigenvalues within the imaginary interval $[0,iA]$.  Indeed, the
rational function $P(\lambda)$ has been replaced by
$e^{L(\lambda)/\hbar}$ at the cost of a factor $S(\lambda)$, which is
unifomly approximated by $1$ on $K_{-1}$.  It turns out (see
\S~\ref{sec:GenusTwoNumerics}) that these equations admit relevant
solutions that evolve in $x$ and $t$ in such a way that the contours
become inconsistent with the fundamental assumption that the region
$D_{-1}$ contains all of the discrete eigenvalues (poles of ${\bf
  m}(\lambda)$ in the upper half-plane).  This can happen because the
function $L(\lambda)$ is defined relative to the contour $C$ which
while having the same endpoints as $[0,iA]$, is otherwise arbitrary,
and the endpoint equations are analytic as long as the endpoint
variables are distinct and avoid $C$.  Thus, it can (and does) happen
that a band $I$ evolves in $x$ and $t$ so as to come into contact with
the locus of accumulation of eigenvalues.  When this occurs, we have
to reconcile the facts that on the one hand the solution of the
endpoint equations may be continued (by choice of $C$) in $x$ and $t$
so that the band $I$ passes through the interval $[0,iA]$ completely,
while on the other hand the function $S(\lambda)$ can no longer be
controlled and the continuum-limit approximation can no longer be
justified in the same way.

\begin{figure}[htbp]
\begin{center}
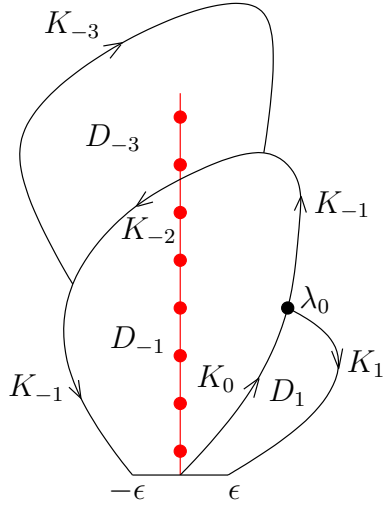
\end{center}
\caption{\em The three regions $D_1$, $D_{-1}$, and $D_{-3}$, and the
boundary contour arcs $K_1$, $K_0$, $K_{-1}$, $K_{-2}$, and $K_{-3}$.}
\label{fig:ModifiedConfiguration}
\end{figure}

The situation can be rectified in the following way.  Returning to the
matrix ${\bf m}(\lambda)$ solving the discrete Riemann-Hilbert problem
characterizing the $N$-soliton, we remove the poles by taking into account
three different interpolants of residues rather than just two.  Consider
the disjoint regions $D_1$, $D_{-1}$, and $D_{-3}$ in the upper half-plane
shown in Figure~\ref{fig:ModifiedConfiguration}.
We then remove the poles by going from ${\bf m}(\lambda)$ to ${\bf
  M}(\lambda)$ by exactly the same formulae as before, except in the
region $D_{-3}$ and its complex conjugate image in the lower
half-plane.  In $D_{-3}$ we define instead,
\begin{equation}
{\bf M}(\lambda):={\bf m}(\lambda)\left[\begin{array}{cc}
1 & 0 \\ -iP(\lambda)e^{[2iQ(\lambda)-3i\theta^0(\lambda)]/\hbar} & 1
\end{array}\right]\,,\hspace{0.2 in}\text{for $\lambda\in D_{-3}$}\,,
\label{eq:Dm3interp}
\end{equation}
and define ${\bf M}(\lambda)$ for $\lambda\in D_{-3}^*$ by ${\bf
  M}(\lambda) = \sigma_2{\bf M}(\lambda^*)^*\sigma_2$.

\begin{figure}[htbp]
\begin{center}
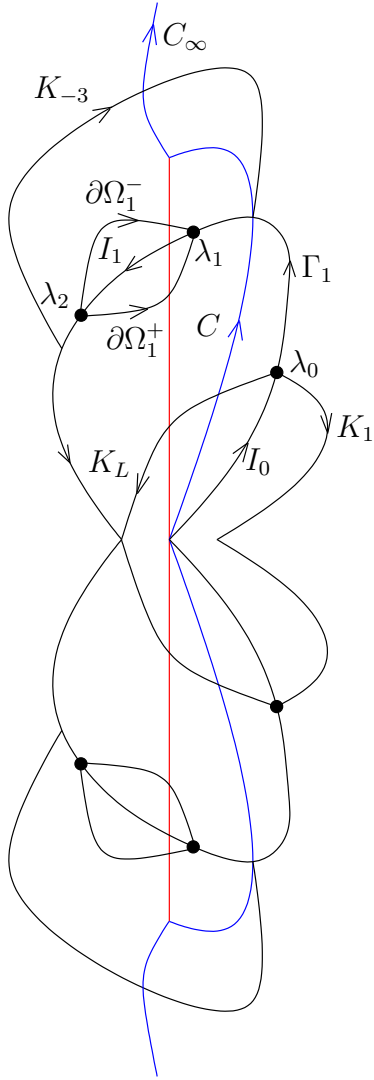
\end{center}
\caption{\em The discontinuity contours for ${\bf P}(\lambda)$ in
the modified configuration.  The arc connecting $\lambda_1$ and $\lambda_2$
between $\partial\Omega_1^\pm$ is the band $I_1$ which may now cross the
interval $[0,iA]$.}
\label{fig:ModifiedConfiguration2}
\end{figure}

Once again it can be checked directly that ${\bf M}(\lambda)$ is
analytic at all of the poles of ${\bf m}(\lambda)$ in the open domains
$D_{-1}$ and $D_{-3}$.  The same argument works if it happens that a
pole of ${\bf m}(\lambda)$ lies exactly on the arc $K_{-2}$ separating
the two domains.  Next, we introduce the function $g(\lambda)$, which
we assume is analytic except on the contours $K_{-2}$, $K_{-1}$,
$K_0$, and their complex conjugates, and we define a new unknown as
before by setting ${\bf N}(\lambda):={\bf
  M}(\lambda)e^{-g(\lambda)\sigma_3/\hbar}$.
On the arcs $K_1$, $K_0$, and $K_{-1}$ the jump conditions
relating the boundary values taken by ${\bf N}(\lambda)$ are exactly
as before, while on the arc $K_{-3}$ we have
\begin{equation}
{\bf N}_+(\lambda)=
{\bf N}_-(\lambda)\left[\begin{array}{cc}
1 & 0 \\ iP(\lambda)
e^{[2iQ(\lambda)-3i\theta^0(\lambda)-2g(\lambda)]/\hbar} & 1
\end{array}\right]\,,\hspace{0.2 in}\lambda\in K_{-3}\,,
\label{eq:jumpNKm3}
\end{equation}
and on the arc $K_{-2}$,
\begin{equation}
{\bf N}_+(\lambda)={\bf N}_-(\lambda)\left[\begin{array}{cc}
e^{-[g_+(\lambda)-g_-(\lambda)]/\hbar} & 0 \\
iT(\lambda)
e^{[2iQ(\lambda)-2i\theta^0(\lambda)+\overline{L}(\lambda)
-g_+(\lambda)-g_-(\lambda)]/\hbar} & e^{[g_+(\lambda)-g_-(\lambda)]/\hbar}
\end{array}\right]\,,
\hspace{0.2 in}\lambda\in K_{-2}\,.
\end{equation}
We choose the contour $C$, relative to which the functions
$\overline{L}(\lambda)$ and $T(\lambda)$ are defined, such that the
contour $K_{-2}$ forming the common boundary between the domains
$D_{-1}$ and $D_{-3}$ lies entirely in the left half-plane according
to $C\cup C_\infty\cup C^*\cup C_\infty^*$.  For convenience we will
take $C$ to lie in the closure of $D_{-1}\cup D_{-3}$ and to connect
$\lambda=0$ to $\lambda=iA$ passing once through the first
intersection point of $K_{-1}$ and $K_{-2}$ (in the direction of their
orientation, beginning with $\lambda=0$).  We also assume (without
loss of generality) that this intersection point turns out to lie in a
gap of $K_{-1}\cup K_{-2}$.  See also
Figure~\ref{fig:ModifiedConfiguration2}.

The next step is to remove the jump discontinuities on the real
intervals $(-\epsilon,0)$ and $(0,\epsilon)$ by exactly the same
factorization as was used previously (see \eqref{eq:K0factorization}
and the subsequent definition of ${\bf O}(\lambda)$ in terms of ${\bf
  N}(\lambda)$ and the discussion thereof).  Here the contour arc
$K_L$ lies in the region $D_{-1}$ as before.

The jump relations satisfied by ${\bf O}(\lambda)$ for $\lambda$ on
$K_0$ and $K_{-1}$ are of exactly the same form as before (see
\eqref{eq:OjumpK0} and \eqref{eq:OjumpKm1}).  Furthermore, because
$K_{-2}$ is in the left half-plane according to $C\cup C_\infty\cup
C^*\cup C_\infty^*$, we may write the jump relation for ${\bf O}(\lambda)$
for $\lambda$ on $K_{-2}$ as
\begin{equation}
{\bf O}_+(\lambda)={\bf O}_-(\lambda)\left[\begin{array}{cc}
e^{-[g_+(\lambda)-g_-(\lambda)]/\hbar} & 0\\
iT(\lambda)e^{[2iQ(\lambda)+L(\lambda)-i\theta^0(\lambda)-g_+(\lambda)-g_-(\lambda)]/\hbar} & e^{[g_+(\lambda)-g_-(\lambda)]/\hbar}\end{array}\right]\,,
\hspace{0.2 in}\lambda\in K_{-2}\,.
\label{eq:jumpOKm2}
\end{equation}
We note here that the exponents of the jump matrix elements are of
exactly the same form as for $\lambda\in K_{-1}$.  Since
$T(\lambda)=1+O(\hbar)$ holds uniformly on $K_{-2}$ as long as we
arrange that this contour is bounded away from the endpoints of $C$,
we may asymptotically analyze the Riemann-Hilbert problem for ${\bf
  O}(\lambda)$ by choosing $g(\lambda)$ in exactly the same way as we
did earlier, {\em with the same integral conditions on the endpoints
  of the bands}.  In particular, the characteristic velocities in the
Whitham equations are exactly the same analytic functions of the
elements of $\vec{v}$ as in the simpler configuration.  The contours
of discontinuity for ${\bf P}(\lambda)$ in the modified configuration
are shown in black in Figure~\ref{fig:ModifiedConfiguration2}.

With regard to the determination of the function $g(\lambda)$, the
only essential difference between this approach and the former
approach is that here a new inequality is required to hold for
$\lambda\in K_{-3}$ in order that the corresponding jump matrix for
${\bf O}(\lambda)$ be a small perturbation of the identity matrix.
From \eqref{eq:jumpNKm3}, we see that
\begin{equation}
{\bf O}_+(\lambda)={\bf O}_-(\lambda)\left[\begin{array}{cc}
1 & 0\\
iS(\lambda)e^{[2iQ(\lambda)+L(\lambda)-i\theta^0(\lambda)-2g(\lambda)
-2i\theta^0(\lambda)]/\hbar} & 1\end{array}\right]\,,
\hspace{0.2 in}\lambda\in K_{-3}\,.
\end{equation}
The exponent is an analytic function in $D_{-3}$, and taking a boundary value
on $K_{-2}$ we see that
\begin{equation}
\begin{array}{rcl}
2iQ(\lambda)+L(\lambda)-i\theta^0(\lambda)-2g(\lambda)-2i\theta^0(\lambda)&=&
2iQ(\lambda)+L(\lambda)-i\theta^0(\lambda)-2g_-(\lambda)-2i\theta^0(\lambda)\\\\&=&\phi(\lambda)-i\theta(\lambda)-2i\theta^0(\lambda)\,.
\end{array}
\end{equation}
By assumption, the arc $K_{-2}$ contains a part of a gap, so we may evaluate
the boundary value in that gap, where $\theta(\lambda)$ is a real constant
and find that the relevant inequality is
\begin{equation}
\Re(2iQ(\lambda)+L(\lambda)-i\theta^0(\lambda)-2g(\lambda)-2i\theta^0(\lambda))
=\Re(\phi(\lambda)-2i\theta^0(\lambda))<0\,,\hspace{0.2 in}\lambda\in K_{-3}\,.
\label{eq:newinequality}
\end{equation}
Here $\phi(\lambda)$ denotes the analytic continuation from any gap on
$K_{-2}$ to $K_{-3}$ (while different analytic functions for each gap
on $K_{-2}$, they all differ by imaginary constants which play no role
in \eqref{eq:newinequality}).

The term $2\Im(\theta^0(\lambda))$ represents a nontrivial
modification to the inequality $\Re(\phi(\lambda))<0$ that must be
satisfied in the gaps of $K_{-1}\cup K_{-2}$.  As we will see in
\S~\ref{sec:GenusTwoNumerics}, the violation of this inequality is
exactly the mechanism for the phase transition from $G=2$ (genus two)
to $G=4$ (genus four).  It should be stressed that this inequality is
fundamentally an artifact of the discrete spectral nature of the
$N$-soliton.  In other words, if we simply propose (as described in
\S~\ref{sec:formallimit}) a model Riemann-Hilbert problem for an
unknown $\tilde{\bf P}(\lambda)$ in the original scheme with all jump
matrices modified simply by setting $S(\lambda)=T(\lambda)\equiv 1$,
we (i) will not care about the implications of a portion of the
contour $K_{-1}$ meeting the interval $[0,iA]$ (which has meaning only
as an artifical branch cut of an analytic function) and (ii) will only
need the inequality $\Re(\phi(\lambda))<0$ and ascribe no particular
meaning to the modified quantity
$\Re(\phi(\lambda)-2i\theta^0(\lambda))$.

\section{The Genus-Zero Region}\label{sec:genus0} 
\subsection{Validity of the genus-zero ansatz.  Ansatz failure and primary caustic.}
\label{sec:validityofG0}
In Chapter~6 of \cite{KMM}, the genus-zero ansatz for the $g$-function
of the $N$-soliton is constructed and it is shown that the inequality
$\Re(\phi(\lambda))<0$ is satisfied throughout the sole (terminal) gap
for $0<t<t_1(x)$, where the curve $t=t_1(x)$ is the primary caustic
curve asymptotically separating the smooth and oscillatory regions in
Figures~\ref{fig:5soliton}--\ref{fig:40soliton}.  The primary caustic
curve is described by obtaining the endpoints $\lambda_0(x,t)$ and
$\lambda_0^*(x,t)$ from the genus-zero ansatz for $g(\lambda)$ and
then eliminating $\hat{\lambda}$ from the equations
\begin{equation}
\phi'(\hat{\lambda}) = 0\,,\hspace{0.2 in} \Re(\phi(\hat{\lambda}))=0\,.
\label{eq:G0break}
\end{equation}
In other words, the first equation says that
$\hat{\lambda}=\hat{\lambda}(x,t)$ is a critical point of
$\phi(\lambda)$ with endpoints $\lambda_0(x,t)$ and
$\lambda_0^*(x,t)$, and then the second equation is a real relation
between $x$ and $t$, that is, a curve in the $(x,t)$-plane.  The
equations \eqref{eq:G0break} describe the existence of a critical
point $\hat{\lambda}$ for $\phi(\lambda)$ that lies on the level curve
$\Re(\phi(\lambda))=0$, which is the boundary of the region of
existence for the terminal gap $\Gamma_1$.  The existence of such a
point on the level indicates a singularity of the level curve,
generically a simple crossing of two perpendicular branches.  In other
words, if $\delta>0$ is small then for $t_1(x)-\delta<t<t_1(x)$ the
region where $\Re(\phi(\lambda))<0$ holds is connected albeit via a
thin channel, delineated by approximate branches of a hyperbola,
through which the gap contour $\Gamma_1$ must pass, whereas for
$t_1(x)<t<t_1(x)+\delta$ the region where $\Re(\phi(\lambda))<0$ holds
becomes disconnected, and it is no longer possible to choose
$\Gamma_1$ so that $\Re(\phi(\lambda))<0$ holds throughout.  When
$t=t_1(x)$ the hyperbolic branches degenerate to lines crossing at
$\lambda=\hat{\lambda}(x,t)$.  This mechanism for the formation of the
primary caustic is illustrated in the spectral $\lambda$-plane in
Figures~6.11 and 6.12 of \cite{KMM}.  In particular, it is known in
the case of the $N$-soliton that $t_1(x)$ is an even function and
$\inf_{x\in\mathbb{R}}t_1(x) = t_1(0)=(2A)^{-1}$.  In \cite{KMM} it is
shown that if $|x|$ decreases with $t$ fixed from a point $(x_{\rm
  crit},t_{\rm crit})$ at which \eqref{eq:G0break} holds, then the
genus-two ansatz takes over.  A new band is born with two new
endpoints $\lambda_1(x,t)$ and $\lambda_2(x,t)$ emerging from the
double point $\hat{\lambda}$ where the singularity of the level curve
for $\Re(\phi(\lambda))=0$ occurs.  Therefore, the primary caustic is
a phase transition between genus zero and genus two.

In fact, the conditions \eqref{eq:G0break} are not specific to the
failure of the genus-zero ansatz, but may be considered for any genus,
being as they simply describe the change in connectivity of the region
where the essential inequality $\Re(\phi(\lambda))<0$ holds.  Thus, we
may consider the conditions \eqref{eq:G0break} for general even genus
with the aim of determining whether phase transitions beyond the
primary caustic can occur for the $N$-soliton.  Simultaneous solutions
of the two equations in \eqref{eq:G0break} can be interpreted as
collisions between branches of the level curve $\Re(\phi(\lambda))=0$,
and these branches may emanate from band/gap endpoints
$\lambda_k(x,t)$, or from the branch cut $C$, or they may be
noncompact.  (As level curves of a harmonic function they cannot
close on themselves unless they enclose a singularity of
$\phi(\lambda)$.)  As shown in Figures~6.11 and 6.12 of \cite{KMM},
the primary caustic is caused by the collision of a noncompact branch
coming in from infinity with a branch joining $\lambda_0(x,t)$ to the
origin.  It is not difficult to analyze the function $\phi(\lambda)$
for general even genus to determine its singular points and asymptotic
behavior as $\lambda\rightarrow\infty$, and such analysis leads to the
conclusion that for genus two no further branches exist for large
$\lambda$ that can eventually collide with branches connecting
band/gap endpoints.  This fact suggests that further phase transitions
(higher-order caustics), if they exist, may come about for a different
reason than the primary caustic.  The main point of this paper is to
show that a secondary caustic indeed occurs for the $N$-soliton in the
large $N$ limit, and that its mechanism is indeed different from that
leading to the formation of the primary caustic.

The validity of the genus-zero ansatz for $g(\lambda)$ for small time
is in fact established in \cite{KMM} for a general class of
``reflectionless'' initial data for the focusing nonlinear
Schr\"odinger equation that leads to a discrete Riemann-Hilbert
problem for ${\bf m}(\lambda)$ of the sort relevant in the study of
the $N$-soliton.  Similar results have been found by Tovbis,
Venakides, and Zhou \cite{TVZ} for certain special-function initial
conditions that are not reflectionless.

\subsection{Fourier power spectrum and supercontinuum 
  frequency generation.}  
\label{sec:Fourier}
\begin{figure}[htbp]
\begin{center}
\includegraphics[width=3 in]{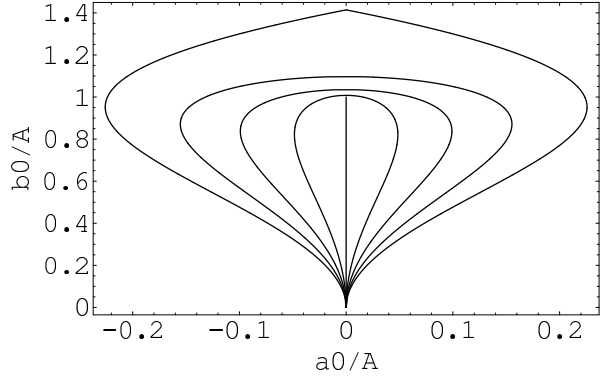}
\end{center}
\caption{\em The upper lobe of the compact part of the curve \eqref{eq:turningpointcurve} shown for $At=0,1/8,1/4,3/8,1/2$ from inside out.  For $t>0$
the relation \eqref{eq:genuszeroendpointII} parametrizes the curve from the
origin to itself in the clockwise direction.}
\label{fig:turningpointcurve}
\end{figure}

In applications of the focusing nonlinear
Schr\"odinger equation to nonlinear fiber optics, the limit
$\hbar\downarrow 0$ is interesting because it corresponds to weak
dispersion, a situation in which nonlinear effects are dominant over
linear dispersive effects.  While conservative linear effects leave
the Fourier power spectrum of a signal unchanged, strongly nonlinear
effects can change the spectrum significantly, and one application of
weak dispersion in fiber optics is the generation of broadband spectra
from input signals with a narrow power spectrum.  This is known in the
literature as {\em supercontinuum generation}.  While it is sometimes
thought that supercontinuum generation occurs as a result of the wave
breaking at the primary caustic, in fact the effect is already present
for $t<\inf_{x\in\mathbb{R}}t_1(x)$.  Indeed, in this region, the
$N$-soliton has, for each $x\in\mathbb{R}$ the representation
\eqref{eq:psiasymp} corresponding to the genus-zero ansatz for
$g(\lambda)$.  Moreover (see \cite{KMM}) in the genus-zero case,
\begin{equation}
2i\lim_{\lambda\rightarrow\infty}\lambda \dot{P}_{12}(\lambda) = b_0
e^{-i\kappa_0/\hbar}\,,
\label{eq:genuszeroasymp}
\end{equation}
where $\lambda_0=a_0(x,t)+ib_0(x,t)$ and $\kappa_0=\kappa_0(x,t)$ are
related by
\begin{equation}
\frac{\partial\kappa_0}{\partial x} = 2a_0\,.
\end{equation}
For the $N$-soliton with $t>0$, the endpoint coordinates $a_0(x,t)$
and $b_0(x,t)$ satisfy
\begin{equation}
a_0^2 = t^2b_0^4\frac{A^2-b_0^2+t^2b_0^4}{A^2 + t^2b_0^4}
\label{eq:turningpointcurve}
\end{equation}
and
\begin{equation}
x = -2ta_0+\Re\left(\text{arcsinh}\left(\frac{a_0+iA}{b_0}\right)\right)\,.
\label{eq:genuszeroendpointII}
\end{equation}
Note that if $0<t<(2A)^{-1}=\inf_{x\in\mathbb{R}}t_1(x)$, the relation
\eqref{eq:turningpointcurve} defines a curve in the real
$(a_0,b_0)$-plane with a compact component having a ``figure-8'' shape
with two lobes joined at the origin, one enclosing the interval
$(0,iA)$ and the other enclosing the interval $(-iA,0)$.  For such
$t$, the correct solution (in the sense of leading to topological
conditions for contours in the complex plane admitting the genus-zero
ansatz) of \eqref{eq:turningpointcurve} and
\eqref{eq:genuszeroendpointII} necessarily lies on the upper lobe of
this compact component, on which it is uniquely determined by
\eqref{eq:genuszeroendpointII} as a function of $x$.  In other words,
\eqref{eq:genuszeroendpointII} parametrizes the upper lobe by
$x\in\mathbb{R}$, taking $\mathbb{R}_-$ to the left half of the lobe,
$\mathbb{R}_+$ to the right half, and $x=0$ to the point above $iA$
where the upper lobe intersects the imaginary axis.  As $x\rightarrow
-\infty$, $\lambda_0(x,t)$ tends to zero from the second quadrant, and
as $x\rightarrow +\infty$, $\lambda_0(x,t)$ tends to zero from the
first quadrant.  See Figure~\ref{fig:turningpointcurve}.

\begin{figure}[htbp]
\begin{center}
\includegraphics[width=3 in]{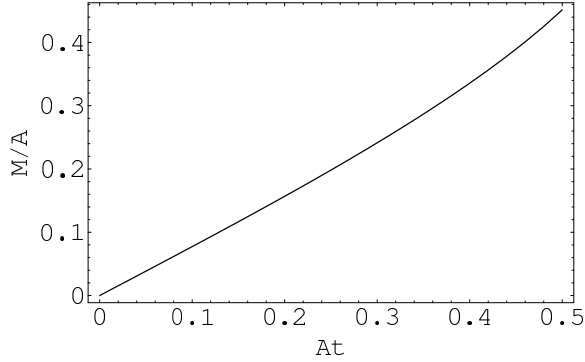}
\end{center}
\caption{\em The scaled cutoff frequency is a monotone increasing function
of $At$.}
\label{fig:cutoff}
\end{figure}

The power spectrum of $\psi_N(x,t)$ in the globally genus-zero regime
(for $t<(2A)^{-1}$) can be approximated by using the method of
stationary phase to analyze the Fourier transform of the approximate
solution \eqref{eq:genuszeroasymp}.  To use the method of stationary
phase to study the approximate power spectrum
\begin{equation}
F(\omega,t):=\left|\frac{1}{2\pi}\int_{-\infty}^\infty b_0(x,t)e^{-i\kappa_0(x,t)/\hbar}e^{-i\omega x}\,dx\right|^2\,,
\end{equation}
it is useful to introduce the scaling $\omega=\Omega/\hbar$ and consider
$\Omega$ fixed as $\hbar\downarrow 0$.  Then,
\begin{equation}
F(\omega,t)=\left|\frac{1}{2\pi}\int_{-\infty}^\infty b_0(x,t)e^{-i(\kappa_0(x,t)+\Omega x)/\hbar}\,dx\right|^2\,,
\end{equation}
and the stationary phase points satisfy
\begin{equation}
0=\frac{\partial\kappa_0}{\partial x}+\Omega = 2a_0(x,t)+\Omega\,.
\end{equation}
According to the parametrization of the upper lobe of the compact
component of the curve \eqref{eq:turningpointcurve} given by
\eqref{eq:genuszeroendpointII}, there exists a finite number $M>0$
(the maximum of $2|a_0|$ over the upper lobe) such that
\begin{itemize}
\item There are no stationary phase points for $|\Omega|>M$.
\item There are two simple stationary phase points $x_+(\Omega)<x_-(\Omega)<0$
for $0<\Omega<M$.
\item There are two simple stationary phase points $0<x_-(\Omega)<x_+(\Omega)$
for $-M<\Omega<0$.
\item For $|\Omega|=M$ there is a double (degenerate) stationary
  phase point.
\item For $\Omega=0$ there is only one finite simple stationary phase point
at $x=0$ but the phase is also stationary at $x=\infty$.
\end{itemize}
Thus, the power spectrum is negligible for frequencies $\omega$ of
magnitude greater than $M/\hbar$.  The scaled cutoff frequency $M$
depends on both $A$ and $t$, but from dimensional analysis of
\eqref{eq:turningpointcurve} it can be seen that $M/A$ is a function
of the combination $At$ alone.  The scaled cutoff frequency is plotted
in Figure~\ref{fig:cutoff} over the interval $0<At<1/2$ in which the
genus-zero ansatz holds for all $x\in\mathbb{R}$.

In the bulk of the spectrum
$|\Omega|<M$, the power spectrum is proportional to $\hbar$.
Indeed, the relevant stationary phase formula
for the generic case of $|\Omega|<M$ and for nonzero fixed $\Omega$
is
\begin{equation}
F(\omega,t)=\frac{\hbar}{4\pi}\left[A_+(\Omega)^2 + A_-(\Omega)^2 - 
2A_+(\Omega)A_-(\Omega)\sin\left(\frac{\Phi(\Omega)}{\hbar}\right)\right] + 
O(\hbar^2)\,,
\end{equation}
where
\begin{equation}
A_\pm(\Omega):=\frac{b_0(x_\pm(\Omega),t)}{\sqrt{|\partial_xa_0(x_\pm(\Omega),t)|}}\,,
\end{equation}
and
\begin{equation}
\Phi(\Omega):=-\kappa_0(x_+(\Omega),t)+\kappa_0(x_-(\Omega),t) -\Omega x_+(\Omega) + \Omega x_-(\Omega)\,.
\end{equation}
Considered as a function of $\Omega$ in the bulk $|\Omega|<M$, the power
spectrum is a rapidly oscillatory function such that 
\begin{equation}
\frac{\hbar}{4\pi}(A_+(\Omega)-A_-(\Omega))^2+O(\hbar^2)<F(\omega,t)<
\frac{\hbar}{4\pi}(A_+(\Omega)+A_-(\Omega))^2+O(\hbar^2)\,.
\end{equation}
The rapid oscillation in the spectrum is a kind of ``ripple'' that has
a characteristic spectral width on the $\Omega$ scale given by
\begin{equation}
\Delta\Omega = \frac{\hbar}{|\Phi'(\Omega)|}  = 
\frac{\hbar}{|x_+(\Omega)-x_-(\Omega)|}\,.
\end{equation}
Since $|x_+(\Omega)-x_-(\Omega)|$ tends to zero as
$|\Omega|\rightarrow M$ and grows without bound as
$\Omega\rightarrow 0$, the ripple becomes less pronounced near the
edge of the spectrum and more violent near the center of the spectrum.
Near the edge of the spectrum the characteristic spectral width still
tends to zero with $\hbar$ but at a slower rate; in this regime more
detailed stationary phase analysis of the power spectrum would reveal
Airy-like behavior on a scale where $|\Omega|-M$ is proportional to
an appropriate fractional power of $\hbar$.  In any case, uniformly over
the region $|\Omega|<M$, the ripple is rapid enough to give meaning
to the local average (weak limit with $\Omega$ fixed):
\begin{equation}
\langle \hbar^{-1}F(\omega,t)\rangle :=
\frac{1}{4\pi}\left[A_+(\Omega)^2 + A_-(\Omega)^2\right]\,.
\end{equation}
The weak convergence of the power spectrum to a broad plateau of
spectral width (on the $\omega$ scale) of $\hbar^{-1}M$ can be
considered as evidence of supercontinuum frequency generation for the
$N$-soliton in the genus-zero region, before any wave breaking occurs.
Furthermore, these calculations show that the broad spectral plateau
develops in absence of the Raman effect or higher-order dispersion,
which are not part of the basic nonlinear Schr\"odinger model but are
frequently thought to play an important role in supercontinuum
generation.

\section{Numerical Computation of Endpoints and
Contours in the Genus-Two Region} 
\label{sec:GenusTwoNumerics}
Our main interest in this paper is to analyze the $N$-soliton in the
genus-two region to determine and explain the mechanism for the
secondary caustic curve that can be observed in
Figures~\ref{fig:5soliton}--\ref{fig:40soliton}.  In the genus-two
region, there are three complex endpoints $\lambda_0(x,t)$,
$\lambda_1(x,t)$, and $\lambda_2(x,t)$ that satisfy the six real
equations
\begin{equation}
\begin{array}{rcl}
V_1&=&0\\\\
R_1&=&0\\\\
M_0&=&0\\\\
M_1&=&2t\\\\
M_2&=&\displaystyle x+2t\sum_{n=0}^2a_n\\\\
M_3&=&\displaystyle x\sum_{n=0}^2a_n+2t\sum_{n=0}^2
\left(a_n^2-\frac{1}{2}b_n^2\right)+2t\sum_{n=0}^2\sum_{m=n+1}^2a_ma_n\,.
\end{array}
\label{eq:G2eqns}
\end{equation}
We wrote a Matlab code that continues a solution of these equations
along a path from point $(x_0,t_0)$ at which a solution is known for
which the Jacobian of the system is nonsingular to a nearby point
$(x_1,t_1)$.  To have a starting point, we need to choose a point
$(x_{\rm crit},t_{\rm crit})$ on the primary caustic, to which there
corresponds a single endpoint $\lambda_0(x_{\rm crit},t_{\rm crit})$
of the genus-zero ansatz, and a critical point $\hat{\lambda}$ at
which \eqref{eq:G0break} holds.  In \cite{KMM} it is shown that for
$x=x_{\rm crit}$ and $t=t_{\rm crit}$ the equations \eqref{eq:G2eqns}
admit the solution given by $\lambda_2=\lambda_1=\hat{\lambda}$ and
$\lambda_0$ taking the same value as in the genus-zero case.  Since
the Jacobian of \eqref{eq:G2eqns} is singular when endpoints coalesce,
we need to begin with an analytical perturbation calculation to move
slightly away from the primary caustic into the genus-two region.
We did the calculations for $x_{\rm crit}=0$ and $t_{\rm
  crit}=(2A)^{-1}$, where $\lambda_0 = \hat{\lambda}=A\sqrt{2}$ and
analytically determined asymptotic formulae for $\lambda_0(x,t)$,
$\lambda_1(x,t)$, and $\lambda_2(x,t)$ in the limit $t\downarrow
(2A)^{-1}$ with $x=0$.  With the help of these formulae, we could
employ our continuation code beginning with a point slightly beyond
the primary caustic where the three endpoints are distinct.  Thus, we
may obtain numerically, for any $(x,t)$ beyond the primary caustic,
the analytic solution of the equations \eqref{eq:G2eqns} that
correctly emerges from the primary caustic.

We also wrote a Matlab code that takes a point $(x,t)$ beyond the
primary caustic and the corresponding endpoints $\lambda_0(x,t)$,
$\lambda_1(x,t)$, and $\lambda_2(x,t)$ as input, and computes the band
$I_0$ connecting $\lambda_0(x,t)$ to the origin and the band $I_1$
connecting $\lambda_1(x,t)$ and $\lambda_2(x,t)$ by integrating
numerically the equation $\Re(R(\eta)Y(\eta)\,d\eta)=0$ to determine
the level curves.  The code also finds the regions of the complex
plane where the inequality $\Re(\phi(\lambda))<0$ holds; the gaps
$\Gamma_1$ connecting $\lambda_0(x,t)$ to $\lambda_1(x,t)$ and
$\Gamma_2$ connecting $\lambda_2(x,t)$ to an $\epsilon$ neighborhood
of the origin must lie in these regions.  By plotting this
information, we can visualize the dynamics of the genus-two ansatz in
the spectral plane, and hopefully determine the nature of the
secondary caustic.

\begin{figure}[htbp]
\begin{center}
\includegraphics[width=2.75 in]{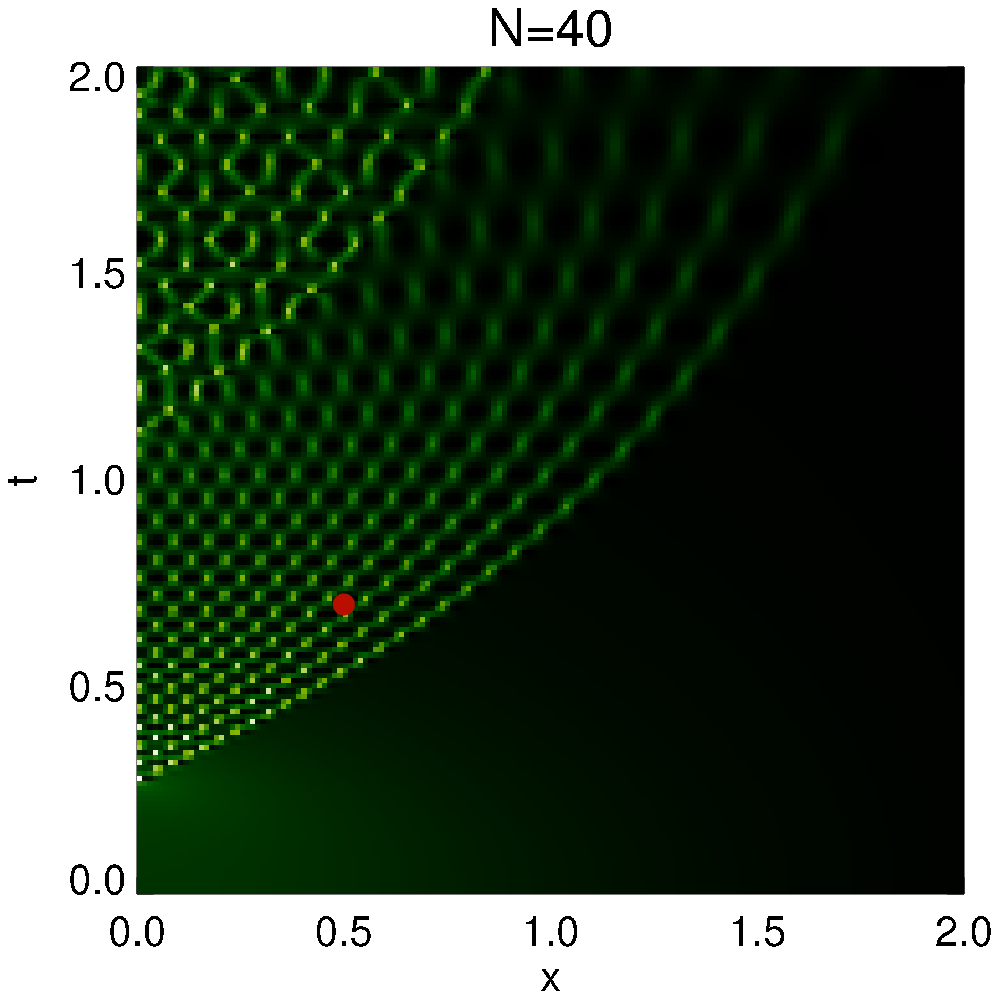}
\includegraphics[width=2.75 in]{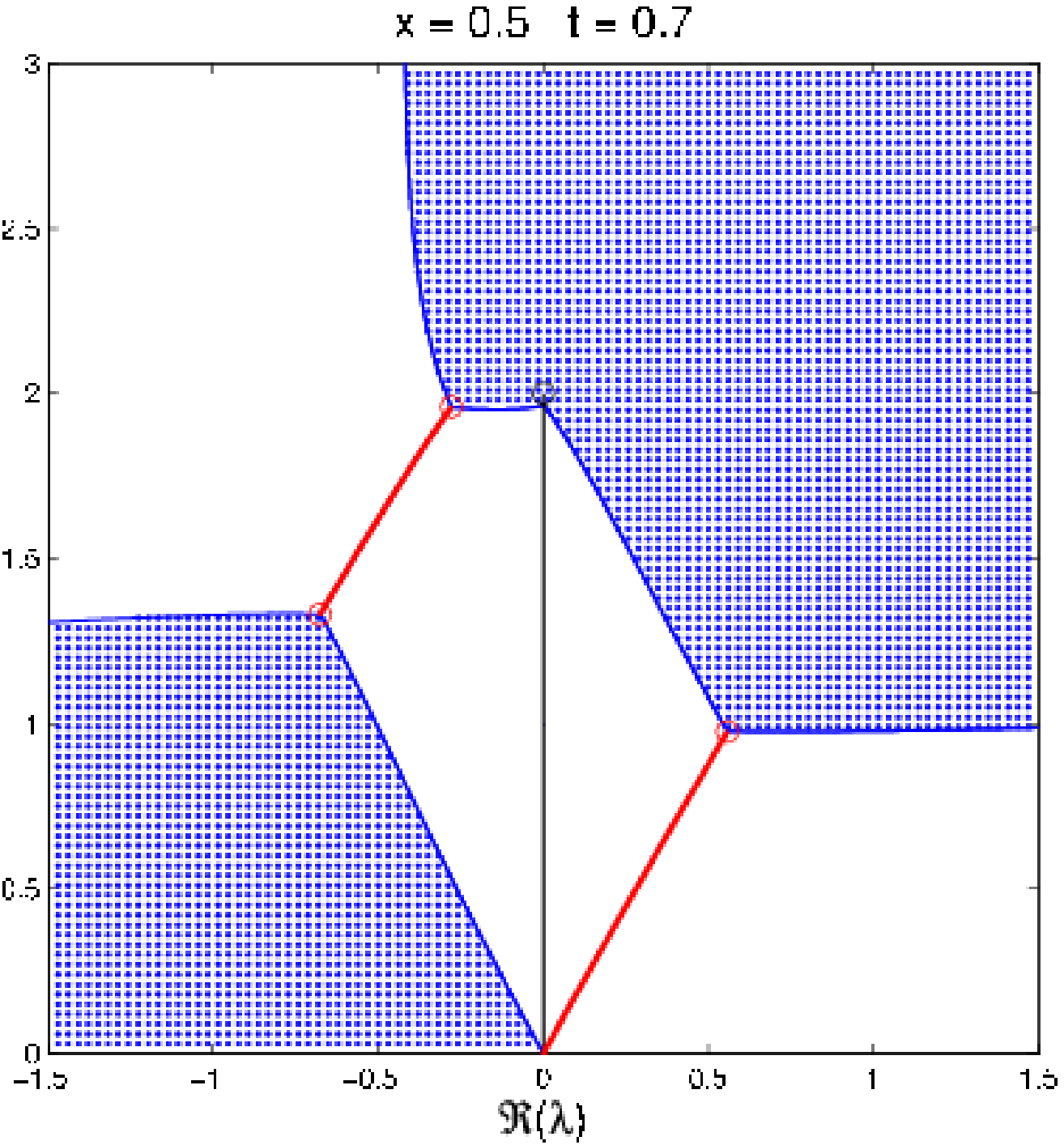}
\end{center}
\vspace{-0.1 in}
\begin{center}
\includegraphics[width=2.75 in]{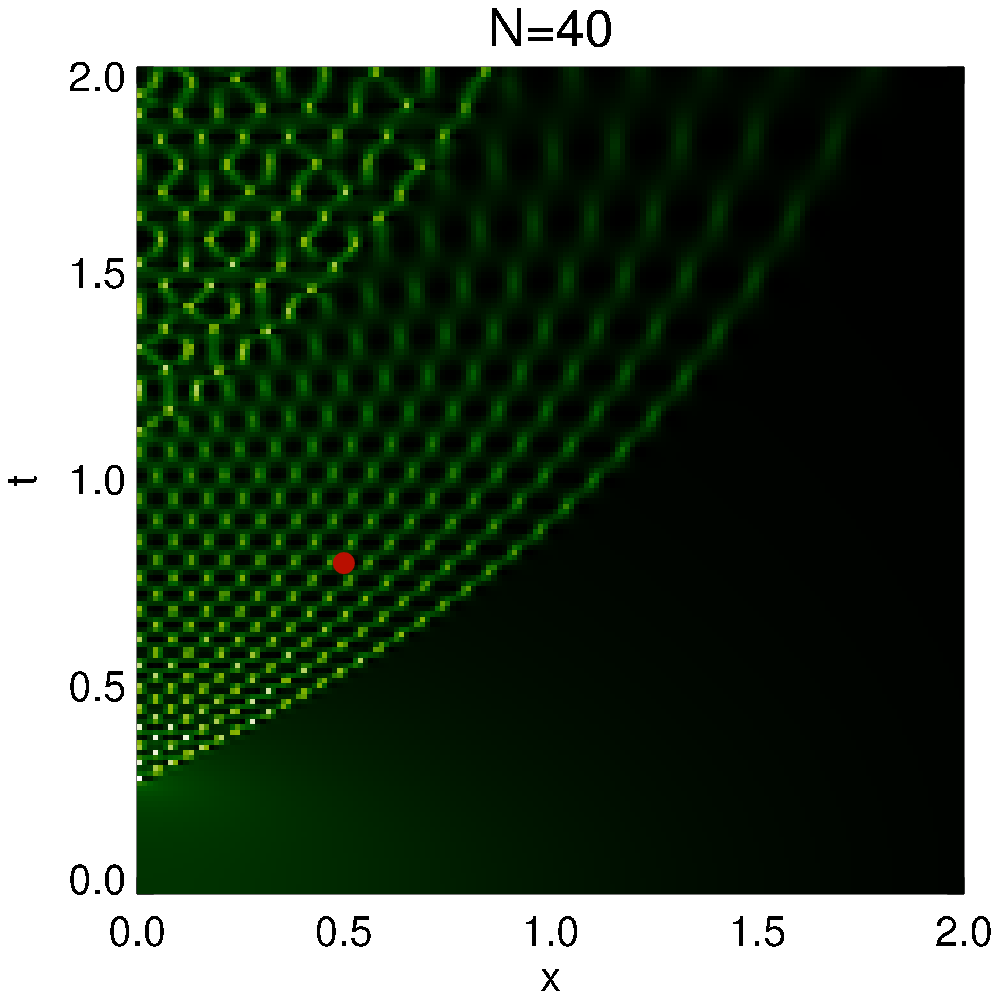}
\includegraphics[width=2.75 in]{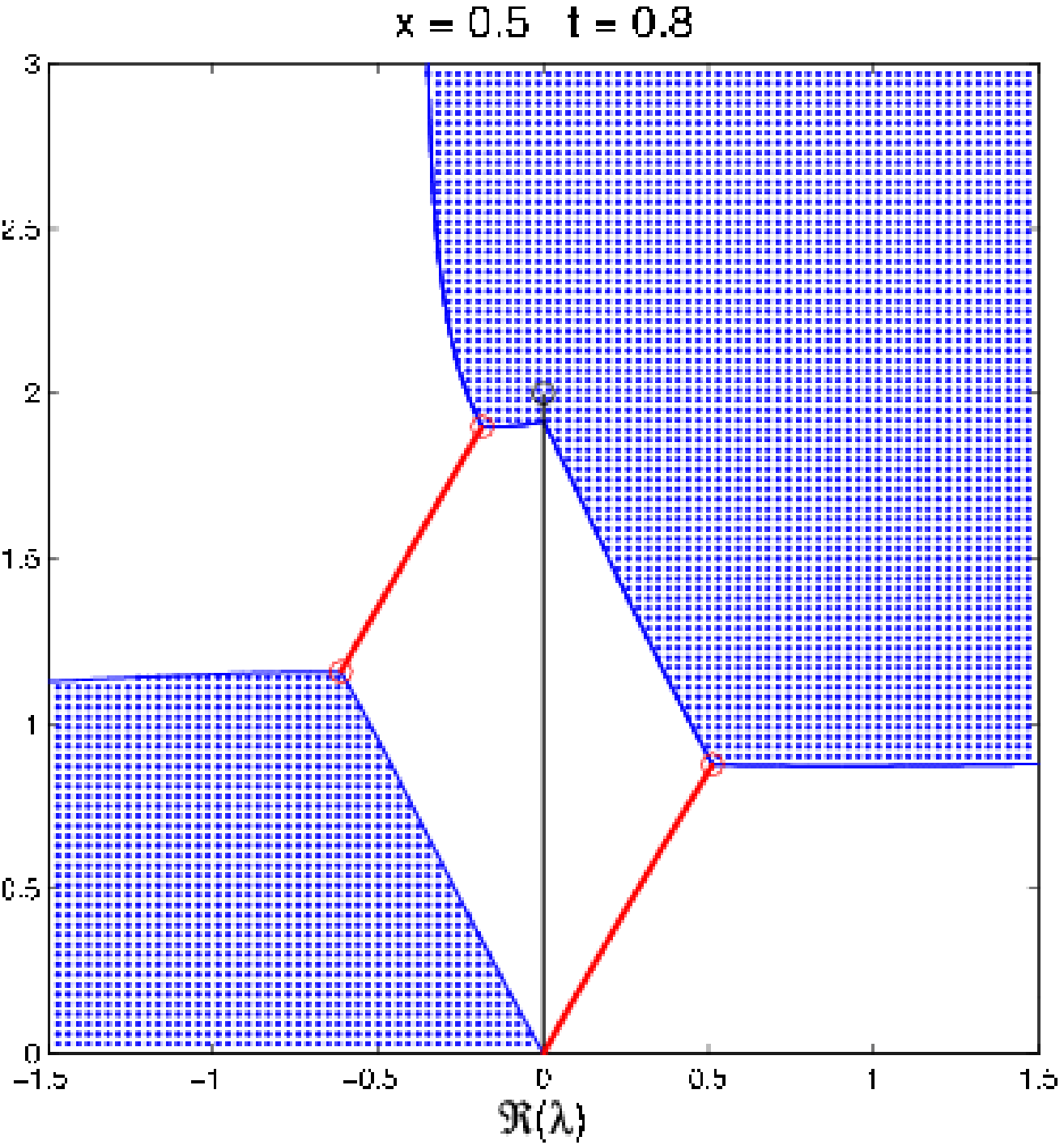}
\end{center}
\vspace{-0.1 in}
\begin{center}
\includegraphics[width=2.75 in]{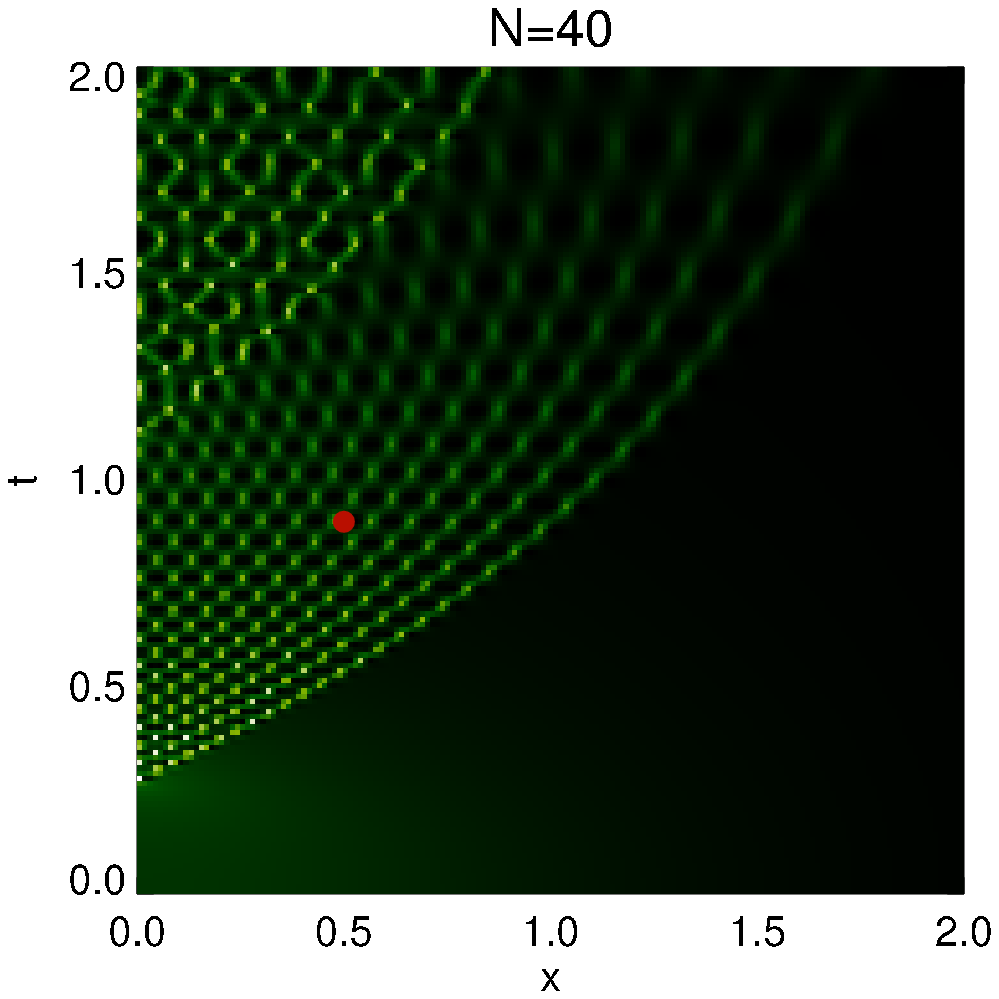}
\includegraphics[width=2.75 in]{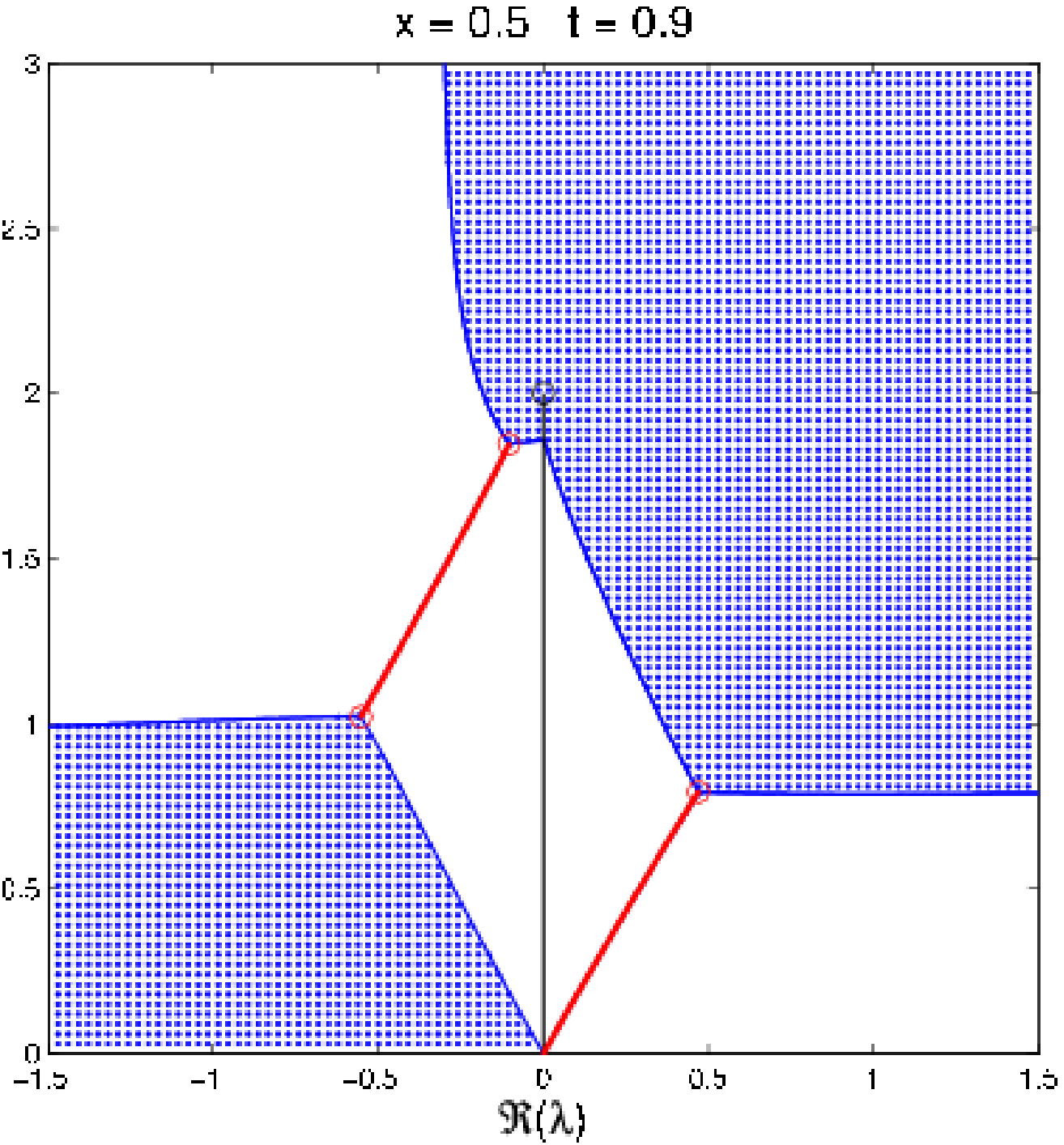}
\end{center}
\vspace{-0.2 in}
\caption{\em Evolution of the contour configuration for $G=2$.  
Here $x=0.5$ and $t=0.7$ through $t=0.9$ in steps of $\Delta t=0.1$.
See the text for a full explanation.}
\label{fig:sequence1}
\end{figure}

A representative sequence of such plots is given in
Figure~\ref{fig:sequence1}. Here we see the bands of the $G=2$ contour
in red and the region where the inequality $\Re(\phi(\lambda))<0$
holds shaded in blue. For the contour $C$ we simply take the imaginary
interval $[0,iA]$, shown with a black line.  Each spectral plot is
paired with a plot like Figure~\ref{fig:40soliton} but with a small
red dot indicating the corresponding point in the $(x,t)$-plane.  The
first plots show the genus-two configuration for a point in the
$(x,t)$-plane just beyond the primary caustic.  The band $I_1$ in the
left half-plane has just been born from a point $\hat{\lambda}$ where
\eqref{eq:G0break} holds.  The bands can clearly be joined together
with gaps $\Gamma_1$ and $\Gamma_2$ lying in the shaded regions to
complete a contour $K_0\cup K_{-1}$ that encircles the locus of poles
for ${\bf m}(\lambda)$.  Notice, however, that according to the
subsequent plots in this figure the band $I_1$ approaches the locus of
accumulation of poles for ${\bf m}(\lambda)$ as $t$ increases.
Although a collision of one of the endpoints with the interval
$[0,iA]$ would imply a failure of the genus-two ansatz in its original
formulation (that is, without the modification described in
\S~\ref{sec:dual}), it is clear by comparison with the corresponding
$(x,t)$-plane plots that the secondary caustic is still far off.

\begin{figure}[htbp]
\begin{center}
\includegraphics[width=2.75 in]{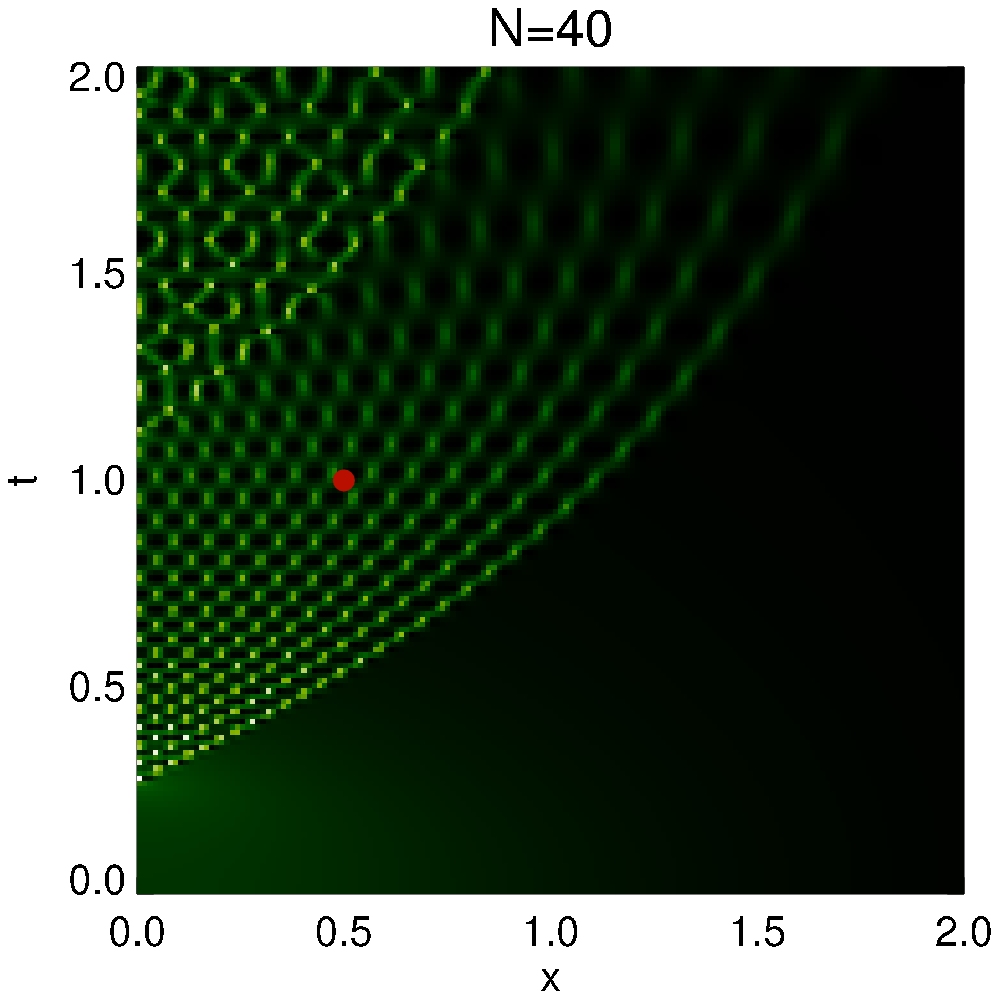}
\includegraphics[width=2.75 in]{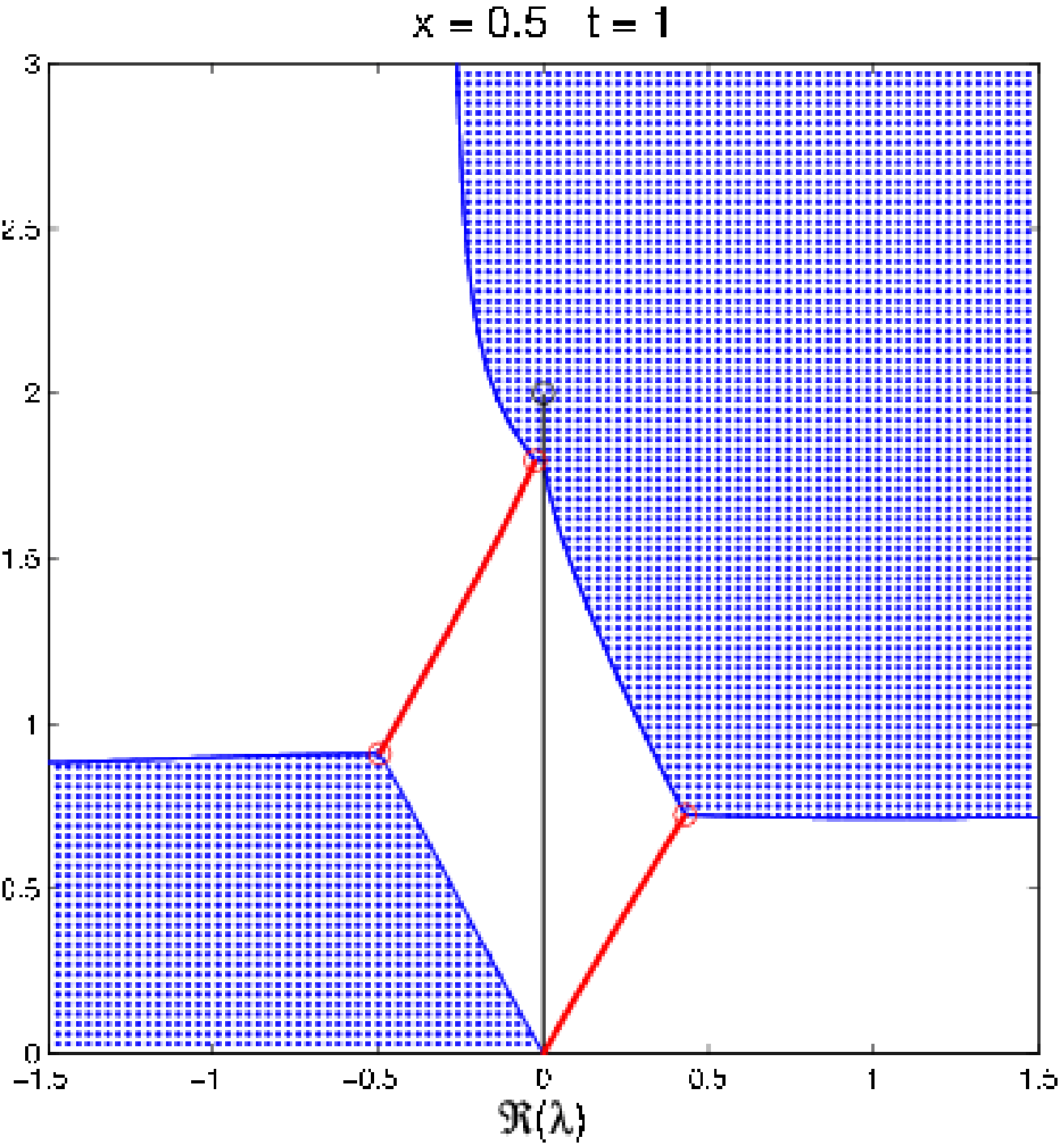}
\end{center}
\vspace{-0.1 in}
\begin{center}
\includegraphics[width=2.75 in]{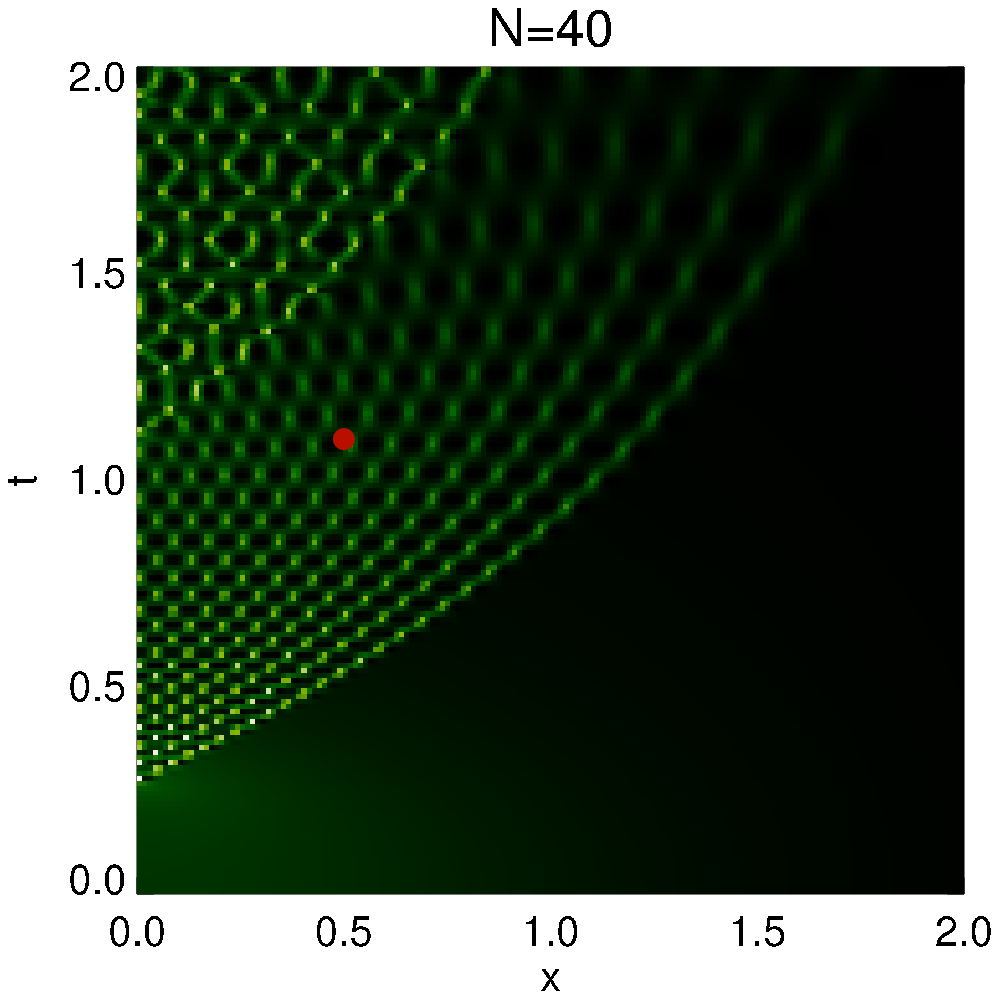}
\includegraphics[width=2.75 in]{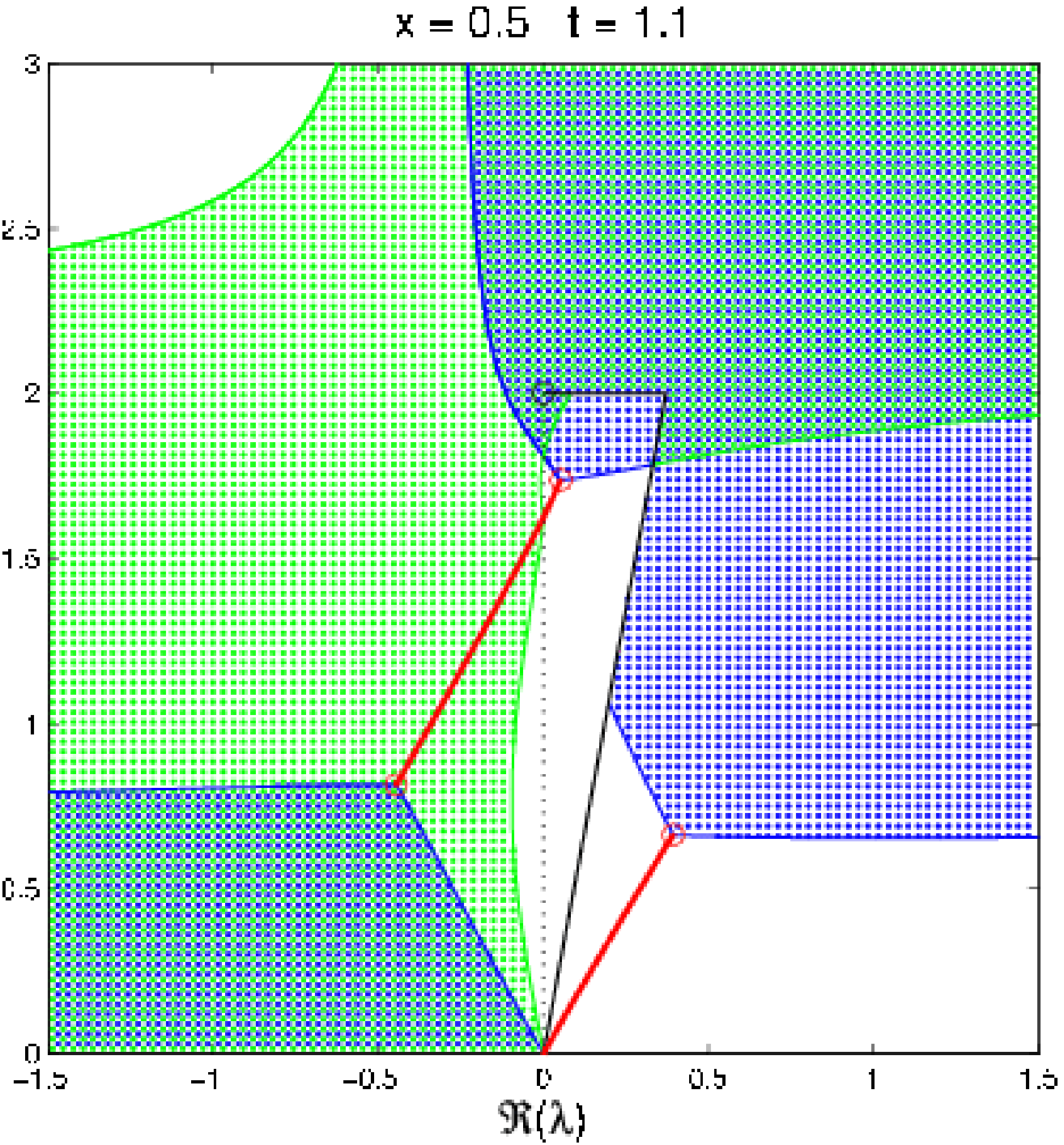}
\end{center}
\vspace{-0.1 in}
\begin{center}
\includegraphics[width=2.75 in]{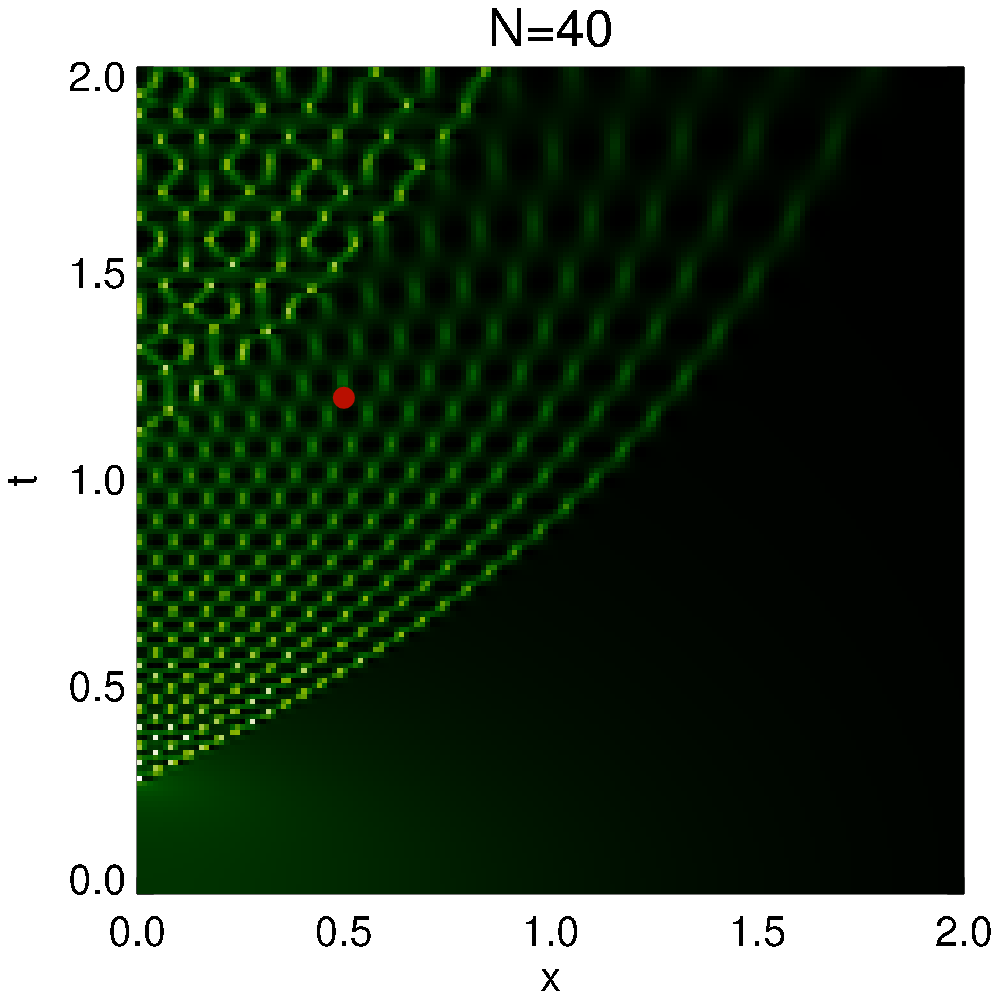}
\includegraphics[width=2.75 in]{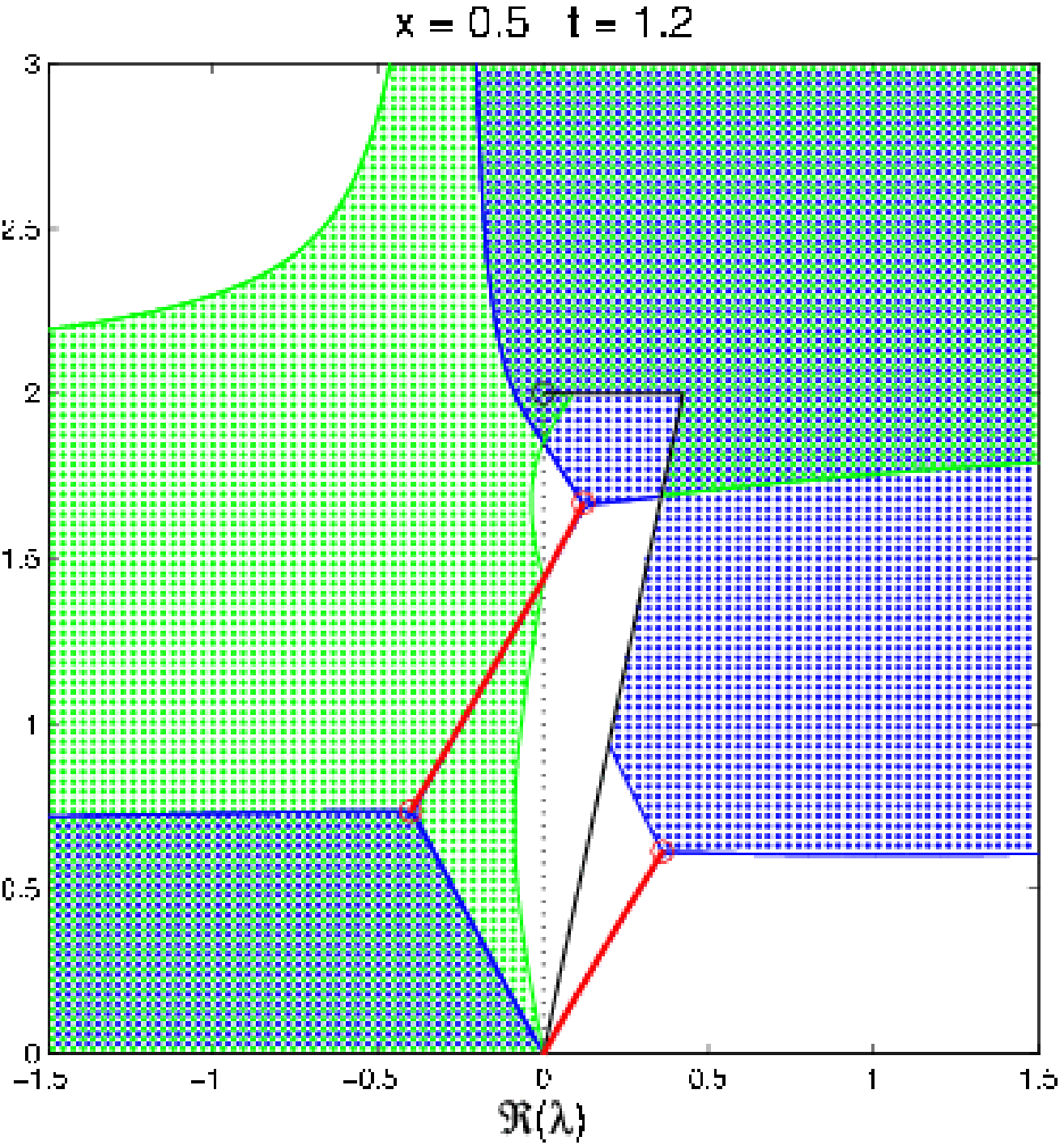}
\end{center}
\vspace{-0.2 in}
\caption{\em Same as in Figure~\ref{fig:sequence1} but with
$t=1.0$ through $t=1.2$ in steps of $\Delta t=0.1$.  See the text for
a full explanation.}
\label{fig:sequence2}
\end{figure}

Figure~\ref{fig:sequence2} continues the evolution.  Note that between
the first two frames, the band $I_1$ collides with the interval
$[0,iA]$, so we must switch to the modified version of the genus-two
ansatz described in \S~\ref{sec:dual}.  Thus, a new contour arc
$K_{-3}$ must be included and a new inequality
$\Re(\phi(\lambda)-2i\theta^0(\lambda))<0$ must be satisfied along
this contour.  The region where this inequality is satisfied is shaded
in green.  Thus, the contour arc $K_{-3}$ must lie in the green shaded
region, while the gaps of $K_{-1}\cup K_{-2}$ must lie in the blue
shaded regions.  Furthermore, the contour $C$ must from this time
onward be deformed somewhat to the right to admit the passage of the
band endpoint $\lambda_1(x,t) $ ``through the branch cut''.  The
deformed contour $C$ we choose here, somewhat arbitrarily, is the
union of the two indicated solid black line segments.
It is clear that in each case the bands may be complemented with gaps
in the blue regions to form (along with a small interval
$(-\epsilon,0)$) a closed loop.  In cases where this loop necessarily
cannot enclose the poles of ${\bf m}(\lambda)$ accumulating in
$[0,iA]$, it is clear that a curve $K_{-3}$ lying entirely in the
green shaded region may be added that passes over the top of $[0,iA]$
and forms along with parts of the above closed loop the boundary of a
region containing the poles.

\begin{figure}[htbp]
\begin{center}
\includegraphics[width=2.75 in]{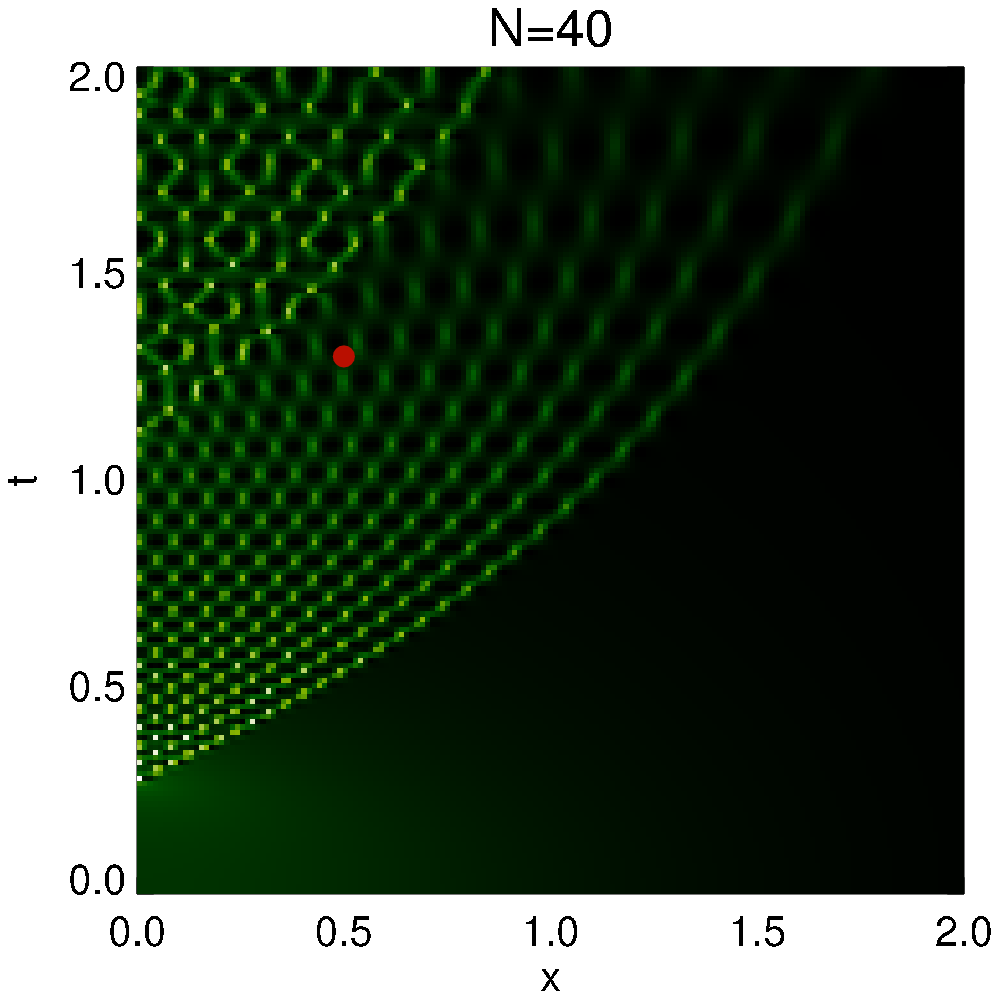}
\includegraphics[width=2.75 in]{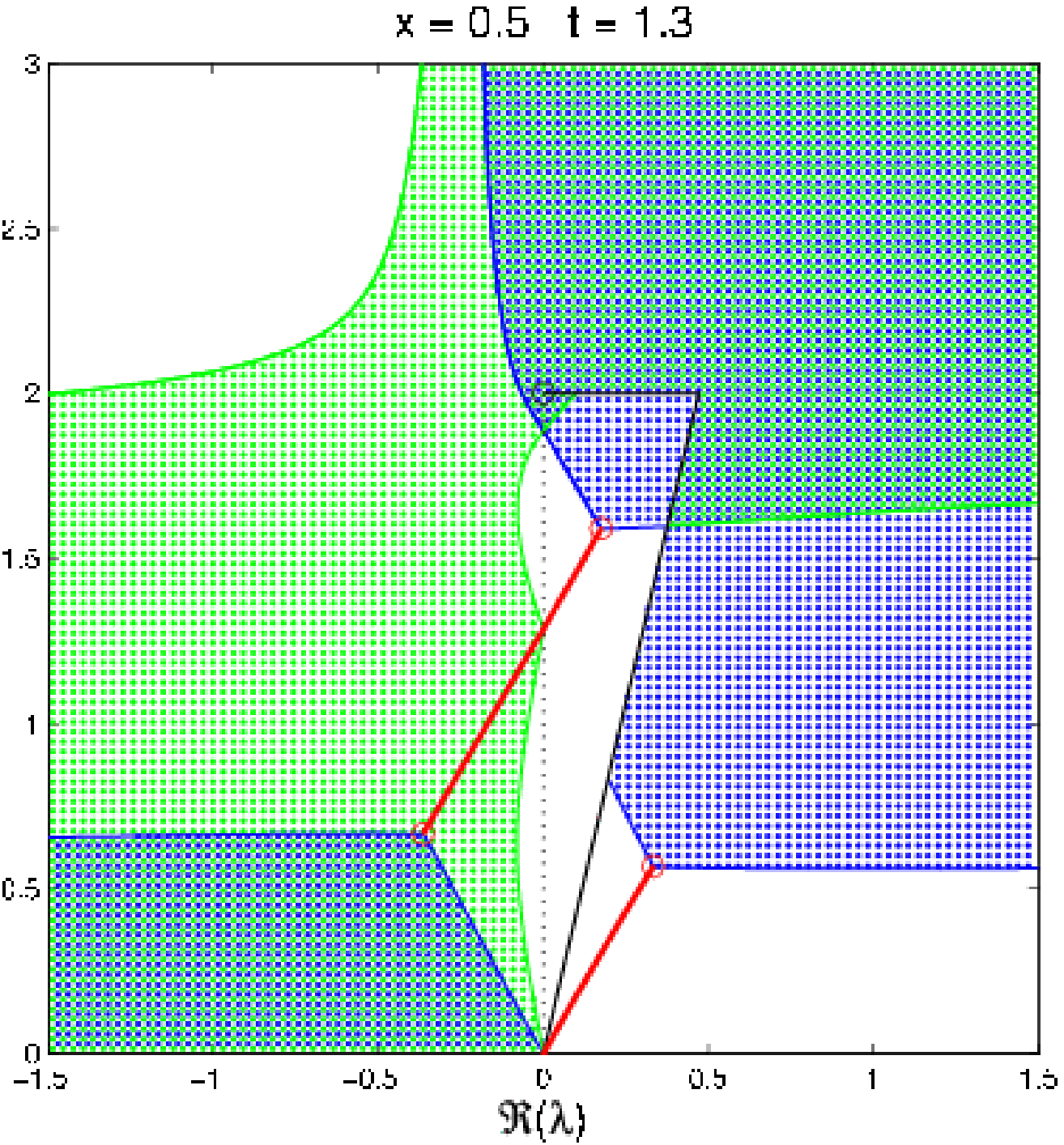}
\end{center}
\vspace{-0.1 in}
\begin{center}
\includegraphics[width=2.75 in]{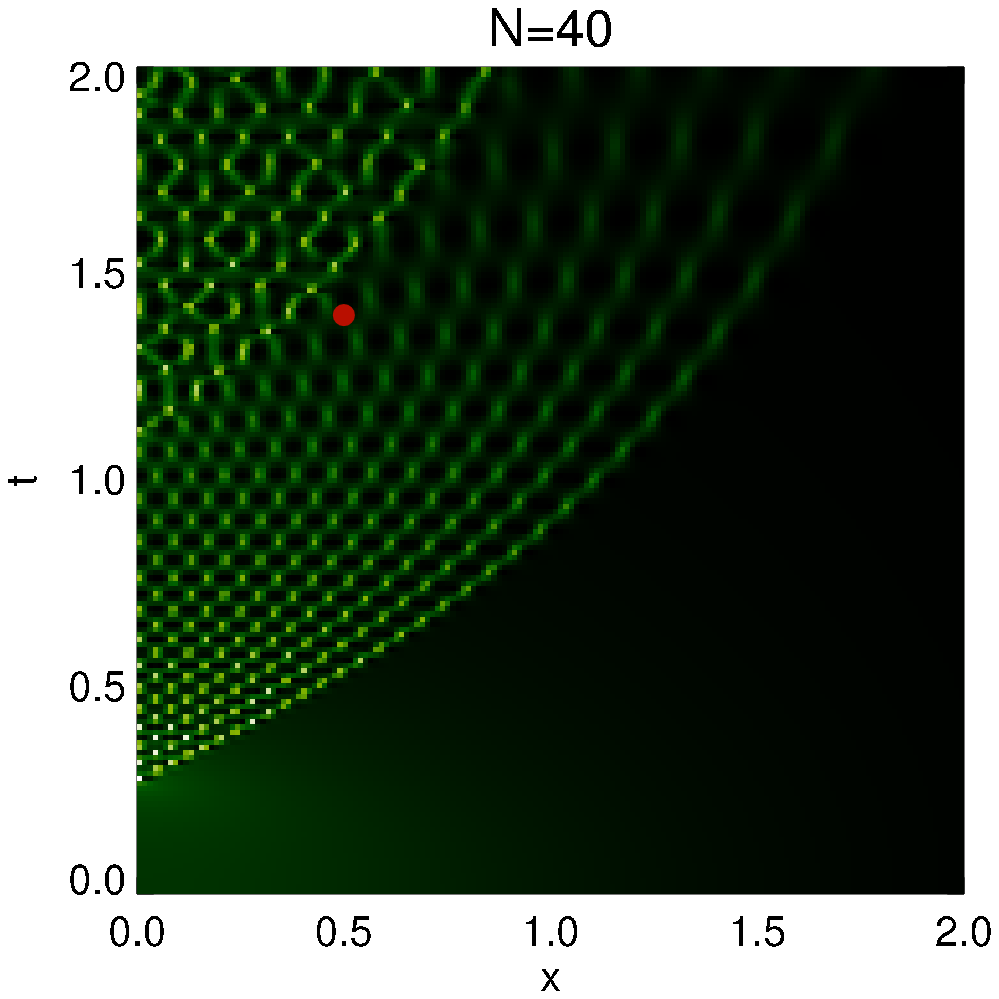}
\includegraphics[width=2.75 in]{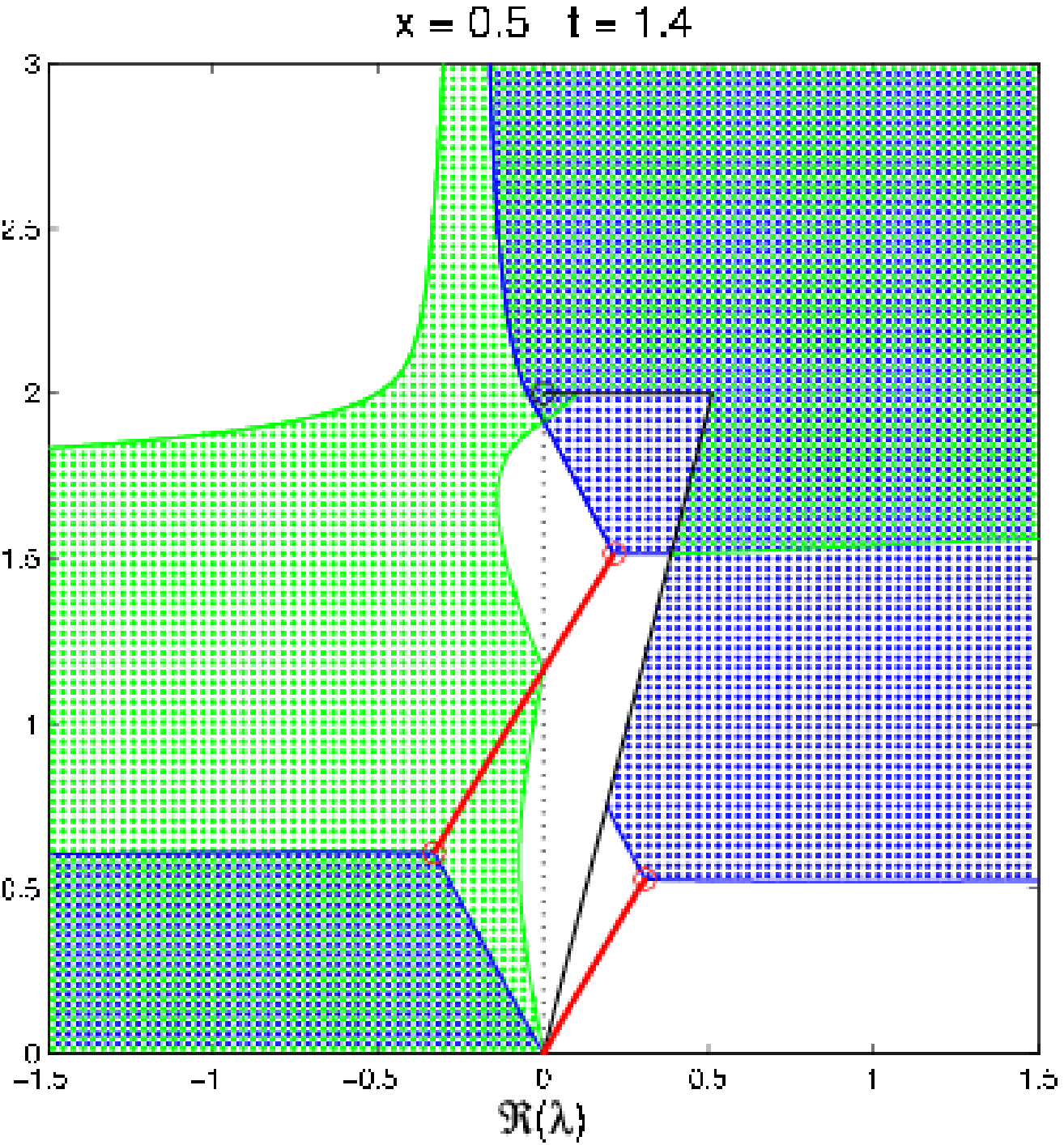}
\end{center}
\vspace{-0.1 in}
\begin{center}
\includegraphics[width=2.75 in]{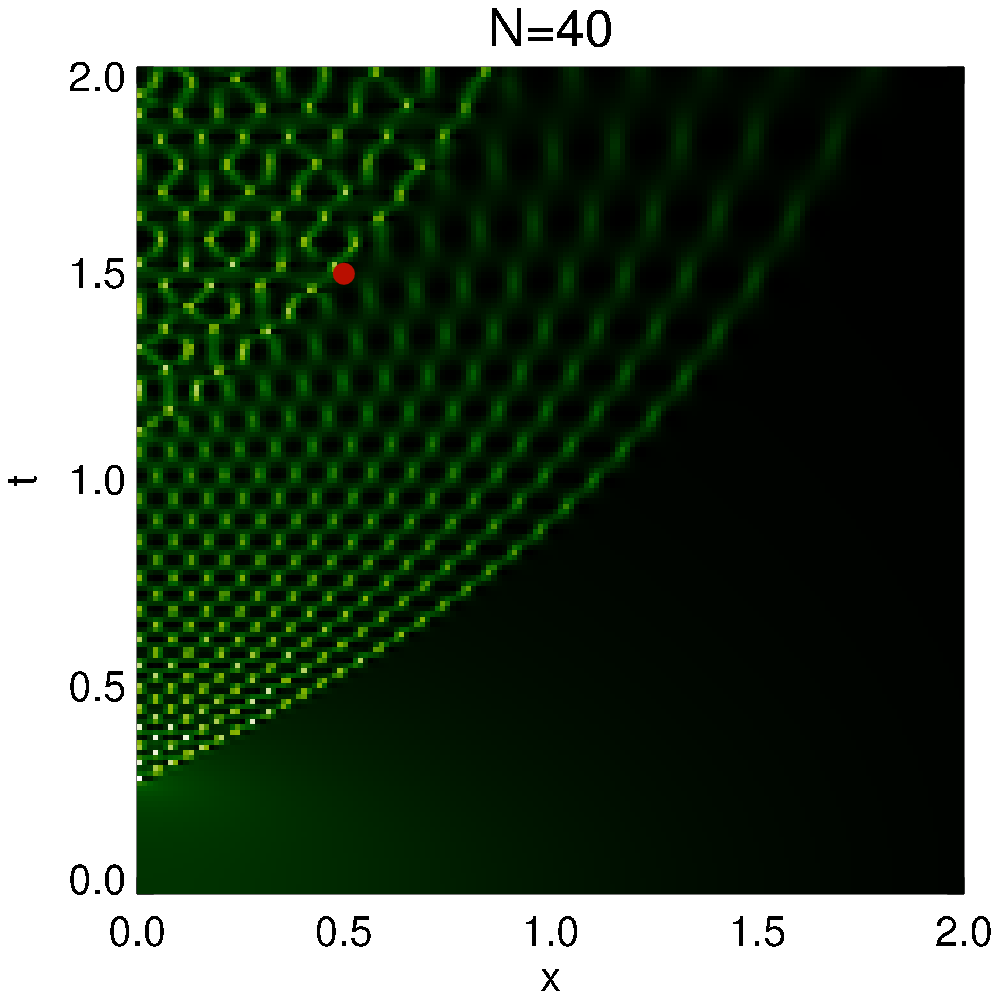}
\includegraphics[width=2.75 in]{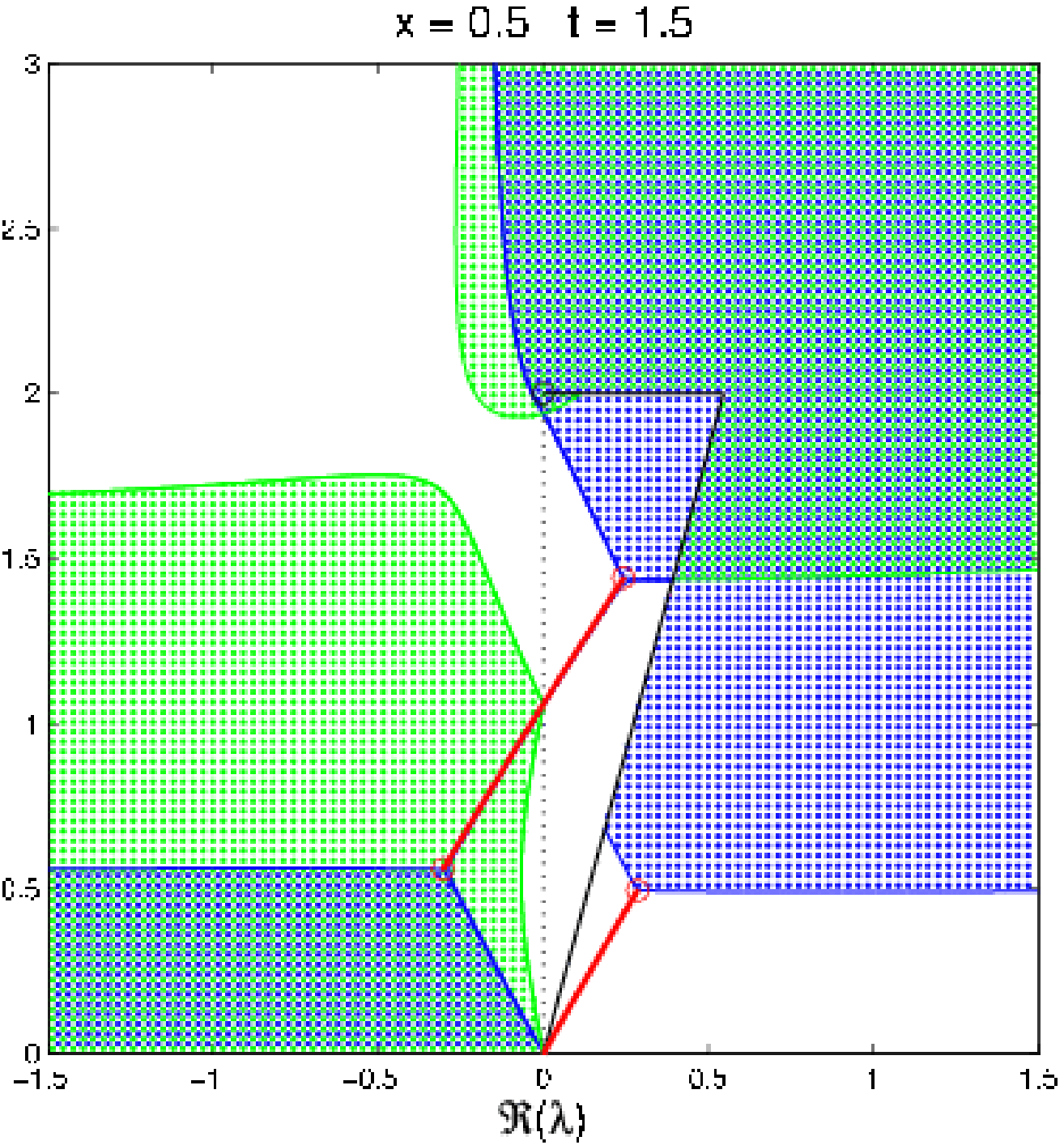}
\end{center}
\vspace{-0.2 in}
\caption{\em Same as in Figure~\ref{fig:sequence2}, but with
  $t=1.3$ through $t=1.5$ in steps of $\Delta t=0.1$.  See the text for
a full explanation.}
\label{fig:sequence3}
\end{figure}

In Figure~\ref{fig:sequence3} we continue the evolution toward a curve
in the $(x,t)$-plane at which the asymptotic behavior abruptly changes
for the second time, where we expect to find a mathematical
explanation for the secondary caustic.  Note that between the last two
snapshots there is a pinch-off of the green shaded region where
$\Re(\phi(\lambda)-2i\theta^0(\lambda))<0$.  When this happens, a
contour lobe $K_{-3}$ that serves to bound the region containing the
poles of ${\bf m}(\lambda)$ (soliton eigenvalues) and also along which
the inequality $\Re(\phi(\lambda)-2i\theta^0(\lambda))<0$ holds can no
longer exist.  {\em This is the mathematical mechanism for the
  development of the secondary caustic.}

\begin{figure}[htbp]
\begin{center}
\includegraphics[width=2.75 in]{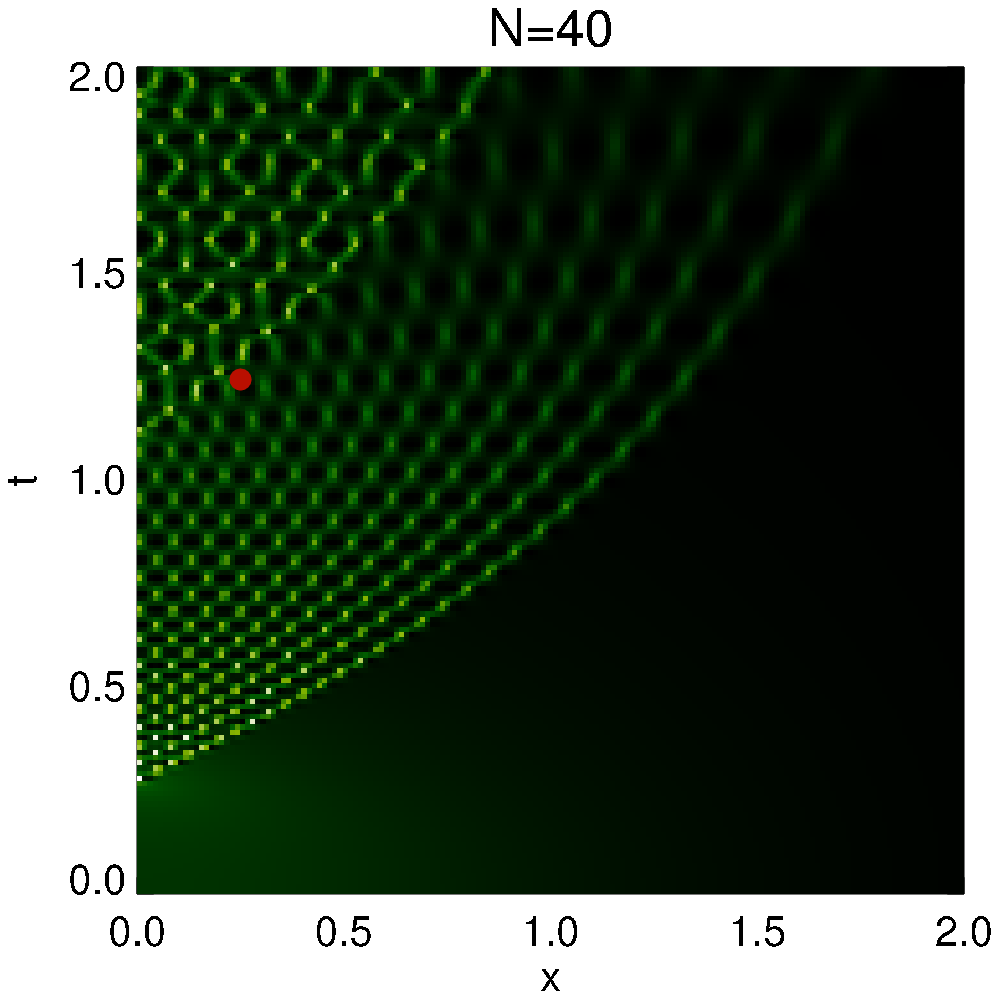}
\includegraphics[width=2.75 in]{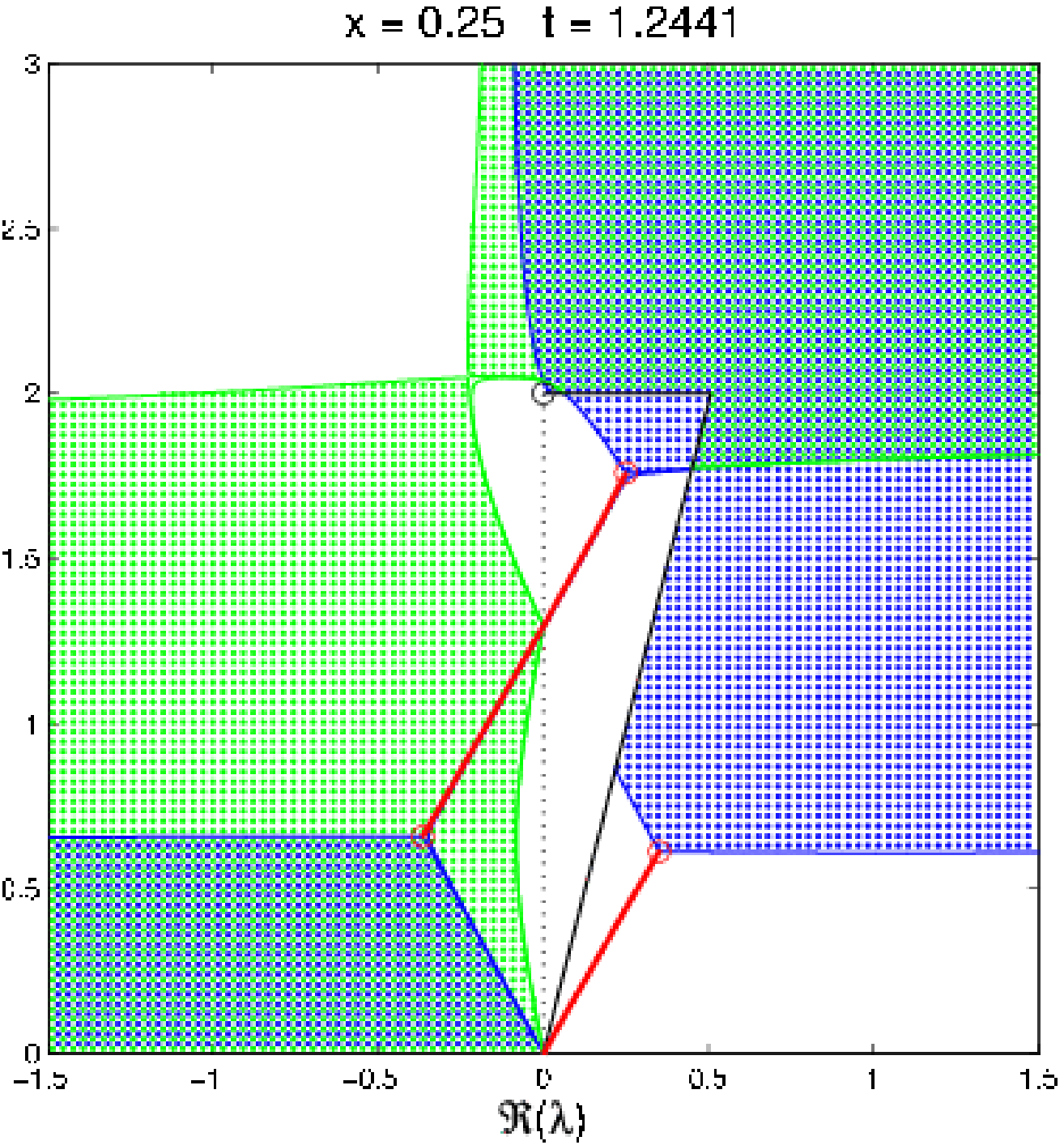}
\end{center}
\vspace{-0.1 in}
\begin{center}
\includegraphics[width=2.75 in]{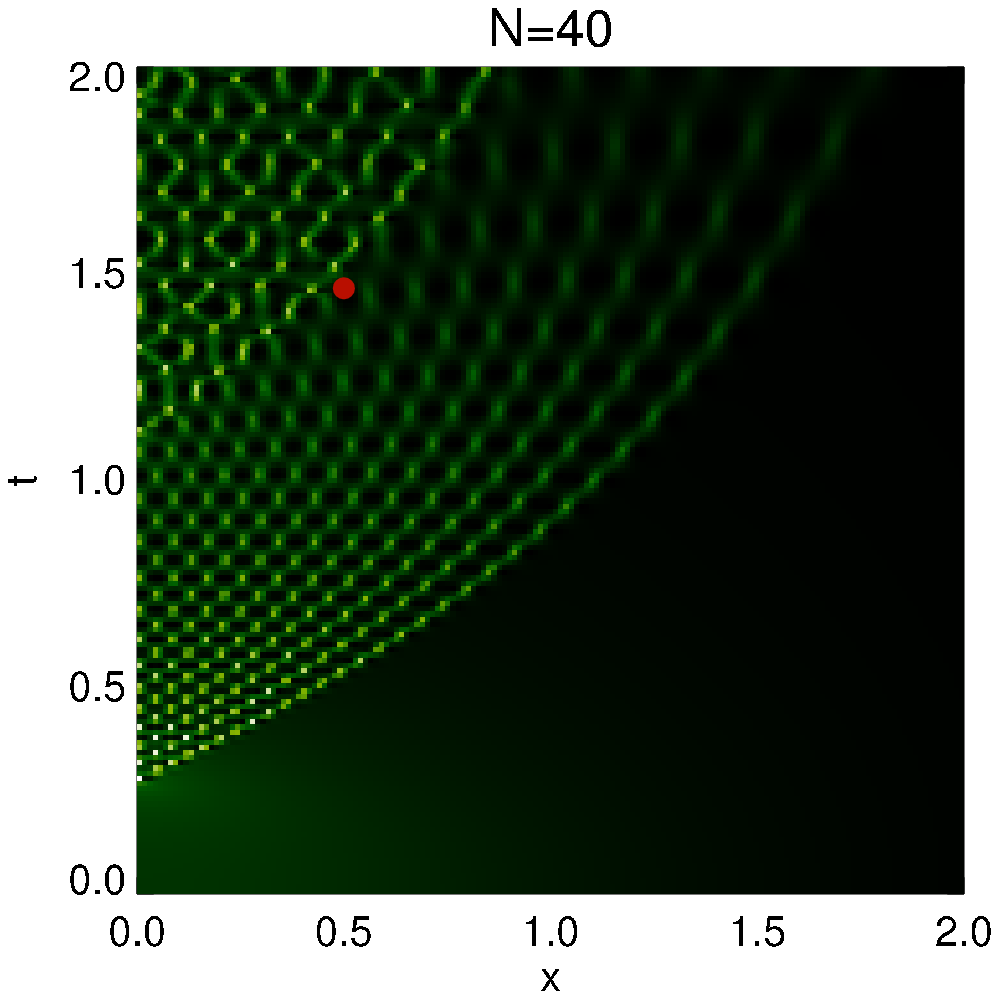}
\includegraphics[width=2.75 in]{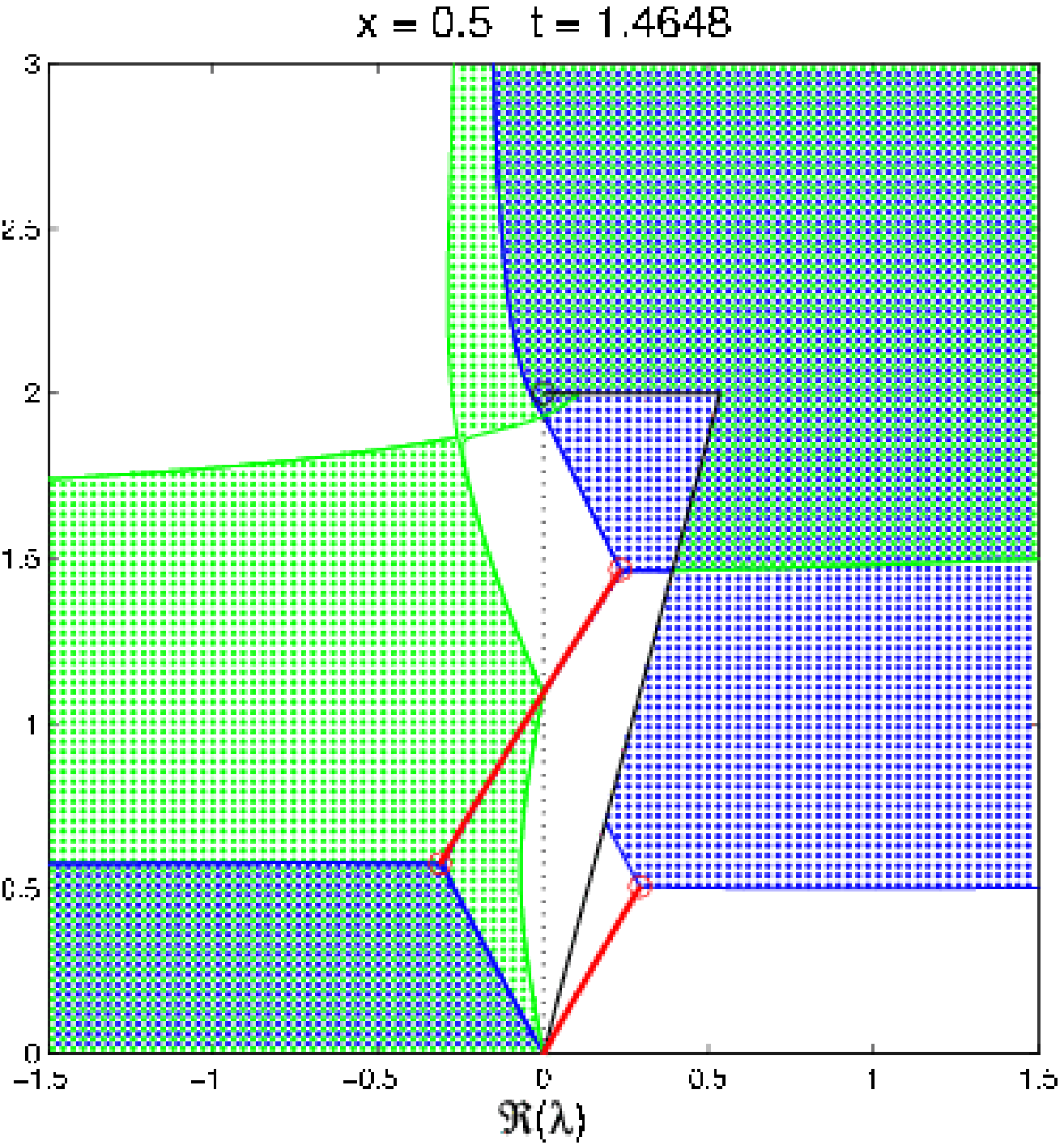}
\end{center}
\vspace{-0.1 in}
\begin{center}
\includegraphics[width=2.75 in]{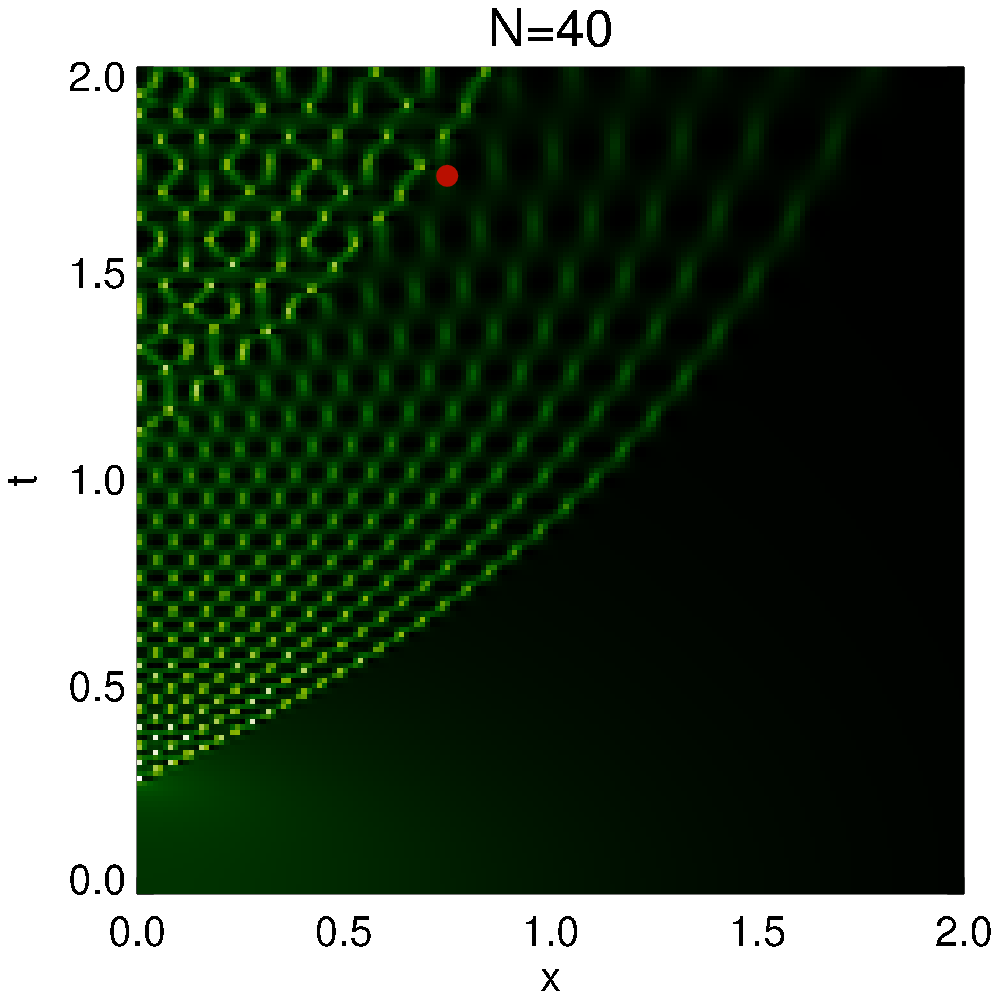}
\includegraphics[width=2.75 in]{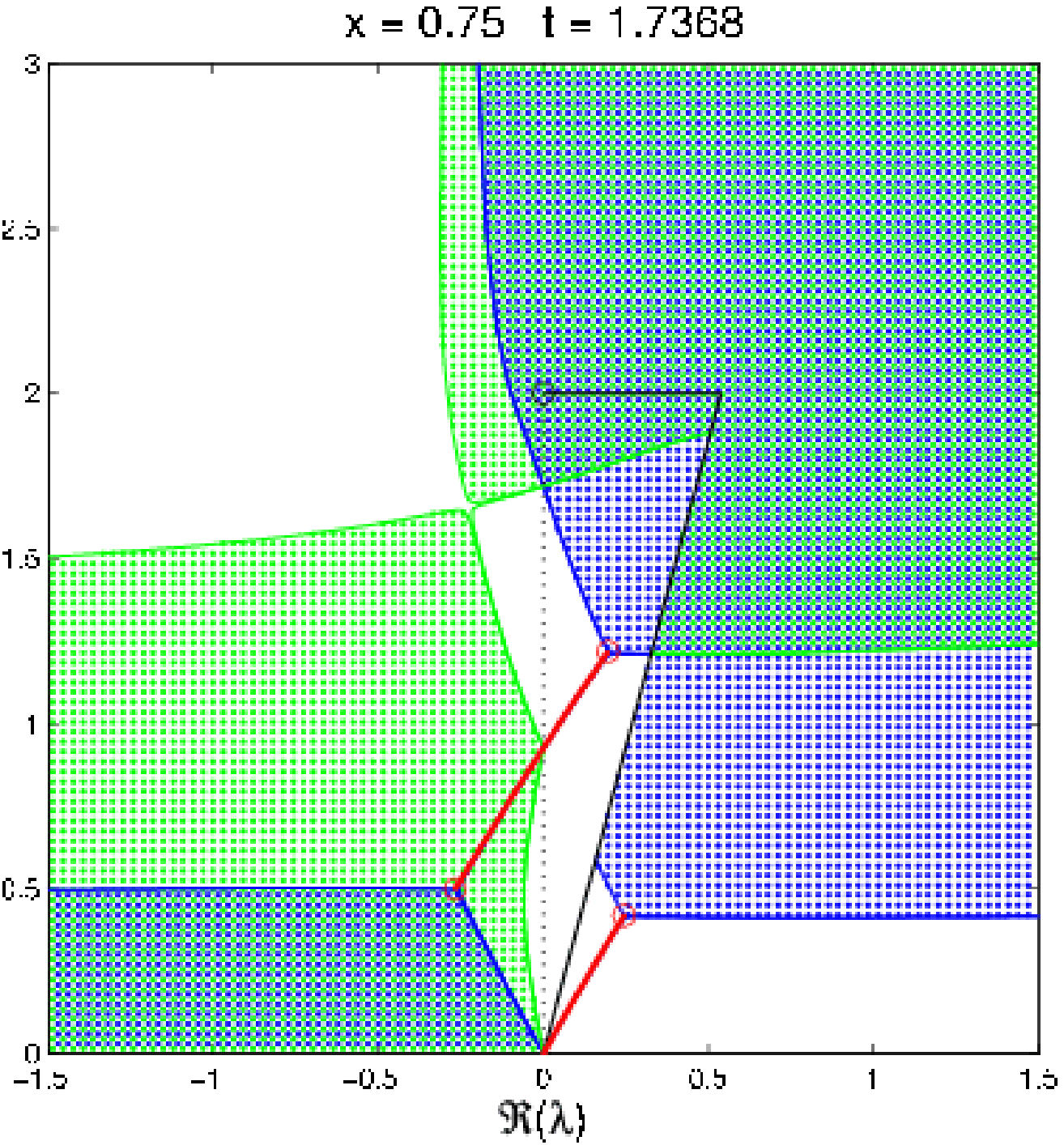}
\end{center}
\vspace{-0.2 in}
\caption{\em Configurations in which the region where the
inequality $\Re(\phi(\lambda)-2i\theta^0(\lambda))<0$ holds is ``pinched-off''
delineate the secondary caustic.}
\label{fig:causticsequence}
\end{figure}

To demonstrate more convincingly that the secondary caustic in the
numerical reconstructions of the $N$-soliton solution for large $N$
actually corresponds to the curve in the $(x,t)$-plane along which the
region in which the inequality
$\Re(\phi(\lambda)-2i\theta^0(\lambda))<0$ is pinched off at a point,
we used a secant method to find, for a given $x$, the $t$-value at
which $\Re(\phi(\hat{\lambda})-2i\theta^0(\hat{\lambda}))=0$ where
$\hat{\lambda}$ is a complex root of
$\phi'(\lambda)-2i\theta^{0\prime}(\lambda)$.  Level curves of
$\Re(\phi(\lambda)-2i\theta^0(\lambda))$ passing through simple
critical points like $\hat{\lambda}$ necessarily exhibit a
characteristic perpendicular crossing singularity.  This method yields
(approximate) coordinates of points in the $(x,t)$-plane at which the
configuration of level curves makes a transition between the two types
of configurations illustrated in the final two frames of
Figure~\ref{fig:sequence3}.  These points lie along a curve $t=t_2(x)$
that defines the location of the secondary caustic.  Configurations
corresponding to three of the points we obtained in this way are
illustrated in Figure~\ref{fig:causticsequence}.

\section{Transition to Higher Genus}
\label{sec:highergenus}
\subsection{Violation of the new inequality.}
According to the numerics explained in \S~\ref{sec:GenusTwoNumerics},
the failure of the genus-two ansatz at the secondary caustic
corresponds to the equations (compare with \eqref{eq:G0break}):
\begin{equation}
\phi'(\hat{\lambda})-2i\theta^{0\prime}(\hat{\lambda})=0\,,\hspace{0.2 in}
\Re(\phi(\hat{\lambda})-2i\theta^0(\hat{\lambda}))=0\,.
\label{eq:G2break}
\end{equation}
We expect the critical point $\hat{\lambda}$ arising from these
equations to be the origin of a new band, and to show this we need to
develop the genus-four ansatz within the framework of the modified
approach given in \S~\ref{sec:dual}, but where the new band appears on
the contour $K_{-3}$.  This gives the equations implicitly defining
the endpoints as functions of $x$ and $t$ a different appearance than
if the new band were assumed to lie on $K_{-1}\cup K_{-2}$.  These
equations turn out to define a previously unknown branch of solutions
of the quasilinear system of Whitham equations (as defined in
\S~\ref{sec:Whitham}).
\subsection{On and beyond the secondary caustic.}
\subsubsection{Endpoint equations for genus four with a band on $K_{-3}$.}
\input{section6-2.tex}

\subsubsection{Birth of a new band upon crossing the secondary caustic.}
Here we show that when $(x,t)$ is a point on the secondary caustic, so
that the equations \eqref{eq:G2break} hold, the equations derived
above that determine the endpoints of the genus-four ansatz with band
$I_2$ on the outer contour lobe $K_{-3}$ admit the solution where
$\lambda_0(x,t)$, $\lambda_1(x,t)$, and $\lambda_2(x,t)$ are all equal
to the corresponding endpoints for the genus-two ansatz at the same point
in the $(x,t)$-plane, while $\lambda_3(x,t)=\lambda_4(x,t)=\hat{\lambda}$.

To show this we first note that the conditions defining the function
$g'(\lambda)$ in terms of the endpoints yield the same function in
both the genus-two case (ignoring $\lambda_3$ and $\lambda_4$) and in
the genus-four case described above (assuming $\lambda_3=\lambda_4$).
Since $\lambda_0$, $\lambda_1$, and $\lambda_2$ are such that
$g'(\lambda) = O(\lambda^{-2})$ as $\lambda\rightarrow\infty$ (because
the genus-two version of $g'(\lambda)$ satisfies the appropriate suite
of four moment conditions), we also know that the genus-four version
of $g'(\lambda)$ satisfies the corresponding suite of six moment
conditions (see \eqref{eq:g4moment012}--\eqref{eq:g4moment5}).

It therefore remains to verify the remaining integral conditions
\eqref{eq:g4reality12}--\eqref{eq:g4vanish2} involving $g'(\eta)$.
Since the function $g'(\eta)$ is the same whether we consider the
genus-two or degenerate genus-four version, the conditions
$\tilde{R}_1=0$ and $\tilde{V}_1=0$ are automatically satisfied, as
these are equivalent to the conditions $R_1=0$ and $V_1=0$ satisfied
by the genus-two version of $g'(\eta)$.  Furthermore, since $g'(\eta)$
is analytic in a neighborhood of $\hat{\lambda}$, we have by Cauchy's
Theorem
\begin{equation}
\oint_{A_2}g'(\eta)\,d\eta = 0\,,
\end{equation}
which in particular implies that the real part is zero, so
$\tilde{R}_2=0$ is satisfied as well.  Finally, we observe that the
second vanishing condition \eqref{eq:g4vanish2} is, when the point
$\hat{\lambda}$ is taken as the second endpoint of the gap $\Gamma_2$,
satisfied as a consequence of the definition \eqref{eq:G2break} of the
critical point characteristic of the secondary caustic.  This
completes the proof that each point on the secondary caustic, as
defined by \eqref{eq:G2break} for the genus-two ansatz, corresponds
to a degeneration of an admissible genus-four configuration with
$\lambda_3=\lambda_4=\hat{\lambda}$.

Showing that the points $\lambda_3$ and $\lambda_4$ separate in a
direction admitting a new band $I_2$ between them as $t$ increases
beyond the secondary caustic is slightly complicated by the fact that
the Jacobian of the system of endpoint equations is singular in this
case.  However, a direct calculation shows that the $8\times 8$
Jacobian of all endpoint equations with the exception of
\eqref{eq:g4reality12} for $n=2$ and \eqref{eq:g4vanish2} taken with
respect to the independent variables $\lambda_0$, $\lambda_1$,
$\lambda_2$, $(\lambda_3+\lambda_4)/2$, and their complex conjugates
is nonsingular in the degenerate configuration when the remaining
independent variables $\delta:=(\lambda_4-\lambda_3)/2$ and
$\delta^*:=(\lambda_4^*-\lambda_3^*)/2$ both vanish.  This reduces the
problem to the study of two equations and two unknowns.  It follows
that near the degenerate configuration, the eliminated variables are
all analytic functions of $x$, $t$, $\delta$, and $\delta^*$, and it can
be checked that all are invariant under $\delta\rightarrow -\delta$
and $\delta^*\rightarrow -\delta^*$ independently.  This actually
makes them analytic functions of $x$, $t$, $\delta^2$, and $\delta^{*2}$.

The remaining equations are $\tilde{R}_2=0$ and $\tilde{V}_2=0$, and
the left-hand sides of these are now to be considered functions of
$x$, $t$, $\delta$, and $\delta^*$.  Knowing that they admit the
solution $\delta=\delta^*=0$ when $(x,t)$ is a point on the secondary
caustic, we wish to solve them for the variables $\delta$ and
$\delta^*$ for nearby $(x,t)$.  We will show that there is a unique
solution satisfying the condition that $\delta$ and $\delta^*$ are
complex conjugates of each other for real $x$ and $t$ (there are other
solutions that bifurcate from $\delta=\delta^*=0$ that do not have
this symmetry, which is the primary source of difficulty).  We define
polar coordinates by
\begin{equation}
\delta = re^{i\theta}\,,\hspace{0.2 in}\delta^*=re^{-i\theta}\,,
\end{equation}
and consider the equations $\tilde{R}_2=0$ and $\tilde{V}_2=0$
naturally complexified (that is, the left-hand sides extended as
analytic functions of four complex variables $x$, $t$, $r$, and
$\theta$.  Being as the complexification of $\tilde{R}_2$ can be written as
the average of two integrals, one on a path surrounding the two
endpoints $\lambda_3$ and $\lambda_4$ and the other on a path
surrounding the two endpoints $\lambda_3^*$ and $\lambda_4^*$, it is
easy to see that $\tilde{R}_2$ is an even function of $\delta$ and $\delta^*$
individually that vanishes when $\delta=\delta^*=0$, and therefore the
equation $\tilde{R}_2=0$ has the form
\begin{equation}
0=\alpha(x,t)r^2\cos(2(\theta-\beta(x,t))) + 
\text{quartic (and higher order) terms
in $r$.}
\end{equation}
This equation clearly defines two analytic curves crossing orthogonally at the
origin in the real polar plane, with angles
$\theta=\beta(x,t)+\pi/4+n\pi/2$.

Now, the complexified form of $\tilde{V}_2$ defines a multivalued
function of $\delta$ and $\delta^*$, even when we restrict to the real
subspace where these two variables are complex conjugates of each
other.  This comes about because the value of the integral terminating
at $\lambda_3$ in $\tilde{V}_2$ can change its value if $\lambda_3$
and $\lambda_4$ are permuted.  This monodromy effect can, however, be
removed in the real subspace, because the deformation
$(\delta,\delta^*)\rightarrow (\delta e^{i\tau}, \delta^*e^{-i\tau})$
for $\tau$ going from zero to $\pi$ (which exchanges the endpoints and
their conjugates) simply adds to $\tilde{V}_2$ a copy of the function
$\tilde{R}_2$.  This suggests that in place of $\tilde{V}_2=0$ we can
solve an equivalent modified equation that is single-valued in the
real subspace.  Indeed, we opt to replace $\tilde{V}_2=0$ by
$\tilde{V}^\#_2=0$ where
\begin{equation}
\begin{array}{rcl}
\tilde{V}_2^\#&:=&\displaystyle
\Re\left(\int_{\lambda_M}^{\lambda_3}R(\eta)\tilde{Y}(\eta)\,d\eta\right) -
\frac{1}{4\pi i}\log\left(\frac{\delta}{\delta^*}\right)\cdot
\Re\left(\oint_{A_2}R(\eta)\tilde{Y}_+(\eta)\,d\eta\right) \\\\
&&\hspace{0.2 in}
+\,\,\,\displaystyle
\Re\left(\int_{\lambda_2}^{\lambda_M}R(\eta)\tilde{Y}(\eta)\,d\eta
+ 2i\theta^0(\lambda_2)-2\pi\int_{\lambda_2}^{\lambda_M}d\eta\right)\,.
\end{array}
\label{eq:g4vanish2rerewrite}
\end{equation}
Here $\tilde{Y}_+(\eta)$ refers to the boundary value taken by
$\tilde{Y}(\eta)$ on $A_2$ from the interior of this closed contour.
Clearly the equation $\tilde{V}^\#_2=0$ has the same solutions as does
$\tilde{V}_2=0$, subject to the condition $\tilde{R}_2=0$.  But the function
$\tilde{V}^\#_2$ has the advantage of being single-valued in the real
polar plane of $(r,\theta)$.  Moreover, we may write $\tilde{V}^\#_2$ in
the form
\begin{equation}
\begin{array}{rcl}
\tilde{V}^\#_2&=&\displaystyle
 \Re\left(\int_{\lambda_M}^{\lambda_3}R(\eta)\tilde{Y}(\eta)\,d\eta
-\frac{1}{2\pi i}\log(\delta)\oint_{A_2}R(\eta)\tilde{Y}_+(\eta)\,d\eta\right)
+\frac{1}{2\pi}\log(r)\cdot\Im\left(\oint_{A_2}R(\eta)\tilde{Y}_+(\eta)\,d\eta\right)\\\\
&&\hspace{0.2 in}+\,\,\,\displaystyle
\Re\left(\int_{\lambda_2}^{\lambda_M}R(\eta)\tilde{Y}(\eta)\,d\eta 
+2i\theta^0(\lambda_2)-2\pi\int_{\lambda_2}^{\lambda_M}d\eta\right)\,.
\end{array}
\label{eq:V2tilde}
\end{equation}
Here we see that the first term of $\tilde{V}^\#_2$ is complexified as an
analytic function of $\delta^2$ and $\delta^{*2}$ individually.
Therefore it has an expression in the real polar plane of the form
$f(x,t)+O(r^2)$.  The complexification of the second term, while
single-valued when $\delta$ and $\delta^*$ are complex conjugates of
one another, is not single-valued in a full $\mathbb{C}^2$
neighborhood of $\delta=\delta^*=0$.  Therefore it cannot have a
two-variable Taylor expansion; however the factor multiplying
$\log(r)$ clearly is an even analytic function of both $\delta$ and
$\delta^*$ that vanishes when $\delta=\delta^*=0$ (this is of course
just the complexification of the harmonic conjugate $\tilde{R}_2$).
Consequently, in the real polar plane we have
\begin{equation}
\tilde{V}^\#_2 = \frac{\alpha(x,t)}{2\pi}r^2\log(r)\sin(2(\theta-\beta(x,t)))
+ \gamma(x,t) + O(r^2)\,,
\label{eq:curve}
\end{equation}
where $\gamma(x,t)$ is an analytic function of $x$ and $t$ that
vanishes to first order on the secondary caustic; it is the value of
$\tilde{V}^\#_2$ given by \eqref{eq:V2tilde} evaluated with all endpoints
in the degenerate configuration.  Equating this to zero, we see that
the balance must occur between the $r^2\log(r)$ term and the constant
term, so the the distance from the secondary caustic scales as
$r^2\log(r)$ as the new band is born.  The curve described by
\eqref{eq:curve} near the origin in the polar plane consists of two
approximately hyperbolic branches with asymptotes rotated from the
curves described by the equation $\tilde{R}_2=0$ by $\pi/4$.  Thus, we obtain
for each point $(x,t)$ sufficiently close to the secondary caustic a
pair of opposite solutions for $\delta=re^{i\theta}$.  As
$\gamma(x,t)$ changes sign, the two points come together at $\delta=0$
and then go back out along the perpendicular direction.

Determination of the sign of $\gamma(x,t)$ analogous to what is
carried out in \cite{KMM} is then needed to show that the points
$\lambda_3(x,t)$ and $\lambda_4(x,t)$ separate in the direction
conducive to the completion of a genus-four ansatz as long as $x$
decreases (or $t$ increases) as the secondary caustic is crossed.
Combining these steps completes the proof that upon crossing the
secondary caustic the asymptotic description of the $N$-soliton is
given in terms of Riemann theta functions of genus four.

We also note that the equations $\tilde{R}_2=0$ and $\tilde{V}_2=0$
imply that the new band $I_2$ and the boundary curves of the region
where the inequality $\Re(\phi(\lambda)-2i\theta^0(\lambda))<0$ holds
are trajectories of the quadratic differential
$R(\eta)^2\tilde{Y}_+(\eta)^2\,d\eta^2$, while the other two bands and
the boundaries of the region where the inequality
$\Re(\phi(\lambda))<0$ holds are trajectories of
$R(\eta)^2Y_-(\eta)^2\,d\eta^2$.

\subsubsection{Whitham equations.}
Recall from \S~\ref{sec:Whitham} that for arbitrary even genus $G$
(including $G=4$ as a special case), if the endpoints occur along
$K_0\cup K_{-1}$ or $K_0\cup K_{-1}\cup K_{-2}$, then the endpoints
implicitly defined as functions of $(x,t)$ via the relations
$E_1=\cdots=E_{2G+2}=0$ satisfy a universal set of quasilinear partial
differential equations, namely the Whitham equations
\eqref{eq:WhithamI} with complex characteristic velocities given as
functions of the endpoints only via \eqref{eq:WhithamIspeeds}.

We have just seen, however, that with
\begin{equation}
\begin{array}{rcl}
\tilde{E}_1 &:= &\tilde{V}_1\,,\\\\
\tilde{E}_2 &:=& \tilde{R}_1\,,\\\\
\tilde{E}_3 &:=&\tilde{V}_2\,,\\\\
\tilde{E}_4 &:=&\tilde{R}_2\,,\\\\
\tilde{E}_n &:=&\tilde{M}_{n-5}\,,\hspace{0.2 in}\text{for $n=5,6,7$}\,,\\\\\tilde{E}_8 &:=&\tilde{M}_3-2t\,,\\\\
\tilde{E}_9&:=&\displaystyle\tilde{M}_4-x-2t\sum_{n=0}^4a_n\,,\\\\
\tilde{E}_{10}&:=&\displaystyle\tilde{M}_5-x\sum_{n=0}^4a_n-2t\left(
\sum_{n=0}^4\left(a_n^2-\frac{1}{2}b_n^2\right)+\sum_{n=0}^4\sum_{m=n+1}^4
a_ma_n\right)\,,
\end{array}
\end{equation}
the conditions $\tilde{E}_1=\cdots=\tilde{E}_{10}=0$ determining the
endpoints $\lambda_0,\ldots,\lambda_4$ in the genus-four
configuration just beyond the secondary caustic have a different form
than the relations $E_1=\cdots=E_{10}=0$ (for $G=4$) from
\S~\ref{sec:Whitham}.  In particular, a configuration
$\lambda_0(x,t),\ldots, \lambda_5(x,t)$ satisfying
$E_1=\cdots=E_{10}=0$ does not satisfy the equations
$\tilde{E}_1=\cdots=\tilde{E}_{10}=0$.  Nonetheless, these latter
conditions also determine implicitly functions
$\lambda_0(x,t),\ldots,\lambda_5(x,t)$ satisfying exactly the same
Whitham equations.

\begin{theorem}
  Any continuously differentiable solution
  $\lambda_0(x,t),\ldots,\lambda_4(x,t)$ of
  $\tilde{E}_1=\cdots=\tilde{E}_{10}=0$ satisfies the Whitham
  equations \eqref{eq:WhithamI} with complex characteristic velocities
  given by \eqref{eq:WhithamIspeeds} in any region of the
  $(x,t)$-plane where the endpoints are distinct.
\end{theorem}

\begin{proof}
  We complexify the functions $\tilde{E}_n$, letting $\vec{v}$ denote
  the vector
  $(\lambda_0,\ldots,\lambda_4,\lambda_0^*,\ldots,\lambda_4^*)^T$ of
  endpoints and their complex conjugates, viewed as independent
  variables.  Then, direct calculations show that, by analogy with
  \eqref{eq:RYderiv},
\begin{equation}
\frac{\partial}{\partial v_k}(R(\eta)\tilde{Y}(\eta))=-\frac{1}{2}\tilde{Y}(v_k)
\frac{R(\eta)}{\eta-v_k}\,,
\end{equation}
and by analogy with \eqref{eq:YM0relation} we have
\begin{equation}
\tilde{Y}(v_k) = 4i\frac{\partial\tilde{M}_0}{\partial v_k}\,,
\label{eq:YtildeMtilde0relation}
\end{equation}
so that
\begin{equation}
\frac{\partial\tilde{E}_n}{\partial v_k} = -i\frac{\partial \tilde{M}_0}{\partial v_k}\left[\int_{\lambda_{n-1}}^{\lambda_n}\frac{R(\eta)\,d\eta}{\eta-v_k}
-\int_{\lambda_n^*}^{\lambda_{n-1}^*}\frac{R(\eta)\,d\eta}{\eta-v_k}\right]
=J_{nk}(\vec{v})\frac{\partial \tilde{M}_0}{\partial v_k}\,,
\label{eq:firsttildedeqnsderivs}
\end{equation}
for $n=1$, $n=2$, and $n=4$, where $J_{nk}(\vec{v})$ are exactly the
same matrix elements as defined for $G=4$ by
\eqref{eq:integralderivs}.  The corresponding calculation for partial
derivatives of $\tilde{E}_3 = \tilde{V}_2$ is slightly more
complicated due to the appearance of terms explicitly depending on
$\lambda_2$ and $\lambda_2^*$; however since
\begin{equation}
\frac{\partial}{\partial \lambda_2}\left[2i\theta^0(\lambda_2)-2\pi\int_{\lambda_2}^{\lambda_M}d\eta\right] = 0\,,
\end{equation}
in fact \eqref{eq:firsttildedeqnsderivs} holds for $n=3$ as well. Next, by
another direct calculation, we find that by analogy with
\eqref{eq:Momentderivsrelation},
\begin{equation}
\frac{\partial \tilde{E}_n}{\partial v_k} = \frac{1}{2}\tilde{E}_{n-1}+
v_k\frac{\partial\tilde{E}_{n-1}}{\partial v_k}\,,\hspace{0.2 in}
\text{for $6\le n\le 10$}
\end{equation}
so that on a solution we have
\begin{equation}
\frac{\partial\tilde{E}_n}{\partial v_k} = 
v_k^{n-1}\frac{\partial \tilde{E}_5}{\partial v_k}=
v_k^{n-1}\frac{\partial\tilde{M}_0}{\partial v_k} = 
J_{nk}(\vec{v})\frac{\partial \tilde{M}_0}{\partial v}\,,
\hspace{0.2 in}\text{for $5\le n\le 10$\,,}
\end{equation}
where again the matrix elements $J_{nk}(\vec{v})$ are precisely the
same as defined by \eqref{eq:Momentderivs}.  Therefore, the elements
$\partial\tilde{E}_n/\partial v_k$ are assembled into the Jacobian matrix
having the factorized form:
\begin{equation}
\frac{\partial\vec{\tilde{E}}}{\partial \vec{v}}={\bf J}(\vec{v})\cdot
\text{diag}\left(\frac{\partial\tilde{M}_0}{\partial v_1},\dots,
\frac{\partial\tilde{M}_0}{\partial v_{10}}\right)\,,
\end{equation}
where the factor ${\bf J}(\vec{v})$ is exactly the same function of
$\vec{v}$ as in the case studied in \S~\ref{sec:Whitham}.  Noting that
the partial derivatives $\partial \tilde{E}_n/\partial x$ and
$\partial\tilde{E}_n/\partial t$ are all identical expressions in terms of
$\vec{v}$ as in \S~\ref{sec:Whitham}, it then follows by application of
Cramer's rule that
\begin{equation}
\frac{\partial v_k}{\partial t} + c_k(\vec{v})\frac{\partial v_k}{\partial x} = 0\,,\hspace{0.2 in}\text{for $k=1,\ldots,10$\,,}
\end{equation}
where the characteristic speeds are exactly the same functions of
$\vec{v}$ (see \eqref{eq:WhithamIspeeds}) as in \S~\ref{sec:Whitham}.

To reiterate, the only difference between the derivation here and that
in the case when no endpoints lie on $K_{-3}$ is in the diagonal
factor of the Jacobian consisting of the partial derivatives $\partial
\tilde{M}_0/\partial v_k$, which certainly are different functions of
$\vec{v}$ than the derivatives $\partial M_0/\partial v_k$.  However,
this diagonal factor cancels out of the Cramer's rule formula for the
characteristic velocities $c_k(\vec{v})$.  Note that the condition
that the diagonal factor is nonsingular is, due to
\eqref{eq:YtildeMtilde0relation}, equivalent to the function
$R(\eta)\tilde{Y}(\eta)$ which defines the bands vanishing exactly
like a square root at the band endpoints.
\end{proof}

\section{Conclusion}
In this paper we have done the following things.
\begin{itemize}
\item We used a numerical linear algebra method for computing the
  $N$-soliton for several values of $N$ as large as $N=40$, plotted
  the results and described the phenomena of macrostructure,
  microstructure, and phase transitions (caustic curves) that become
  more well-defined as $N\rightarrow\infty$.
\item We have described a new modification of the asymptotic technique first
  used in \cite{KMM} to analyze the $N$-soliton for large $N$ that
  gives an improved error estimate of $O(N^{-1})$ over the cruder
  estimate of $O(N^{-1/3})$ reported in \cite{KMM}.
\item We analyzed the Fourier power spectrum of the $N$-soliton in the
  genus-zero region before the primary caustic and related the
  results to supercontinuum generation of coherent broadband spectra
  from narrow-band sources in optical fibers with weak anomalous
  dispersion.
\item We computed numerically for the first time the bands of
  discontinuity of $g'(\lambda)$ in the genus-two region beyond the
  primary caustic and pointed out the phenomenon of a band crossing
  the locus of accumulation of discrete eigenvalues for the
  $N$-soliton.
\item We introduced a new multi-interpolant asymptotic procedure that
  allows the genus-two ansatz for $g(\lambda)$ to be continued through
  the discrete eigenvalues, proving that this phenomenon is {\em not}
  the cause of the secondary caustic.
\item We observed numerically the failure of a {\em new} inequality associated
with the multi-interpolant approach that becomes necessary once a band
crosses the discrete spectrum, and we gave evidence that this failure is
the mathematical mechanism for the secondary caustic phase transition.
\item We have shown that, under the assumption that the new inequality
  indeed fails as we have observed numerically, the secondary caustic
  represents a change of genus from $G=2$ to $G=4$ but with new
  topological features of the genus-four ansatz for $g(\lambda)$ that have
  not been seen before.  We also proved that despite the unusual
  appearance of the implicit relations defining the band
  endpoints in this case, the endpoints still satisfy the Whitham
  equations.
\end{itemize}
The nature of the mathematical mechanism for the secondary caustic
leads us to hypothesize that there could easily be an infinite number
of caustic curves in the $(x,t)$-plane of the $N$-soliton in the large
$N$ limit.  The scenario we imagine is the following: as $t$ increases
for fixed $x>0$, the new band $I_2$ that we have shown opens up upon
crossing the secondary caustic drifts toward the imaginary interval
$[0,iA]$ of accumulation of discrete eigenvalues.  When $I_2$ meets
this interval, the analysis fails once again until it is rescued (in
analogy with the modification we introduced in \S~\ref{sec:dual}) by
adjoining a new contour lobe, $K_{-5}$, enclosing a region $D_{-5}$
adjacent to $D_{-3}$ in which the correct interpolation formula to use
is the analogue of \eqref{eq:Dm3interp} with the term
$-3i\theta^0(\lambda)$ in the exponent replaced by
$-5i\theta^0(\lambda)$.  Along $K_{-5}$ we require a new inequality,
namely $\Re(\phi(\lambda)-4i\theta^0(\lambda))<0$.  The region in
which this inequality holds eventually pinches off (a tertiary
caustic), and the situation is resolved by introducing a new band
$I_3$ on $K_{-5}$ which results in local dynamics described by Riemann
theta functions of genus six.  The reader can then imagine that the
process repeats again and again.
The $k$th caustic curve is described by eliminating $\hat{\lambda}$ from
the equations
\begin{equation}
\phi'(\hat{\lambda})-2i(k-1)\theta^{0\prime}(\hat{\lambda}) = 0\,,
\hspace{0.2 in}
\Re(\phi(\hat{\lambda})-2i(k-1)\theta^0(\hat{\lambda}))=0\,.
\end{equation}
The formula for $\phi(\lambda)$ of course changes each time a caustic
is crossed and a new band is added.  Thus, a harbinger of a developing
caustic curve is a recently born band meeting the imaginary axis.
This can be detected just by observing the wave field as long as (like
we see in the numerical constructions of $g(\lambda)$ in the genus-two
case) the endpoint crossing the imaginary axis has an imaginary part
much larger than those of all other endpoints.  In this case, the wave
field, although locally described by Riemann theta functions of some
even genus $G>0$, is well-approximated by a complex exponential
function $e^{i(kx-\omega t)/\hbar}$ (because the genus is ``almost''
zero), and the real part of the dominant endpoint is proportional to
the wavenumber $k$.  Thus a sign change of $k$, which could be
detected by local spectral analysis, indicates the dominant endpoint
crossing the imaginary axis, an event that precedes the formation of a
new caustic.

We want to stress that the mechanism for all of the caustics beyond
the primary caustic depends in a crucial way on the discrete nature of
the spectrum for the $N$-soliton, which causes us to reinvent our
method of analysis each time a band wants to cross the discrete
spectrum.  Significantly, if we chose to begin our analysis by
assuming that there is no essential cost in the limit
$\hbar\rightarrow 0$ for analyzing the formal continuum-limit problem
for $\tilde{\bf P}(\lambda)$ (in which we simply neglect the
difference between the discrete spectral measure and its weak limit as
$N\rightarrow\infty$) in place of the exact Riemann-Hilbert problem
for ${\bf P}(\lambda)$, we would not need to modify our analysis if a
band crosses the spectrum, and we would not need to introduce the new
inequality $\Re(\phi(\lambda)-2i\theta^0(\lambda))<0$ that leads to
the secondary caustic.  In other words, the formal continuum-limit
problem corresponds to a solution $\tilde{\psi}_N(x,t)$ of the
focusing nonlinear Schr\"odinger equation
\eqref{eq:semiclassical-fnls} that accurately resembles the
$N$-soliton $\psi_N(x,t)$ for $(x,t)$ in the region before the
secondary caustic, but not beyond.  In particular, the solution
obtained from the formal continuum-limit problem does not experience
any phase transition at all when the $N$-soliton experiences the
secondary caustic.  Of course, as the formal continuum-limit problem
may break the even symmmetry of the $N$-soliton (recall
\S~\ref{sec:aside}) we have no solid grounds for expecting any
similarity between $\psi_N(x,t)$ and $\tilde{\psi}_N(x,t)$ for any $x$
and $t$, and therefore it seems to us that we must therefore regard
the correspondence before the secondary caustic as a lucky
coincidence.

\section{Acknowledgements}
Both authors would like to thank Jinho Baik, Tony Bloch, Jeff
DiFranco, and Guada Lozano for contributing to this work through
discussions in meetings of a weekly Working Group in Integrable
Systems and Asymptotics held at the University of Michigan; these
meetings provided a great forum for our thought process.  The second
author gratefully acknowledges the support of the National Science
Foundation under grants DMS-0103909 and DMS-0354373.

 \end{document}

%% file: section1.tex
It is well known (see, {\em e.g.}, \cite{SY}) that the solution of the
initial-value problem for the focusing nonlinear Schr\"odinger
equation
\begin{equation}
i\psi_t+\frac{1}{2}\psi_{xx}+|\psi|^2\psi =0 \label{eq:fnls}
\end{equation}
with initial data
\begin{equation}
\psi(x,0)=N\,\text{sech}(x)\label{eq:N-initial-data}
\end{equation}
is a special solution called the $N$-soliton. This solution is rapidly
decreasing in $|x|$ and periodic in $t$ with period independent of
$N$. To further describe this solution, we recall that \eqref{eq:fnls}
is exactly solvable via the inverse-scattering framework introduced by
Zakharov and Shabat in \cite{ZS}. There are three steps in the
procedure:
\begin{enumerate}
\item[(i)] forward scattering --- the initial data generates the
  scattering data which consists of
eigenvalues, proportionality
constants, and a reflection coefficient; 
\item[(ii)] time evolution --- the scattering data have a simple
evolution in time; 
\item[(iii)] inverse scattering --- the solution of the partial differential
equation at later times is reconstructed from the time-evolved scattering
data. 
\end{enumerate}
For the special initial data \eqref{eq:N-initial-data}, Satsuma and
Yajima~\cite{SY} have shown that the reflection coefficient is
identically zero and there are $N$ purely imaginary eigenvalues. 
Such a reflectionless solution whose eigenvalues have a common real
part is an $N$-soliton. 

Here, using the the inverse-scattering framework for 
\eqref{eq:fnls}--\eqref{eq:N-initial-data}, 
we study the $N$-soliton in the limit $N\to\infty$.
For the first few positive integer values of $N$,
it is possible to write down explicit formulas for these exact
solutions of the nonlinear partial differential equation
\eqref{eq:fnls}. 
However, these formulae rapidly become unwieldy. 
Even for $N=3$, the formula is already
quite complicated and hard to analyze. Amazingly, while these formulae
grow increasingly complicated as $N$ increases, it is also true that
certain orderly features emerge in the limit $N\to\infty$.

As a first step, we rescale 
\(\psi\) and \(t\) in 
\eqref{eq:fnls}--\eqref{eq:N-initial-data} to make the initial
amplitude independent of $N$ and the period proportional to $N$, and 
we arrive at 
%
\begin{align}
&i\hbar\psi_t+\frac{\hbar^2}{2}\psi_{xx}
+|\psi|^2\psi =0, \label{eq:semiclassical-fnls}\\
&\psi(x,0)=A\,\text{sech}(x),\label{eq:initial-data}
\end{align}
where \(\hbar=\hbar_N:=A/N\) and \(A>0\). Studying the large-$N$ limit
of the $N$-soliton is thus equivalent to studying the semiclassical
(\(\hbar\downarrow 0\)) limit of
\eqref{eq:semiclassical-fnls}--\eqref{eq:initial-data}. A feature that
emerges in the limit is the sharpening boundaries in the $(x,t)$-plane
that separate different behaviors of the solution. Two such ``phase
transitions'' were noticed in the numerical experiments of Miller and
Kamvissis \cite{MK}, and the first, a so-called {\em primary caustic}
curve in the $(x,t)$-plane, was rigorously explained by Kamvissis,
McLaughlin, and Miller \cite{KMM}.  Here, our main result is a
description of the mechanism for the observed second phase transition;
it differs from the first one.

We use two formulations of the inverse-scattering step to study the
limiting behavior of \eqref{eq:semiclassical-fnls}--\eqref{eq:initial-data}.
When, as is the case here, 
the reflection coefficient is absent, the algebraic-integral system
that one expects to solve to reconstruct the solution of
\eqref{eq:semiclassical-fnls} for generic initial data  
reduces to a linear algebraic system. We derive such a system in
\S~\ref{sec:linearsystem} below. We also show, in \S~\ref{sec:drhp}, 
that the reconstruction
step can be cast as a discrete Riemann-Hilbert problem for a
meromorphic $2\times 2$ matrix unknown. We use the linear algebraic
system as a basis for numerically computing the $N$-soliton while
the Riemann-Hilbert formulation provides the starting point for our
asymptotic analysis. In either case, the reconstruction of the
solution begins with the eigenvalues
\begin{equation}
\lambda_{N,k}:=iA-i\left(k+\frac{1}{2}\right)\hbar_N,\quad
k=0,\ldots,N-1,
\label{eq:evals}
\end{equation}
and the proportionality constants
\begin{equation}
\gamma_{N,k}:=(-1)^{k+1},\quad k=0,\ldots,N-1.
\label{eq:proportionalityconstants}
\end{equation}
%

In \S~\ref{sec:asymptoticanalysis} we describe how to explicitly
modify the discrete Riemann-Hilbert problem we obtain in
\S~\ref{sec:drhp} to arrive at an equivalent problem that is conducive
to rigorous asymptotic analysis in the limit $N\rightarrow\infty$ (or
equivalently $\hbar\rightarrow 0$).  In \S~\ref{sec:nopoles} we
convert the discrete Riemann-Hilbert problem into a conventional
Riemann-Hilbert problem for an unknown with specified jump
discontinuities across contours in the complex plane.  Next, we show
how to ``precondition'' the resulting Riemann-Hilbert problem for the
limit $N\rightarrow\infty$ by introducing a scalar function
$g(\lambda)$ that is used to capture the most violent asymptotic
behavior so that the residual may be analyzed rigorously.  In some
cases, the function $g(\lambda)$ leads to an asymptotic analysis in
the limit $N\rightarrow\infty$ based on the theta functions of
hyperelliptic Riemann surfaces of even genus $G$, and in such cases
the function $g(\lambda)$ can be constructed explicitly, as we show in
\S~\ref{sec:choiceofg}.  This construction forces certain dynamics in
$(x,t)$ on the moduli (branch points) of this Riemann surface, and in
\S~\ref{sec:Whitham} we show that the moduli satisfy a universal
system of quasilinear partial differential equations in
Riemann-invariant form, the Whitham equations.  The success or failure
of the function $g(\lambda)$ constructed in this way in capturing the
essential dynamics as $N\rightarrow\infty$ hinges upon certain
inequalities and topological conditions on level curves that are
described in \S~\ref{sec:inequalities}.  With a function $g(\lambda)$
in hand that satisfies all of these conditions, we may proceed with
the asymptotic analysis, which is based on a steepest descent
technique for matrix Riemann-Hilbert problems developed by Deift and
Zhou \cite{DZ}.  We summarize these steps in
\S~\ref{sec:steepestdescent}, \S~\ref{sec:parametrix}, and
\S~\ref{sec:error}.

The theory outlined above provides, through a handful of technical
modifications, a refinement of results previously obtained by
Kamvissis, McLaughlin, and Miller \cite{KMM}.  The key techincal
improvements are the avoidance, through a dual interpolant approach
developed in \cite{MIMRN} and \cite{DOP}, of a local parametrix near
the origin in the complex plane and the explicit and careful tracking
of the errors in replacing a distribution of point masses representing
condensing soliton eigenvalues by its weak continuum limit, encoded in
certain functions $S(\lambda)$ and $T(\lambda)$.  These extra steps
allow us to deduce the same asymptotic formulae obtained in \cite{KMM}
governing the semiclassical limit up to and just beyond the primary
caustic curve in the $(x,t)$-plane, but with an improved error
estimate that is $O(\hbar)$ (which beats the $O(\hbar^{1/3})$ estimate
in \cite{KMM}).

It is fair to say that much of the work in analyzing the $N$-soliton
in the limit of large $N$ comes about from vanquishing the poles
representing the soliton eigenvalues from the matrix-valued
Riemann-Hilbert problem that represents the inverse-scattering step.
In some sense, the poles disappear with the interpolation step that
converts the discrete Riemann-Hilbert problem with poles but no jump
discontinuities into one with jump discontinuities and no poles.  On
the other hand, this is only a partial solution, since the jump matrix
relating the boundary values taken along a curve of jump discontinuity
extends from this curve as an analytic function with poles at the
soliton eigenvalues.  However, {\em as long as the jump contour
  remains bounded away from these poles}, it is a reasonable
approximation that can be controlled rigorously to replace the jump
matrix by another one in which the distribution of poles is
``condensed'' into a continuous distribution with an analytic density.
Making this replacement on an {\em ad hoc} basis changes the problem.
We refer to this changed inverse-scattering problem as the {\em formal
  continuum-limit problem} and we discuss it briefly in
\S~\ref{sec:formallimit}.  The formal continuum-limit problem indeed
gives the correct asymptotics of the $N$-soliton and related initial
data for the semiclassical focusing nonlinear Schr\"odinger equation
\eqref{eq:semiclassical-fnls} subject to the above caveat, and it even
forms the starting point for the analysis of \cite{TVZ}.

Interestingly, the choice of the function $g(\lambda)$ turns out to
have the following two properties:
\begin{itemize}
\item It depends only on the analytic weak limit of the discrete distribution
of soliton eigenvalues.
\item It determines the contours of the Riemann-Hilbert problem of
inverse scattering, and in particular their relation to the locus
of accumulation of the discrete soliton eigenvalues.
\end{itemize}
So, from the first property, $g(\lambda)$ does not know about the poles
of the jump matrix, and from the second property it chooses the contours
appropriate for the asymptotic analysis.  It is therefore possible that
the contours selected by choice of $g(\lambda)$ could cross the locus of
accumulation of discrete eigenvalues.  And once the contours are no longer
bounded away from the poles, the formal continuum-limit problem is not
the correct model for inverse scattering.  

It turns out that, in the region beyond the primary caustic curve,
this actually happens.  In other words, the dependence of the function
$g(\lambda)$ on $(x,t)$ predicts a contour that as $t$ increases
passes through the branch cut that is the continuum limit of the pole
distribution.  At this point, the rigorous analysis must break down.
Significantly, however, this is {\em not} the mechanism for further
phase transitions.  It turns out that the difficulty is a technical
one that can be removed with the further use of the
multi-interpolation method developed in \cite{MIMRN} and \cite{DOP}.
The additional steps that are required to surmount this crisis, and to
therefore prove that there is no phase transition when curves
determined from the function $g(\lambda)$ cross the pole locus, are
described in \S~\ref{sec:dual}.

The analysis described in \S~\ref{sec:dual} is new, and one of the new
features is the appearance of a {\em new inequality} that must be
satisfied by the function $g(\lambda)$ on certain contours.  It turns
out that the failure of this inequality is the mathematical mechanism
for the next phase transition of the $N$-soliton, a {\em secondary
  caustic} curve in the $(x,t)$-plane.

Before studying the secondary caustic, we discuss briefly in
\S~\ref{sec:genus0} the quiescent region of the $(x,t)$-plane before
the primary caustic curve.  We recall in \S~\ref{sec:validityofG0}
some information from \cite{KMM} about the relation between the
dynamics of the $N$-soliton in this region and a family of
hyperelliptic Riemann surfaces of genus zero.  We also discuss the
mathematical mechanism behind the primary caustic phase transition,
which turns out to correspond to an instantaneous jump from genus zero
to genus two.  Then, in \S~\ref{sec:Fourier} we analyze the Fourier
power spectrum of the $N$-soliton in the region before the primary caustic,
and show that its evolution is consistent with the {\em supercontinuum
generation} phenomenon of current interest in optical science (see below).

In carrying out a study of the $N$-soliton for large $N$, three different
computational techniques come to mind:
\begin{enumerate}
\item[(i)] Direct numerical simulation of the initial-value problem for the
  focusing nonlinear Schr\"odinger equation
  \eqref{eq:semiclassical-fnls} in the semiclassical limit.  By
  adapting discretization methods to this problem, it is possible to
  study the dynamics of the $N$-soliton as well as much more general
  initial data, and also non-integrable variants of
  \eqref{eq:semiclassical-fnls}.  However, these methods are severely
  constrained in the limit of interest due to numerical stiffness, and
  worse yet, modulational instability that is exponentially strong in
  $N$.  See \cite{BK,CM,Cen,CT}.
\item[(ii)] Numerical solution of the inverse-scattering problem for the
  focusing nonlinear Schr\"odinger equation.  This method applies only
  to the integrable equation \eqref{eq:semiclassical-fnls}, and like
  the direct numerical simulation method is also ill-conditioned in
  the semiclassical limit.  However, it affords an important advantage
  over direct numerical simulation, namely that the solution is
  calculated independently for each $(x,t)$, and therefore roundoff
  errors do not accumulate.
\item[(iii)] Numerical construction of the function $g(\lambda)$.  This
  method also applies only to the integrable equation
  \eqref{eq:semiclassical-fnls}, and it is further constrained in that
  it is only meaningful in the semiclassical limit.  However, it 
  specifically takes advantage of the mathematical structure of this
  limit, and consequently the numerical calculation is extremely
  well-conditioned.
\end{enumerate}  
In \S~\ref{sec:GenusTwoNumerics}, we take the third approach and use
numerical methods to solve for the function $g(\lambda)$ in the region
between the primary and secondary caustics for the $N$-soliton.  While
a numerical calculation, this is clearly fundamentally different from
either the numerical solution of linear algebra problems equivalent to
inverse scattering for the $N$-soliton as in
\S~\ref{sec:linearsystem}, or direct numerical simulation of the
initial-value problem for \eqref{eq:semiclassical-fnls} corresponding
to the $N$-soliton.  It is with the help of these calculations that we
observe the crossing of the pole locus (which does not correspond to
the secondary caustic, as can be seen once the analysis is modified as
in \S~\ref{sec:dual}) and ultimately the violation of the new
inequality introduced in \S~\ref{sec:dual} (which causes the secondary
caustic phase transition).

In \S~\ref{sec:highergenus} we take up the question of exactly what
happens to the $N$-soliton immediately beyond the secondary caustic
curve in the $(x,t)$-plane.  We characterize the secondary caustic
(failure of the new inequality from \S~\ref{sec:dual}) as the
occurance of a critical point on a level curve defining the boundary
of the region in the complex plane where the relevant inequality
holds.  We then make the guess that the failure of $g(\lambda)$
corresponding to a Riemann surface of genus two in this new way is
resolved by going over to a formula for $g(\lambda)$ corresponding to
a Riemann surface of genus four, but with significantly different
topological features than in the (hypothetical) case that genus two
fails due to the violation of the same inequality that leads to the
primary caustic.  We construct $g(\lambda)$ in this new situation, and
obtain an implicit description of the moduli of the corresponding
genus four Riemann surface.  These are new formulae that have a
different character than those obtained earlier.  We then prove that
these new formula provide another solution to the same type of Whitham
equations that govern the moduli in the case of simpler contour
topology.

We conclude the paper with some hypotheses regarding further phase
transitions (higher-order caustics) for the $N$-soliton and related
initial-value problems for \eqref{eq:semiclassical-fnls}, and by
stressing once again that the phenomenon of the secondary caustic (and
possibly further phase transitions) is fundamentally linked to the
discrete nature of the eigenvalue distribution, which forces
modifications to the analysis as described in \S~\ref{sec:dual}, and
introduces coincident inequalities to be satisfied by $g(\lambda)$.
These inequalities are simply not part of the asymptotic theory of the
formal continuum-limit problem discussed in \S~\ref{sec:formallimit},
and therefore the latter problem is unable to correctly predict the
secondary caustic.  This came as some surprise to us, as we and other
authors had always assumed that either the formal continuum-limit problem
governs the dynamics for all time or that the secondary caustic occurs
upon crossing the locus of eigenvalues.  Both of these are incorrect.

The problem \eqref{eq:semiclassical-fnls} for small $\hbar$ is
relevant as a model for the propagation of light pulses in optical
fibers that have the property of weak (because $\hbar$ is small)
anomalous dispersion.  There has always been great interest in optical
fibers with weak dispersion. Initially this was because when such
fibers are operated in the linear regime (small amplitude) the
propagation is immune to dispersive spreading of pulses that is
considered to degrade a data stream.  For long fiber links, however,
the theory of linear propagation becomes inadequate due to the
accumulation of weakly nonlinear effects.  More recently these fibers
have become important again because it became clear that advantage
could be taken specifically of the nonlinearity.  Indeed, when
operated in the weakly nonlinear regime (moderate amplitude) nonlinear
effects are much stronger compared to linear effects and therefore
propagation in such fibers can drastically alter the power spectrum of
a signal, possibly in a useful way.  Indeed, one of the applications
envisioned for weakly dispersive fibers is so-called {\em
  supercontinuum generation} (see, for example, \cite{Crist,Dudley})
in which a nearly monochromatic laser source is coupled into the fiber
and spectrally broadened under propagation so that the output is a
coherent source of white light.  The output can then be filtered to
produce coherent light of virtually any frequency, which is desirable
for wavelength division multiplexing telecommunication systems.  The
most promising current technology for creating optical fibers with
very low dispersion, in both the normal and anomalous regimes, is
based on {\em photonic crystal fibers}, which are made from fiber
preforms with extremely complicated cross sections.  The complexity of
the cross section is preserved to the microscopic level upon drawing
the fiber (carefully stretching the cylindrical fiber preform along
its axis until an optical fiber results), and as the possibilities for
cross sections go far beyond the traditional core/cladding/jacket
step-index model it is possible to engineer fibers with properties
thought impossible until recently \cite{Ferrando}.

In this paper, we make frequent use of the three Pauli matrices:
\begin{equation}
\sigma_1:=\left[\begin{array}{cc} 0 & 1\\ 1 & 0\end{array}\right]\,,
\hspace{0.2 in}
\sigma_2:=\left[\begin{array}{cc} 0 & -i \\ i & 0\end{array}\right]\,,
\hspace{0.2 in}
\sigma_3:=\left[\begin{array}{cc} 1 & 0 \\  0 & -1\end{array}\right]\,.
\end{equation}
We use boldface notation ({\em e.g.} ${\bf m}$ or ${\bf P}$)
throughout for square matrices, with the exception of the identity
matrix $\mathbb{I}$, and arrow notation for column vectors ({\em e.g.}
$\vec{v}$, with row-vector transpose $\vec{v}^T$).  Complex
conjugation is indicated with an asterisk, and ${\bf P}^*$ means the
matrix whose elements are the complex conjugates of the corresponding
elements of ${\bf P}$ (no transpose).

%% file: section2-2.tex
In general, numerical integration of the initial-value problem
\eqref{eq:semiclassical-fnls}--\eqref{eq:initial-data} for the
semiclassically scaled focusing nonlinear Schr\"odinger equation is
dificult; the problem is notoriously stiff due to multiple scales.
The presence of small oscillations (microstructure) of wavelength and
period of order $\hbar$ requires that one use a timestep proportional
to \(\hbar\) while resolving the larger scale structures
(macrostructure) necessitates that the number of timesteps must be of
order \(\hbar^{-1}\).  In addition, to accurately compute the spatial
microstructure, the number of gridpoints/Fourier modes must also be
proportional to \(\hbar^{-1}\). Thus, simulating the semiclassical
limit is computationally intensive, and the accumulation of roundoff
errors is a serious issue. In spite of these difficulties, some
careful numerical experiments have been carried out
\cite{BK,CM,Cen,CT}.

For the special case of initial data \(\psi(x,0)=A\,\text{sech}(x)\),
the calculation described in the previous section provides a useful
alternative approach. Rather than using some numerical integration
scheme to directly compute an approximation to the solution of the
partial differential equation, we may take advantage of the fact that
the $N$-soliton can be recovered from the solution of the linear
algebraic system~\eqref{system}. That is, for each fixed pair
\((x,t)\), we numerically solve \eqref{system}. An approach like this
was first used by Miller and Kamvissis~\cite{MK} using instead the
larger system of equations \eqref{neweq1} and \eqref{neweq2}.

In addition to being limited to particular initial data, the approach
of solving \eqref{system} suffers from the fact that the matrix
\(\mathbb{I}+\mathbf{W}_N\) is ill-conditioned for large $N$ and
high-precision arithmetic is necessary for accurate computations.  As
the derivation leading to \eqref{system} involves Lagrange
interpolation on equi-spaced points, this is perhaps not surprising.
On the other hand, an advantage of this approach is that the
calculation for each \((x,t)\) is independent of all the other
calculations, so errors do not propagate in time and accumulate.
Figures~\ref{fig:5soliton}--\ref{fig:40soliton} show density plots of
\(|\psi_N(x,t)|^2\) for \(N=5,10,20,\) and \(40\) computed by solving
\eqref{system} independently for a large number of $(x,t)$ values
using high-precision arithmetic.

%% file: section2-3.tex
In each of Figures~\ref{fig:5soliton}--\ref{fig:40soliton}, the
plotted solutions share some common features.
After an initial period of smoothness, the solution
changes over to a form with a central oscillatory region and
quiescent tails. What is particularly striking in this sequence of figures is how
the boundary between these two behaviors of the solution appears to
become increasingly sharp as $N$ increases. This initial boundary curve is
called the primary caustic. 
The solution up to and just beyond the primary caustic has been
carefully studied in \cite{KMM}. See also the discussion in
\S~\ref{sec:genus0} below.

In Figure~\ref{fig:20soliton} and Figure~\ref{fig:40soliton}, a second
transition of solution behavior is clearly visible. It is this
\emph{secondary caustic curve} which is the main focus of this paper. 
As our subsequent analysis makes clear, this second phase transition
is linked to the presence of the discrete soliton eigenvalues
\(\{\lambda_{N,k}\}\cup\{\lambda_{N,k}^*\}\), and the mechanism for
this transition is different that of the transition across the primary caustic.

We note that Tovbis,
Venakides, and Zhou \cite{TVZ} have studied the semiclassically scaled
focusing nonlinear Schr\"odinger equation for a one-parameter family of special
initial data. The forward-scattering procedure for this family \cite{TV}
generates a reflection coefficent and --- depending on the value of the
parameter --- some discrete soliton eigenvalues. (By contrast, recall
that our
initial data is reflectionless.) For those values of the parameter for which
there are no discrete eigenvalues, they prove that the
solution undergoes only a single phase transition and that there
is no secondary caustic. When the value of the parameter is such that
there are both solitons and reflection, they leave the possibility
of a second phase transition as an open question.

%% file: InitialConfiguration.eps_t
\begin{picture}(0,0)%
\includegraphics{InitialConfiguration.eps}%
\end{picture}%
\setlength{\unitlength}{3947sp}%
\begingroup\makeatletter\ifx\SetFigFont\undefined%
\gdef\SetFigFont#1#2#3#4#5{%
  \reset@font\fontsize{#1}{#2pt}%
  \fontfamily{#3}\fontseries{#4}\fontshape{#5}%
  \selectfont}%
\fi\endgroup%
\begin{picture}(2718,3150)(1237,-2969)
\put(2701,-2911){\makebox(0,0)[lb]{\smash{{\SetFigFont{12}{14.4}{\rmdefault}{\mddefault}{\updefault}{\color[rgb]{0,0,0}$\epsilon$}%
}}}}
\put(1951,-2911){\makebox(0,0)[lb]{\smash{{\SetFigFont{12}{14.4}{\rmdefault}{\mddefault}{\updefault}{\color[rgb]{0,0,0}$-\epsilon$}%
}}}}
\put(1378,-865){\makebox(0,0)[lb]{\smash{{\SetFigFont{12}{14.4}{\rmdefault}{\mddefault}{\updefault}{\color[rgb]{0,0,0}$K_{-1}$}%
}}}}
\put(2509,-916){\makebox(0,0)[lb]{\smash{{\SetFigFont{12}{14.4}{\rmdefault}{\mddefault}{\updefault}{\color[rgb]{0,0,0}$D_{-1}$}%
}}}}
\put(2965,-2272){\makebox(0,0)[lb]{\smash{{\SetFigFont{12}{14.4}{\rmdefault}{\mddefault}{\updefault}{\color[rgb]{0,0,0}$D_1$}%
}}}}
\put(3151,-1696){\makebox(0,0)[lb]{\smash{{\SetFigFont{12}{14.4}{\rmdefault}{\mddefault}{\updefault}{\color[rgb]{0,0,0}$\lambda_0$}%
}}}}
\put(3448,-2116){\makebox(0,0)[lb]{\smash{{\SetFigFont{12}{14.4}{\rmdefault}{\mddefault}{\updefault}{\color[rgb]{0,0,0}$K_1$}%
}}}}
\put(2545,-2272){\makebox(0,0)[lb]{\smash{{\SetFigFont{12}{14.4}{\rmdefault}{\mddefault}{\updefault}{\color[rgb]{0,0,0}$K_0$}%
}}}}
\end{picture}%

%% file: Configuration2.eps_t
\begin{picture}(0,0)%
\includegraphics{Configuration2.eps}%
\end{picture}%
\setlength{\unitlength}{3947sp}%
\begingroup\makeatletter\ifx\SetFigFont\undefined%
\gdef\SetFigFont#1#2#3#4#5{%
  \reset@font\fontsize{#1}{#2pt}%
  \fontfamily{#3}\fontseries{#4}\fontshape{#5}%
  \selectfont}%
\fi\endgroup%
\begin{picture}(2718,6404)(1237,-5703)
\put(1378,-865){\makebox(0,0)[lb]{\smash{{\SetFigFont{12}{14.4}{\rmdefault}{\mddefault}{\updefault}{\color[rgb]{0,0,0}$K_{-1}$}%
}}}}
\put(3151,-1696){\makebox(0,0)[lb]{\smash{{\SetFigFont{12}{14.4}{\rmdefault}{\mddefault}{\updefault}{\color[rgb]{0,0,0}$\lambda_0$}%
}}}}
\put(3448,-2116){\makebox(0,0)[lb]{\smash{{\SetFigFont{12}{14.4}{\rmdefault}{\mddefault}{\updefault}{\color[rgb]{0,0,0}$K_1$}%
}}}}
\put(2545,-2272){\makebox(0,0)[lb]{\smash{{\SetFigFont{12}{14.4}{\rmdefault}{\mddefault}{\updefault}{\color[rgb]{0,0,0}$K_0$}%
}}}}
\put(1840,421){\makebox(0,0)[lb]{\smash{{\SetFigFont{12}{14.4}{\rmdefault}{\mddefault}{\updefault}{\color[rgb]{0,0,0}$C_\infty$}%
}}}}
\put(1942,-1340){\makebox(0,0)[lb]{\smash{{\SetFigFont{12}{14.4}{\rmdefault}{\mddefault}{\updefault}{\color[rgb]{0,0,0}$C$}%
}}}}
\put(1849,-2355){\makebox(0,0)[lb]{\smash{{\SetFigFont{12}{14.4}{\rmdefault}{\mddefault}{\updefault}{\color[rgb]{0,0,0}$K_L$}%
}}}}
\end{picture}%

%% file: Configuration3.eps_t
\begin{picture}(0,0)%
\includegraphics{Configuration3.eps}%
\end{picture}%
\setlength{\unitlength}{3947sp}%
\begingroup\makeatletter\ifx\SetFigFont\undefined%
\gdef\SetFigFont#1#2#3#4#5{%
  \reset@font\fontsize{#1}{#2pt}%
  \fontfamily{#3}\fontseries{#4}\fontshape{#5}%
  \selectfont}%
\fi\endgroup%
\begin{picture}(3564,6404)(391,-5703)
\put(3151,-1696){\makebox(0,0)[lb]{\smash{{\SetFigFont{12}{14.4}{\rmdefault}{\mddefault}{\updefault}{\color[rgb]{0,0,0}$\lambda_0$}%
}}}}
\put(3448,-2116){\makebox(0,0)[lb]{\smash{{\SetFigFont{12}{14.4}{\rmdefault}{\mddefault}{\updefault}{\color[rgb]{0,0,0}$K_1$}%
}}}}
\put(1840,421){\makebox(0,0)[lb]{\smash{{\SetFigFont{12}{14.4}{\rmdefault}{\mddefault}{\updefault}{\color[rgb]{0,0,0}$C_\infty$}%
}}}}
\put(1942,-1340){\makebox(0,0)[lb]{\smash{{\SetFigFont{12}{14.4}{\rmdefault}{\mddefault}{\updefault}{\color[rgb]{0,0,0}$C$}%
}}}}
\put(1849,-2355){\makebox(0,0)[lb]{\smash{{\SetFigFont{12}{14.4}{\rmdefault}{\mddefault}{\updefault}{\color[rgb]{0,0,0}$K_L$}%
}}}}
\put(1351,-961){\makebox(0,0)[lb]{\smash{{\SetFigFont{12}{14.4}{\rmdefault}{\mddefault}{\updefault}{\color[rgb]{0,0,0}$I_1$}%
}}}}
\put(1726,-361){\makebox(0,0)[lb]{\smash{{\SetFigFont{12}{14.4}{\rmdefault}{\mddefault}{\updefault}{\color[rgb]{0,0,0}$\lambda_1$}%
}}}}
\put(1351,-1861){\makebox(0,0)[lb]{\smash{{\SetFigFont{12}{14.4}{\rmdefault}{\mddefault}{\updefault}{\color[rgb]{0,0,0}$\lambda_2$}%
}}}}
\put(391,-1306){\makebox(0,0)[lb]{\smash{{\SetFigFont{12}{14.4}{\rmdefault}{\mddefault}{\updefault}{\color[rgb]{0,0,0}$\partial\Omega_1^-$}%
}}}}
\put(781,-871){\makebox(0,0)[lb]{\smash{{\SetFigFont{12}{14.4}{\rmdefault}{\mddefault}{\updefault}{\color[rgb]{0,0,0}$\Omega_1^-$}%
}}}}
\put(1681,-1681){\makebox(0,0)[lb]{\smash{{\SetFigFont{12}{14.4}{\rmdefault}{\mddefault}{\updefault}{\color[rgb]{0,0,0}$\partial\Omega_1^+$}%
}}}}
\put(2605,-2317){\makebox(0,0)[lb]{\smash{{\SetFigFont{12}{14.4}{\rmdefault}{\mddefault}{\updefault}{\color[rgb]{0,0,0}$I_0$}%
}}}}
\put(3331,-661){\makebox(0,0)[lb]{\smash{{\SetFigFont{12}{14.4}{\rmdefault}{\mddefault}{\updefault}{\color[rgb]{0,0,0}$\Gamma_1$}%
}}}}
\put(1441,-1246){\makebox(0,0)[lb]{\smash{{\SetFigFont{12}{14.4}{\rmdefault}{\mddefault}{\updefault}{\color[rgb]{0,0,0}$\Omega_1^+$}%
}}}}
\end{picture}%

%% file: NewModifiedConfiguration.eps_t
\begin{picture}(0,0)%
\includegraphics{NewModifiedConfiguration.eps}%
\end{picture}%
\setlength{\unitlength}{3947sp}%
\begingroup\makeatletter\ifx\SetFigFont\undefined%
\gdef\SetFigFont#1#2#3#4#5{%
  \reset@font\fontsize{#1}{#2pt}%
  \fontfamily{#3}\fontseries{#4}\fontshape{#5}%
  \selectfont}%
\fi\endgroup%
\begin{picture}(2673,3185)(1321,-2969)
\put(2701,-2911){\makebox(0,0)[lb]{\smash{{\SetFigFont{12}{14.4}{\rmdefault}{\mddefault}{\updefault}{\color[rgb]{0,0,0}$\epsilon$}%
}}}}
\put(1951,-2911){\makebox(0,0)[lb]{\smash{{\SetFigFont{12}{14.4}{\rmdefault}{\mddefault}{\updefault}{\color[rgb]{0,0,0}$-\epsilon$}%
}}}}
\put(3448,-2116){\makebox(0,0)[lb]{\smash{{\SetFigFont{12}{14.4}{\rmdefault}{\mddefault}{\updefault}{\color[rgb]{0,0,0}$K_1$}%
}}}}
\put(2956,-2296){\makebox(0,0)[lb]{\smash{{\SetFigFont{12}{14.4}{\rmdefault}{\mddefault}{\updefault}{\color[rgb]{0,0,0}$D_1$}%
}}}}
\put(3241,-1111){\makebox(0,0)[lb]{\smash{{\SetFigFont{12}{14.4}{\rmdefault}{\mddefault}{\updefault}{\color[rgb]{0,0,0}$K_{-1}$}%
}}}}
\put(2026,-1276){\makebox(0,0)[lb]{\smash{{\SetFigFont{12}{14.4}{\rmdefault}{\mddefault}{\updefault}{\color[rgb]{0,0,0}$K_{-2}$}%
}}}}
\put(1546, 14){\makebox(0,0)[lb]{\smash{{\SetFigFont{12}{14.4}{\rmdefault}{\mddefault}{\updefault}{\color[rgb]{0,0,0}$K_{-3}$}%
}}}}
\put(1801,-691){\makebox(0,0)[lb]{\smash{{\SetFigFont{12}{14.4}{\rmdefault}{\mddefault}{\updefault}{\color[rgb]{0,0,0}$D_{-3}$}%
}}}}
\put(1321,-2266){\makebox(0,0)[lb]{\smash{{\SetFigFont{12}{14.4}{\rmdefault}{\mddefault}{\updefault}{\color[rgb]{0,0,0}$K_{-1}$}%
}}}}
\put(3151,-1711){\makebox(0,0)[lb]{\smash{{\SetFigFont{12}{14.4}{\rmdefault}{\mddefault}{\updefault}{\color[rgb]{0,0,0}$\lambda_0$}%
}}}}
\put(2503,-2209){\makebox(0,0)[lb]{\smash{{\SetFigFont{12}{14.4}{\rmdefault}{\mddefault}{\updefault}{\color[rgb]{0,0,0}$K_0$}%
}}}}
\put(1960,-1963){\makebox(0,0)[lb]{\smash{{\SetFigFont{12}{14.4}{\rmdefault}{\mddefault}{\updefault}{\color[rgb]{0,0,0}$D_{-1}$}%
}}}}
\end{picture}%

%% file: NewModifiedConfiguration2.eps_t
\begin{picture}(0,0)%
\includegraphics{NewModifiedConfiguration2.eps}%
\end{picture}%
\setlength{\unitlength}{3947sp}%
\begingroup\makeatletter\ifx\SetFigFont\undefined%
\gdef\SetFigFont#1#2#3#4#5{%
  \reset@font\fontsize{#1}{#2pt}%
  \fontfamily{#3}\fontseries{#4}\fontshape{#5}%
  \selectfont}%
\fi\endgroup%
\begin{picture}(2574,6774)(1381,-6148)
\put(3448,-2116){\makebox(0,0)[lb]{\smash{{\SetFigFont{12}{14.4}{\rmdefault}{\mddefault}{\updefault}{\color[rgb]{0,0,0}$K_1$}%
}}}}
\put(1546, 14){\makebox(0,0)[lb]{\smash{{\SetFigFont{12}{14.4}{\rmdefault}{\mddefault}{\updefault}{\color[rgb]{0,0,0}$K_{-3}$}%
}}}}
\put(3151,-1711){\makebox(0,0)[lb]{\smash{{\SetFigFont{12}{14.4}{\rmdefault}{\mddefault}{\updefault}{\color[rgb]{0,0,0}$\lambda_0$}%
}}}}
\put(2875,-2326){\makebox(0,0)[lb]{\smash{{\SetFigFont{12}{14.4}{\rmdefault}{\mddefault}{\updefault}{\color[rgb]{0,0,0}$I_0$}%
}}}}
\put(1888,-2338){\makebox(0,0)[lb]{\smash{{\SetFigFont{12}{14.4}{\rmdefault}{\mddefault}{\updefault}{\color[rgb]{0,0,0}$K_L$}%
}}}}
\put(1585,-1279){\makebox(0,0)[lb]{\smash{{\SetFigFont{12}{14.4}{\rmdefault}{\mddefault}{\updefault}{\color[rgb]{0,0,0}$\lambda_2$}%
}}}}
\put(1873,-646){\makebox(0,0)[lb]{\smash{{\SetFigFont{12}{14.4}{\rmdefault}{\mddefault}{\updefault}{\color[rgb]{0,0,0}$\partial\Omega_1^-$}%
}}}}
\put(3229,-1111){\makebox(0,0)[lb]{\smash{{\SetFigFont{12}{14.4}{\rmdefault}{\mddefault}{\updefault}{\color[rgb]{0,0,0}$\Gamma_1$}%
}}}}
\put(2359,344){\makebox(0,0)[lb]{\smash{{\SetFigFont{12}{14.4}{\rmdefault}{\mddefault}{\updefault}{\color[rgb]{0,0,0}$C_\infty$}%
}}}}
\put(2554,-1012){\makebox(0,0)[lb]{\smash{{\SetFigFont{12}{14.4}{\rmdefault}{\mddefault}{\updefault}{\color[rgb]{0,0,0}$\lambda_1$}%
}}}}
\put(1948,-997){\makebox(0,0)[lb]{\smash{{\SetFigFont{12}{14.4}{\rmdefault}{\mddefault}{\updefault}{\color[rgb]{0,0,0}$I_1$}%
}}}}
\put(2569,-1486){\makebox(0,0)[lb]{\smash{{\SetFigFont{12}{14.4}{\rmdefault}{\mddefault}{\updefault}{\color[rgb]{0,0,0}$C$}%
}}}}
\put(2005,-1567){\makebox(0,0)[lb]{\smash{{\SetFigFont{12}{14.4}{\rmdefault}{\mddefault}{\updefault}{\color[rgb]{0,0,0}$\partial\Omega_1^+$}%
}}}}
\end{picture}%

%% file: section6-2.tex
The modification described in \S~\ref{sec:dual} leads to a new
inequality that must be satisfied for \(\lambda\in K_{-3}\).
Moreover, the numerical computations in \S~\ref{sec:GenusTwoNumerics}
indicate that the failure of the genus-two ansatz is due to the
violation of this new inequality. We then expect, similarly as in the
case of the transition from genus zero to genus two across the
primary caustic, that the pinching off of the inequality region
\(\Re(\phi(\lambda)-2i\theta^0(\lambda))<0\) leads to the birth of a
new band and a phase transition.

\begin{figure}[htbp]
\begin{center}
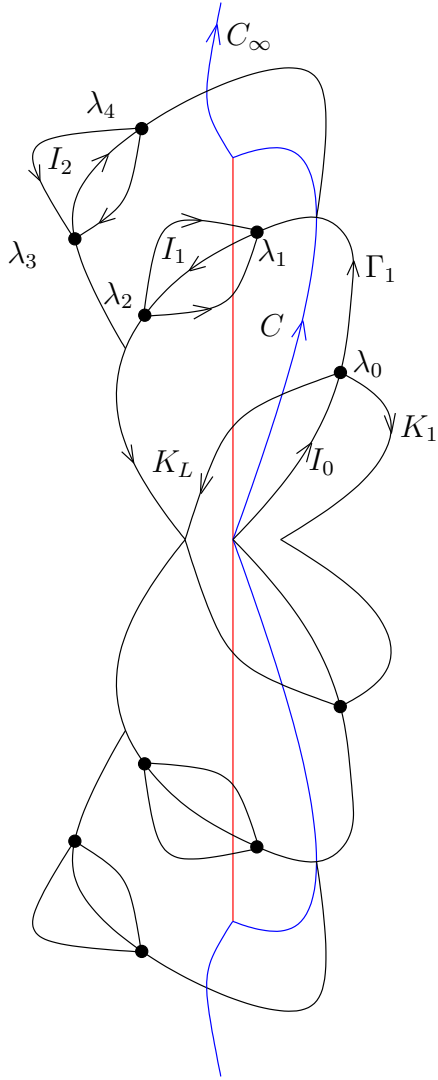
\end{center}
\caption{\em The discontinuity contours for ${\bf P}(\lambda)$ in
the genus-four configuration with a band on the outer contour lobe $K_{-3}$.}
\label{fig:Genus4}
\end{figure}

In this section we describe the modifications to the endpoint
equations that arise when we allow a band to lie on the arc $K_{-3}$.
In particular, we suppose that there is a single band in each of the
arcs $K_0$, $K_{-2}$, and $K_{-3}$, and we develop equations 
corresponding to \eqref{eq:momentsG} and
\eqref{eq:bandints}--\eqref{eq:gapints} in this special case. The
resulting contours of discontinuity for ${\bf P}(\lambda)$ in this
situation are illustrated in Figure~\ref{fig:Genus4}.
To begin, consider the analytic matrix \(\mathbf{M}(\lambda)\) obtained as in
\S~\ref{sec:dual} from the triple interpolation procedure. As before
we introduce the function $g(\lambda)$, but now we suppose that
$g(\lambda)$ is analytic except on $K_0,K_{-1},K_{-2}$, \emph{and}
$K_{-3}$ along with the complex conjugates of these arcs. We then
define a new unknown via
\(\mathbf{N}(\lambda):=\mathbf{M}(\lambda)e^{-g(\lambda)\sigma_3/\hbar}\).
All the jump conditions relating boundary values are as before with
the exception of the condition regarding boundary values on
$K_{-3}$. In place of \eqref{eq:jumpNKm3}, we find
\begin{equation}
\mathbf{N}_+(\lambda)=\mathbf{N}_-(\lambda)
\begin{bmatrix}
e^{-[g_+(\lambda)-g_-(\lambda)]/\hbar} & 0\\
iP(\lambda)e^{[2iQ(\lambda)-3i\theta^0(\lambda)-g_+(\lambda)-g_-(\lambda)]/\hbar}
  & e^{[g_+(\lambda)-g_-(\lambda)]/\hbar}
\end{bmatrix}.
\label{eq:newjumpNKm3}
\end{equation}
Continuing to follow the steps of our earlier analysis, we
arrive at a Riemann-Hilbert Problem for an unknown matrix
\(\mathbf{O}(\lambda)\) which has no jump discontinuities on the real
intervals $(-\epsilon,0)$ and $(0,\epsilon)$, satisfies the
jump conditions \eqref{eq:OjumpK0}--\eqref{eq:OjumpKm1},
\eqref{eq:jumpOKm2}, and also satisfies 
\begin{equation}
\mathbf{O}_+(\lambda)=\mathbf{O}_-(\lambda)
\begin{bmatrix}
e^{-[g_+(\lambda)-g_-(\lambda)]/\hbar} & 0\\
iS(\lambda)e^{[2iQ(\lambda)+L(\lambda)
-i\theta^0(\lambda)-g_+(\lambda)-g_-(\lambda)-2i\theta^0(\lambda)]/\hbar}
  & e^{[g_+(\lambda)-g_-(\lambda)]/\hbar}
\end{bmatrix},\quad\text{for}\,\lambda\in K_{-3}.
\label{eq:newjumpOKm3}
\end{equation}
From \eqref{eq:newjumpOKm3}, the band
and gap conditions for the arc \(K_{-3}\) are slightly different from
those in \eqref{eq:bandconditions} and \eqref{eq:gapconditions} which
continue to hold in the arcs $K_0,K_{-1}$, and $K_{-2}$. Namely, for a
band in \(K_{-3}\), we require
\begin{equation}
\phi(\lambda)-2i\theta^0(\lambda)\equiv\,\text{imaginary constant},\quad
\theta(\lambda)=\,
\text{real decreasing function,}
\label{eq:newbandcondition}
\end{equation}
while for a gap in \(K_{-3}\), there must hold
\begin{equation}
\Re(\phi(\lambda)-2i\theta^0(\lambda))<0,\quad \theta(\lambda)\equiv\,\text{real constant}.
\label{eq:newgapcondition}
\end{equation}
Then, we note that $g'(\lambda)$ must satisfy \eqref{eq:gprimebands}
in the bands not on \(K_{-3}\), \eqref{eq:gprimegaps} in all gaps,
\eqref{eq:gprimesymmetry}, \eqref{eq:gprimedecay}, and 
\begin{equation}
g_+'(\lambda)+g_-'(\lambda)  =
2iQ'(\lambda)+L'(\lambda)-3i\theta^{0\prime}(\lambda),\quad\text{for 
$\lambda$ in the band on $K_{-3}$}.\label{eq:newgprimebandKm3}
\end{equation}
These conditions 
amount to a scalar Riemann-Hilbert Problem for
$g'(\lambda)$. We (re)define $C_f$ to be the contour $K_0\cup K_{-1}\cup
K_{-2}\cup (-\epsilon,0)\cup C$, and we define $f(\lambda)$ as before
by \eqref{eq:fU} and \eqref{eq:ffromfU}. We then make the change of
 variables
\begin{equation}
h(\lambda):=g'(\lambda)+f(\lambda).
\label{eq:hgprimef}
\end{equation}
We see that the function $h(\lambda)$ is characterized by
\eqref{eq:hbands} for bands not in \(K_{-3}\),
\eqref{eq:hgaps}--\eqref{eq:hdecay}, and
\begin{equation}
h_+(\lambda)+h_-(\lambda)=2iQ'(\lambda)-3i\theta^{0\prime}(\lambda),\quad\text{in the
  band on $K_{-3}$}.\label{eq:hbands2}
\end{equation}
We have thus traded our scalar Riemann-Hilbert Problem to determine
the unknown
$g'(\lambda)$ for one in which the unknown is $h(\lambda)$. We suppose that the
endpoints of the bands
\[\lambda_0,\lambda_1,\ldots,\lambda_4\]
are known, and we define $R(\lambda)$ by
\begin{equation}
R(\lambda)^2=\prod_{n=0}^4(\lambda-\lambda_n)(\lambda-\lambda_n^*),
\label{eq:genus4Rdef}
\end{equation}
and the conditions (i) the branch cuts are the bands in $K_0$, $K_{-2}$,
and $K_{-3}$ and (ii) $R(\lambda)=\lambda^5+O(\lambda^4)$ as
$\lambda\to\infty$.  We denote the bands by $I_0$, $I_1$, and $I_2$.
Note that $I_2$ has the same orientation as $K_{-3}$.  Finally, we are
able to solve for $h(\lambda)$ by writing it as the product of
$R(\lambda)$ and a new unknown function $k(\lambda)$ as in
\eqref{eq:hRk}. From the definition of $R(\lambda)$, the conditions on
$h(\lambda)$ translate to conditions on $k(\lambda)$ as in
\eqref{eq:kcond1}--\eqref{eq:kdecay} with
\begin{equation}
k_+(\lambda)-k_-(\lambda)=\frac{2iQ'(\lambda)-3i\theta^{0\prime}(\lambda)}{R_+(\lambda)},
\quad\text{for $\lambda$ in the
  band on $K_{-3}$}.\label{eq:kbands2}
\end{equation}
Now, using the differences of boundary values of $k(\lambda)$ in
\eqref{eq:kcond1}--\eqref{eq:kcond3} and \eqref{eq:kbands2}, 
we can write down an explicit formula for $k(\lambda)$.
We define \(k_U(\lambda):=k_U^{(1)}(\lambda)+k_U^{(2)}(\lambda)\) where
\begin{align}
k_U^{(1)}(\lambda)&:=\frac{1}{\pi}
\int_{\text{bands}\subset C_f}
\frac{Q'(\eta)}{R_+(\eta)(\eta-\lambda)}\,d\eta
+\frac{1}{\pi}\int_{I_2}
\frac{Q'(\eta)-\frac{3}{2}\theta^{0\prime}(\eta)}{R_+(\eta)(\eta-\lambda)}\,d\eta,
\label{eq:newKu1def} \\
k_U^{(2)}(\lambda)&:=-\frac{1}{2i}
\int_{C_f\setminus\text{bands}}\frac{d\eta}{R(\eta)(\eta-\lambda)}.
\label{eq:newKu2def}
\end{align}
We note that we can rewrite $k_U^{(1)}$ as 
\begin{equation*}
k_U^{(1)}(\lambda)=\frac{1}{\pi}\int_{\text{bands}}
\frac{Q'(\eta)}{R_+(\eta)(\eta-\lambda)}\,d\eta\underbrace{-
\frac{3}{2\pi}\int_{I_2}
\frac{\theta^{0\prime}(\eta)}{R_+(\eta)(\eta-\lambda)}\,d\eta}_{\mathcal{I}_U(\lambda)}
\end{equation*}
Finally, $k(\lambda)$ must have the following form:  
\begin{equation}
k(\lambda):=k_U(\lambda)-k_U(\lambda^*)^*.
\label{eq:kdef}
\end{equation}
A residue calculation shows that 
\begin{equation*}
k_U^{(1)}(\lambda)-k_U^{(1)}(\lambda^*)^*
=\frac{iQ'(\lambda)}{R(\lambda)}+\mathcal{I}_U(\lambda)-\mathcal{I}_U(\lambda^*)^*,
\end{equation*}
hence
\begin{equation}
k(\lambda)=\frac{iQ'(\lambda)}{R(\lambda)}
+\mathcal{I}_U(\lambda)-\mathcal{I}_U(\lambda^*)^*+k_U^{(2)}(\lambda)-k_U^{(2)}(\lambda^*)^*.
\label{eq:newkformula}
\end{equation}
Finally, we derive the analogues of \eqref{eq:momentsG} by enforcing
the decay condition \eqref{eq:kdecay}. We expand the integrals
appearing in \eqref{eq:newkformula} for large $\lambda$.  We note that
$\theta^{0\prime}(\lambda)\equiv i\pi$, so
\begin{equation*}
\mathcal{I}_U(\lambda)=
\sum_{p=0}^5
\left[
\frac{3i}{2}\int_{I_2}\frac{\eta^p}{R_+(\eta)}\,d\eta
\right]\lambda^{-1-p}+O(\lambda^{-7})\quad\text{as}\;\lambda\to\infty.
\end{equation*}
Starting from the notation \eqref{eq:Mpdef}, we introduce modified
moments
\begin{equation}
\tilde{M}_p(\lambda_0,\ldots,\lambda_4):=M_p(\lambda_0,\ldots,\lambda_4) -
3\Re\left(\int_{I_2}\frac{\eta^p}{R_+(\eta)}\,d\eta\right)\,,
\end{equation}
and we thus find that
\begin{equation}
0=\tilde{M}_p(\lambda_0,\ldots,\lambda_4)\,,
\quad\text{for \(p=0,1,2\)},
\label{eq:g4moment012}
\end{equation}
\begin{equation}
2t=\tilde{M}_3(\lambda_0,\ldots,\lambda_4)\,,
\end{equation}
\begin{equation}
x+2t\sum_{n=0}^4a_n=\tilde{M}_4(\lambda_0,\ldots,\lambda_4)\,,
\end{equation}
\begin{equation}
x\sum_{n=0}^4a_n+2t\left(\sum_{n=0}^4\left(a_n^2-\frac{1}{2}b_n^2\right)
+\sum_{n=0}^4\sum_{m=n+1}^4a_ma_n\right)=
\tilde{M}_5(\lambda_0,\ldots,\lambda_4)\,.
\label{eq:g4moment5}
\end{equation}
These equations amount to real constraints that must be satisfied by
the 5 complex endpoints in the upper half plane.

To ensure that \(\Im(\theta(\lambda))=0\) in the gaps, we require as
before that (recall that $A_n$ is a small counter-clockwise oriented
closed contour surrounding the band $I_n$)
\begin{equation}
\Re\left(\oint_{A_n}g'(\eta)\,d\eta\right)=0,\quad n=1,2.
\label{eq:g4reality12}
\end{equation}
Additionally, we need to enforce that \(\Re(\phi(\lambda))=0\) for
$\lambda$ in the
bands $I_0$ and $I_1$ and that \(\Re(\phi(\lambda)-2i\theta^0(\lambda))=0\) for
$\lambda\in I_2$. The condition
\begin{equation}
\Re\left(\int_{\lambda_0}^{\lambda_1}[2g'(\eta)-2iQ'(\eta)-L'(\eta)+i\theta^{0\prime}(\eta)]\,d\eta\right)=0\,,
\label{eq:g4vanish1}
\end{equation}
where the contour of integration is arbitrary in the domain of
analyticity of the integrand, will guarantee that $\phi(\lambda)$ is
purely imaginary in the band $I_1$. On the other hand, we know that
\(\phi(\lambda)-2i\theta^0(\lambda)\) is guaranteed to be constant on
$I_2$ by our construction. To ensure that this constant has zero real
part, we require that the net change of the real part of
$\phi(\lambda)$ across the second gap is equal to the real part of
\(2i\theta^0(\lambda_3)\). The second vanishing condition is
\begin{equation}
\Re\left(\int_{\lambda_2}^{\lambda_3}
[2g'(\eta)-2iQ'(\eta)-L'(\eta)+i\theta^{0\prime}(\eta)]\,d\eta-2i\theta^0(\lambda_3)\right)
=0\,,
\label{eq:g4vanish2}
\end{equation}
where again the path of integration is arbitrary within the domain of
analyticity of the integrand.  Together with
\eqref{eq:g4moment012}--\eqref{eq:g4moment5}, equations
\eqref{eq:g4reality12}--\eqref{eq:g4vanish2} make up 10 real equations
for the real and imaginary parts of the endpoints
\(\lambda_0,\ldots,\lambda_4\).

The conditions \eqref{eq:g4reality12}--\eqref{eq:g4vanish2} can be
written in a simpler form that is useful for subsequent calculations.
Define
\begin{equation}
\tilde{Y}_U(\lambda):=Y_U(\lambda) + i\oint_{A_2}
\frac{d\eta}{R(\eta)(\eta-\lambda)}\,,
\end{equation}
and then set
$\tilde{Y}(\lambda):=\tilde{Y}_U(\lambda)-\tilde{Y}_U(\lambda^*)^*$.
Therefore $\tilde{Y}(\lambda)$ is a function that is analytic where
$Y(\lambda)$ is with the exception of the contours $A_2$ and $A_2^*$.
Then, the conditions \eqref{eq:g4reality12} are equivalent to
\begin{equation}
\tilde{R}_n(\lambda_0,\ldots,\lambda_4):=\Re\left(\int_{\lambda_{2n-1}}^{\lambda_{2n}}
R(\eta)\tilde{Y}(\eta)\,d\eta\right) = 0\,,
\end{equation}
for $n=1$ and $n=2$.
Similarly, the condition \eqref{eq:g4vanish1} is equivalent to
\begin{equation}
\tilde{V}_1(\lambda_0,\ldots,\lambda_4):=\Re\left(\int_{\lambda_{2n-2}}^{\lambda_{2n-1}}R(\eta)\tilde{Y}(\eta)\,d\eta\right) = 0\,.
\end{equation}
In all of these formulae, the contour of integration is arbitrary as
long as it lies within the domain of analyticity of the integrand.
Let $\lambda_M$ denote a point on $A_2$.  Then, the condition
\eqref{eq:g4vanish2} may be written as
\begin{equation}
\tilde{V}_2(\lambda_0,\ldots,\lambda_4):=
\Re\left(\int_{\lambda_2}^{\lambda_M}R(\eta)\tilde{Y}(\eta)\,d\eta +
\int_{\lambda_M}^{\lambda_3}R(\eta)\tilde{Y}(\eta)\,d\eta + 2i\theta^0(\lambda_2) - 2\pi\int_{\lambda_2}^{\lambda_M}d\eta\right) = 0\,.
\end{equation}
Here the paths of integration lie in each case within the domain of
analyticity of the integrand.

%% file: Genus4.eps_t
\begin{picture}(0,0)%
\includegraphics{Genus4.eps}%
\end{picture}%
\setlength{\unitlength}{3947sp}%
\begingroup\makeatletter\ifx\SetFigFont\undefined%
\gdef\SetFigFont#1#2#3#4#5{%
  \reset@font\fontsize{#1}{#2pt}%
  \fontfamily{#3}\fontseries{#4}\fontshape{#5}%
  \selectfont}%
\fi\endgroup%
\begin{picture}(2970,6774)(985,-6148)
\put(3448,-2116){\makebox(0,0)[lb]{\smash{{\SetFigFont{12}{14.4}{\rmdefault}{\mddefault}{\updefault}{\color[rgb]{0,0,0}$K_1$}%
}}}}
\put(3151,-1711){\makebox(0,0)[lb]{\smash{{\SetFigFont{12}{14.4}{\rmdefault}{\mddefault}{\updefault}{\color[rgb]{0,0,0}$\lambda_0$}%
}}}}
\put(2875,-2326){\makebox(0,0)[lb]{\smash{{\SetFigFont{12}{14.4}{\rmdefault}{\mddefault}{\updefault}{\color[rgb]{0,0,0}$I_0$}%
}}}}
\put(1888,-2338){\makebox(0,0)[lb]{\smash{{\SetFigFont{12}{14.4}{\rmdefault}{\mddefault}{\updefault}{\color[rgb]{0,0,0}$K_L$}%
}}}}
\put(1585,-1279){\makebox(0,0)[lb]{\smash{{\SetFigFont{12}{14.4}{\rmdefault}{\mddefault}{\updefault}{\color[rgb]{0,0,0}$\lambda_2$}%
}}}}
\put(3229,-1111){\makebox(0,0)[lb]{\smash{{\SetFigFont{12}{14.4}{\rmdefault}{\mddefault}{\updefault}{\color[rgb]{0,0,0}$\Gamma_1$}%
}}}}
\put(2359,344){\makebox(0,0)[lb]{\smash{{\SetFigFont{12}{14.4}{\rmdefault}{\mddefault}{\updefault}{\color[rgb]{0,0,0}$C_\infty$}%
}}}}
\put(2554,-1012){\makebox(0,0)[lb]{\smash{{\SetFigFont{12}{14.4}{\rmdefault}{\mddefault}{\updefault}{\color[rgb]{0,0,0}$\lambda_1$}%
}}}}
\put(1948,-997){\makebox(0,0)[lb]{\smash{{\SetFigFont{12}{14.4}{\rmdefault}{\mddefault}{\updefault}{\color[rgb]{0,0,0}$I_1$}%
}}}}
\put(2569,-1486){\makebox(0,0)[lb]{\smash{{\SetFigFont{12}{14.4}{\rmdefault}{\mddefault}{\updefault}{\color[rgb]{0,0,0}$C$}%
}}}}
\put(985,-1027){\makebox(0,0)[lb]{\smash{{\SetFigFont{12}{14.4}{\rmdefault}{\mddefault}{\updefault}{\color[rgb]{0,0,0}$\lambda_3$}%
}}}}
\put(1477,-82){\makebox(0,0)[lb]{\smash{{\SetFigFont{12}{14.4}{\rmdefault}{\mddefault}{\updefault}{\color[rgb]{0,0,0}$\lambda_4$}%
}}}}
\put(1234,-436){\makebox(0,0)[lb]{\smash{{\SetFigFont{12}{14.4}{\rmdefault}{\mddefault}{\updefault}{\color[rgb]{0,0,0}$I_2$}%
}}}}
\end{picture}%